\documentclass[aps,reprint,groupedaddress,floatfix,nofootinbib]{revtex4-1}
\usepackage{float}
\usepackage[final]{graphicx}
\usepackage{natbib}
\usepackage{hyperref}
\usepackage{amssymb}
\usepackage{amsmath}
\usepackage{mathrsfs}
\usepackage{xcolor}
\usepackage{bm}
\makeatletter

\providecommand{\U}[1]{\protect\rule{.1in}{.1in}}

\begin{document}
\title{Interplay between orbital-quantization effects and the Fulde-Ferrell-Larkin-Ovchinnikov instability in multiple-band layered superconductors}
	\author{Kok Wee Song}
\affiliation{
	Materials Science Division,
	Argonne National Laboratory,
	9700 South Cass Avenue, Lemont, Illinois 60639, USA
}
\author{Alexei E. Koshelev}
\affiliation{
	Materials Science Division,
	Argonne National Laboratory,
	9700 South Cass Avenue, Lemont, Illinois 60639, USA
}
\date{\today}

\begin{abstract}
	We explore superconducting instability for a clean two-band layered superconductor with deep and shallow bands in the magnetic field applied perpendicular to the layers. In the shallow band, the quasiclassical approximation is not applicable, and Landau quantization has to be accounted for exactly. 
	The electronic spectrum of this band in the magnetic field is composed of the one-dimensional Landau-level minibands. With increasing magnetic field the system experiences a series of Lifshitz transitions when the chemical potential enters and exits the minibands. These transitions profoundly influence the shape of the upper critical field at low temperatures.
	In addition, the Zeeman spin splitting may cause the nonuniform state with interlayer modulation of the superconducting order parameter [Fulde-Ferrell-Larkin-Ovchinnikov (FFLO) state]. Typically, the quantization effects in the shallow band strongly promote the formation of this state. The uniform state remains favorable only in the exceptional resonance cases when the spin-splitting energy exactly matches the Landau-level spacing. Furthermore, for specific relations between electronic spectrum parameters, the alternating FFLO state may realize, in which the order parameter changes sign between the neighboring layers. For all above cases, the reentrant high-field superconducting states may emerge at low temperatures if the shallow band has significant contribution to the Cooper pairing. 
\end{abstract}

\maketitle

\section{Introduction}

The nature of superconducting instability in the magnetic field is a long-standing fundamental problem. In clean type-II superconductors, the upper critical field $H_{C2}$ and its temperature dependence are sensitive to the electronic band properties as well as to the gap structure which contains crucial information on how the correlated Cooper pairs are formed. The magnitude of $H_{C2}$ is mostly determined by the suppression of superconductivity due to the quasiparticles orbital motion and the Zeeman spin splitting of the Fermi surface by the magnetic field.
The problem of orbital $H_{C2}$ for a single-band system was solved in the seminal papers 
\cite{Helfand:PRev.1966,*Werthamer:PRev147.1966} within the quasiclassical approximation
which neglects the Landau quantization of the orbital motion. This approach works with high accuracy in common metals with large Fermi energies and describes very well most of known conventional superconductors.
Nevertheless, the exact Landau quantization calculation was discussed in Refs.\ \onlinecite{Gunther:SSComm4.1966,*Gruenberg:PRev176.1968,Rajagopal:PLett23.1966} shortly after the quasiclassical work and the topic was further elaborated in great details later\cite{Tesanovic:PRB39.1989,*Tesanovic:PRB43.1991,Reick:PhysicaC170.1990,MacDonald:PRB45.1992,NormanBook92,Maniv:PRB46.1992,Maniv:RMP73.2001}.

Landau-quantization effects are most relevant when the Fermi energy is comparable to the cyclotron frequency so that only a few lowest Landau levels are occupied. Such extreme quantum limit has not been viable in most of known superconducting materials. The situation changed with the recent discovery of the iron-based superconductors (FeSC). It is very common in FeSCs that some of the bands have very small Fermi energies and they can be driven through the Lifshitz transition by chemical dopings\cite{Liu:Nat6.2010,Xu:PhysRevB.2013,Miao:NatComm6.2015,Shi:NatComm8.2017} and, importantly, these materials usually remain superconducting after one of the bands is completely depleted.
As the Fermi energy of the shallow band may be smaller than the pairing energy scale, the proper treatment of this band requires revision of the conventional BCS approach. Such situation was first considered in the context of shape resonances in confined superconductors (thin films and nanowires) \cite{Blatt:PRL10.1963,*Kresin:JETP50.1966,Shanenko:PRB73.2006,*Croitoru:PRB76.2007,CarigliaJSNM2016,ValentinisPhysRevB16}
The influence of the shallow band on the superconducting transition temperature  in multiple-band materials near Lifshitz transition has been considered in Ref.\ \cite{Innocenti:PhysRevB.2010} and, more recently, 
physics of FeSCs motivated detailed investigations of this problem   
\cite{Bang:NJP.2014,Koshelev:PRB90.2014,Chen:PRB92.2015,Valentinis:PRB94.2016,Chubukov:PRB93.2016}. 
It was demonstrated that, in spite of low quasiparticle density, shallow bands may strongly influence the Cooper pairing. 
Another key property facilitating the extreme quantum limit near superconducting instability of FeSCs is that they are characterized by the very high upper critical fields, up to 100 T.

As the orbital upper critical field is inversely proportional to the Fermi energy, it is strongly affected by the shallow bands. Moreover, in the vicinity of the Lifshitz transition, the extreme quantum limit can be reached for these bands near $H_{C2}$ meaning that the quasiclassical consideration cannot be used. 
It has been indeed demonstrated that the Landau quantization has a dramatic effect on superconducting instability in the magnetic field for two-dimensional materials near the Lifshitz transition\cite{Song:PRB95.2017}. The most spectacular prediction is the emergence of the pronounced reentrant states at the magnetic field corresponding to the matching of the Landau levels with the chemical potential.

In addition to the orbital effect, shallow bands also promote the Zeeman spin-splitting suppression of superconductivity. The relative role of the spin and orbital mechanisms is usually characterized by the Maki parameter $\alpha_M$ defined as  $\sqrt{2}H_{C2}^O/H_{C2}^P$, where $H_{C2}^O$ and $H_{C2}^P$ are the orbital and spin upper critical fields, respectively.  In a clean isotropic single-band material with the free-electron $g$ factor, the Maki parameter is proportional to the ratio of the superconducting gap $\Delta$ and the Fermi energy $\epsilon_F$, $\alpha_M\!=\!\pi^2\Delta/4\epsilon_F$.
In this case, unless the band is very shallow, $\alpha_M$ is very small, meaning that the orbital effect strongly dominates. The Maki parameter is greatly enhanced in special cases of weak orbital effect such as quasi-one-dimensional materials and layered superconductors for magnetic field directed along the layers.

An important consequence of the strong Zeeman effect (i.e., large $\alpha_M$) is the emergence of the nonuniform Fulde-Ferrell-Larkin-Ovchinnikov (FFLO) state in very clean materials \cite{Fulde:PRev135.1964,Larkin:JETP20.1965}. In this state, the order parameter is periodically modulated which allows for a gain in the Zeeman energy which may exceed the kinetic-energy loss due to the nonzero center-of-mass momentum pairing. Rich physics of the FFLO state has been extensively investigated in many subsequent theoretical studies within quasiclassical approximation \cite{Gruenberg:PRL16.1966,TakadaPrThPhys1969,Machida:PRB30.1984,BurkhardtAnnPhys94,ShimaharaPhysRevB.50.12760,*Shimahara:JPSJ66.1997,*Shimahara:PRB80.2009,HouzetEPL00,HouzetPhysRevB01,Budzin:PRL95.2005,DaoPhysRevB.87.174509,CroitoruPhysRevB.89.224506}. 
The FFLO instability may only realize in very pure materials with weak scattering of quasiparticles.
That is why,
even though the FFLO state was predicted more than half century ago, the experimental data consistent with this state have been reported only relatively recently in organic superconductors \cite{BeyerLTP13,AgostaPhysRevLett.118.267001} and, less convincingly, in the heavy-fermion compounds \cite{Kumagai:PRL97.2006,Matsuda:JPSJ76.2007}.
The conditions for the FFLO instability with modulation along the magnetic field in \emph{isotropic} material were studied by Gruenberg and Gunther \cite{Gruenberg:PRL16.1966} within the quasiclassical framework. They found that the nonuniform state appears only for very large Maki parameters, $\alpha_M> 1.8$, which is unlikely to realize in any single-band isotropic material. That is why most FFLO studies have been focused on quasi-low-dimensional materials with strongly reduced orbital effects. 

The FFLO state may realize in the iron-based superconductors due to the presence of shallow bands and huge upper critical fields which are likely to be limited by the Zeeman effect. 
This motivated recent investigations of the conditions for the emergence of this state in multiple-band materials within the quasiclassical approach in different situations \cite{Gurevich:PRB82.2010,Takeshi:JPSJ83.2014,*Takahashi:PRB89.2014,AdachiJPSJ15,Ptok:JLowT172.2013,*Ptok:EPJB87.2014,*Ptok:JPhysCM27.2015,*Ptok:NJPhys19.2017}. However, in the presence of very shallow bands, this approach may be insufficient and quantization effects have to be accounted for.

In this paper, we investigate the superconducting instability for a two-band layered material in the magnetic field applied perpendicular to the layers. We consider the case of the material near the Lifshitz transition when one of the bands is very shallow. In this case, the Landau-quantization effects strongly influence the formation of the superconducting state. Even though our consideration is motivated by physics of iron-based superconductors, it is very general, and our goal is not to describe  any particular compound but, instead, to develop a general understanding of how the orbital-quantization effects influence the onset of superconductivity in clean two-band layered materials.  The major new feature in comparison with a pure two-dimensional case \cite{Song:PRB95.2017} is that, due to the large Zeeman effect in the shallow band, this system is prone to the formation of the FFLO state with interlayer modulation of the order parameter. The quantum effects have a profound influence on this FFLO instability.

The interlayer tunneling lifts all the degenerate Landau levels to dispersive minibands along the out-of-plane momentum direction. As a consequence, the system experiences series of Lifshitz transitions with increasing the magnetic field corresponding to crossing of the chemical potential with the miniband edges. Every Landau-level miniband has two van Hove singularities at the reduced $z$-axis momentums $k_z=0$ and $\pi$, at which the density of states (DOS) is enhanced. In special situations, when two such singular points for spin-up and spin-down bands simultaneously match the Fermi level, the pairing strongly enhanced. Two distinct cases of such resonance matching are possible. The first well-known case is realized when spin-splitting energy is equal to the Landau-level spacing \cite{Maniv:PRB46.1992,NormanBook92}. In this case, the van Hove points of the same kind (either $k_z=0$ or $\pi$) may match leading to the standard uniform superconducting state. The second case corresponds to matching of the opposite van Hove points, e.~g., spin-down/$k_z=0$ and spin-up/$k_z=\pi$ points, which may occur only for a certain relation between the electronic band parameter. In this situation, the \textit{alternating} FFLO state may emerge, in which the superconducting order parameter changes sign between the neighboring layers.  These matching effects may generate the high-field reentrant superconducting states which are somewhat less pronounced than for two-dimensional case \cite{Song:PRB95.2017} due to the DOS spreading by the interlayer tunneling.

On the other hand, the Landau-level spreading somewhat mitigates the Zeeman pair-breaking effect, 
since the dispersive spin-up and -down minibands can cross the Fermi level simultaneously within a finite energy range for arbitrary spin splitting. In such generic situation, the shallow band favors the formation of the FFLO state, in which the optimal modulation wave vector equals to the difference between the spin-up and -down Fermi momenta. Such FFLO instability leads to different kind of reentrant states with the field-dependent modulation wave vectors. In contrast to the reentrant states caused by the matching of the van Hove singularities, the latter states can extend over a broad magnetic-field range.

This paper is organized as follows. In Sec.\ \ref{sec:model}, we introduce the microscopic Hamiltonian describing a two-band layered superconductor the in magnetic field and derive the corresponding linearized gap equations. In Sec.\ \ref{sec:TC}, we briefly discuss these equations for the transition temperature in zero magnetic field providing the reference for the further investigation of instability in finite magnetic field. 
In Sec.\ \ref{HC2eqn}, we derive the equations for the upper critical field $H_{C2}$ and evaluate the pairing kernels in these equations. 
In Sec.\ \ref{sec:J1}, we discuss the dependences of the quantum pairing kernel on relevant parameters.
In Sec.\ \ref{sec:HT}, we present the typical magnetic field versus temperature phase diagrams. We conclude the paper in Sec.\ \ref{Sec:Summary}.

\section{The model of a two-band layered superconductor}\label{sec:model}

\begin{figure*}[htbp] 
	\centering
	\includegraphics[width=2.7in]{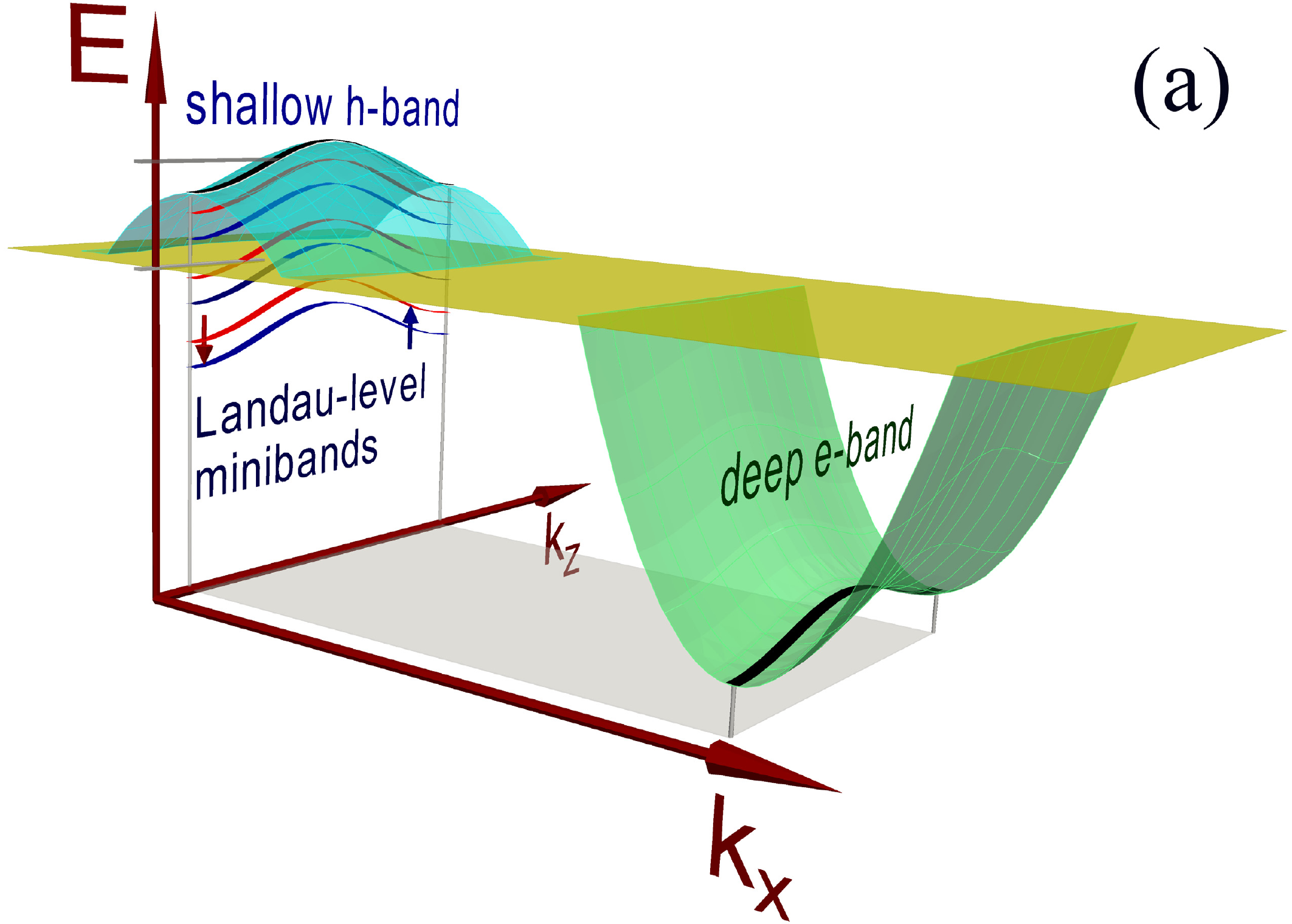} 
	\includegraphics[width=4.1in]{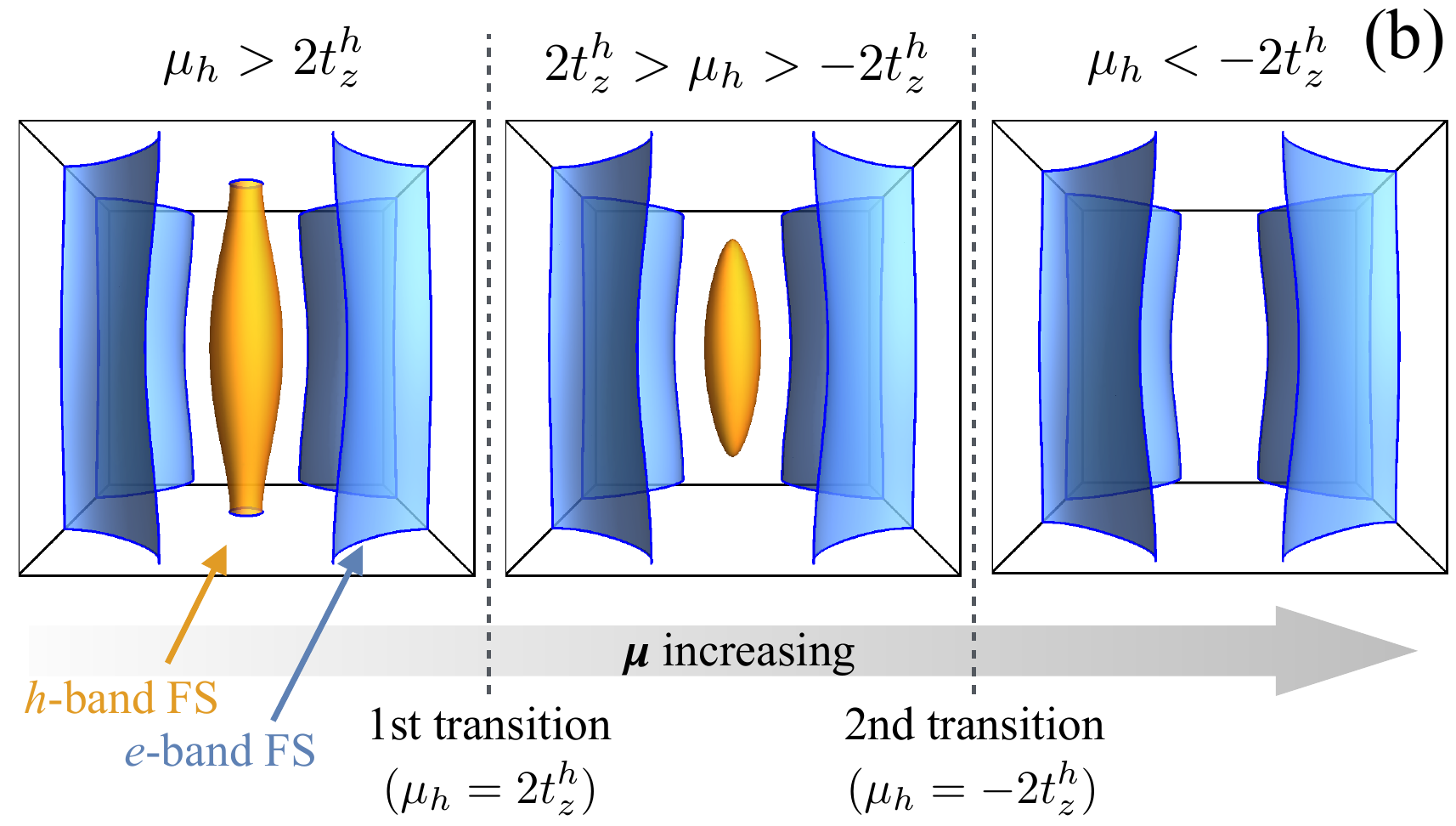}
	\caption{(a) The illustration of electronic spectrum for a layered superconductor with deep and shallow bands. (b) The schematic diagram demonstrating the evolution of the Fermi surfaces of the quasi-2D model in Eq.\ \eqref{eqn:modelH} near the Lifshitz transitions. The first Lifshitz transition occurs at $\mu_h=2t^h_z$, at which the neck at $k_z=0$ interrupts and the FS of the $h$ band changes from open to closed. The second Lifshitz transition occurs at $\mu_h=-2t^h_z$, where  the $h$ band becomes completely depleted.
}
	\label{fig:LT}
\end{figure*}

We will investigate the shallow-band effects in layered superconductors using the simple tight-binding Hamiltonian with only the nearest-neighbor interlayer hopping term,
\begin{align}
	&\mathcal{H}=\sum_{ j\alpha}\int\mathrm{d}^2\mathbf{r}\Bigg[c^\dagger_{\alpha j s}(\mathbf{r})(\varepsilon^\alpha(\hat{\mathbf{k}}) \sigma^0_{ss'}-\mu_zH\sigma^z_{ss'})c_{\alpha js'}(\mathbf{r})\notag\\
	&-t_z^\alpha\exp\left[\frac{\mathrm{i}e}{c}\!\int_{l_{j}}\!\mathrm{d}zA_z\right]c^\dagger_{\alpha j s}(\mathbf{r})c_{\alpha,j+1,s}(\mathbf{r})+\text{H.c.}\notag\\
	&-\sum_{\beta}U_{\alpha \beta}c^\dagger_{\alpha j\downarrow}(\mathbf{r})c^\dagger_{\alpha j\uparrow}(\mathbf{r})
				c_{\beta j\downarrow}(\mathbf{r})c_{\beta j\uparrow}(\mathbf{r})\Bigg],\label{eqn:modelH}
\end{align}
where $\mathbf{r}=(x, y)$ is the in-plane coordinate, $j$ is the layer index, $s$ represents spin, and $\alpha=e$ ($h$) represents the $e$-band ($h$-band). Furthermore, $t^\alpha_z$ is the (nearest) interlayer hopping energy, $\varepsilon^e(\hat{\mathbf{k}})=\hat{\mathbf{k}}^2/(2m_e)$ and $\varepsilon^h(\hat{\mathbf{k}})=-\hat{\mathbf{k}}^2/(2m_h)+\varepsilon_0$ are intralayer energy dispersions with the band masses $m_h$ and $m_e$ and the momentum operator $\hat{\mathbf{k}}=-\mathrm{i}\nabla_{\mathbf{r}}-e\mathbf{A}/c$. In the model, we also consider the Zeeman spin splitting and assume, for simplicity, that the band electron's magnetic moment, $\mu_z$, is the same in each band. 
In the second line of Eq.\ \eqref{eqn:modelH}, $\int_{l_{j}}=\int^{(j+1)a_z}_{j a_z}$ is the line integration along the out-of-plane direction and $a_z$ is the interlayer spacing (set to unity in the later calculation) and $A_z$ is the $z$-component of the vector potential. In zero magnetic field, the three-dimensional energy dispersions are $\epsilon^\alpha(\mathbf{k},k_z)=\varepsilon^\alpha(\mathbf{k})-2t^\alpha_z\cos k_z$, see Fig.\ \ref{fig:LT}(a).

In this paper, we limit ourselves to the case of the magnetic field applied in the $z$ direction. In this case, $A_z=0$ with standard gauge choices and the magnetic phase factor in Eq.\ \eqref{eqn:modelH} tight-binding terms drops out. For definiteness, we consider the case when the hole band is shallow meaning that the chemical potential $\mu$ is located near its edge $\mu_h=\varepsilon_0-\mu \ll \varepsilon_0$. We note that whether the shallow band is hole-like or electron-like 
does not have any influence on the results of this paper.
The two Lifshitz transitions in this model occur at $\mu_h\!=\!\pm 2t^h_z$ [see Fig.\ \ref{fig:LT}(b)] \cite{Lifshitz:JETP11.1960}. At $\mu_h= 2t^h_z$ the neck near $k_z=0$ interrupts and $\mu_h\!=\! -2t^h_z$ the hole band got completely depleted. 

To study the superconducting instabilities in the model Eq.\ \eqref{eqn:modelH}, we follow the standard approach and write the linearized gap equation as
\begin{equation}\label{eqn:gap-eqn}
\Delta^\alpha_{\mathbf{r}, j}\!=\!T\sum_{\omega_n}\sum_{\beta j' } \!U_{\alpha\beta}\int_{\mathbf{r}'}K^\beta_{\omega_n}(\mathbf{r} j,\mathbf{r}' j')\Delta^\beta_{\mathbf{r}',j'},
\end{equation}
where the gap function is defined as $\Delta^\alpha_{\mathbf{r}, j}\!=\!\sum_\beta U_{\alpha\beta}\left\langle c_{\beta, j\downarrow}(\mathbf{r})c_{\beta, j\uparrow}(\mathbf{r})\right\rangle$ and we used the notation $\int_{\mathbf{r}}=\int\mathrm{d}\mathbf{r}$, $\omega_n=2\pi T(n+1/2)$ are the Matsubara frequencies, and the kernel is 
\begin{equation}
K^\alpha_{\omega_n}\!(\mathbf{r} j,\mathbf{r}' j')
\!=\!G^\alpha_{\omega_n,+}(\mathbf{r} j,\mathbf{r}' j')G^\alpha_{-\omega_n,-}(\mathbf{r}' j',\mathbf{r} j),
\label{KernelDef}
\end{equation}
where $G^\alpha_{\omega_n,\pm}(\mathbf{r} j,\mathbf{r}' j')$ is the one-particle Green's function in the normal phase and the subscripts $+$ or $-$ describe spin orientation. Without magnetic field, the normal Green's functions can be expanded into the plane-wave basis 
\begin{equation}\label{G0}
G^\alpha_{\omega_n,\pm}(\mathbf{r} j,\mathbf{r}' j')=\sum_{\mathbf{k}k_z}\frac{\mathrm{e}^{-\mathrm{i}[\mathbf{k}\cdot\bm{\rho}+
k_z( j- j')]}}{\mathrm{i}\omega_n-\xi^\alpha_\mathbf{k}+2t_z^{\alpha}\cos k_z},
\end{equation}
where $\xi^\alpha_\mathbf{k}=\varepsilon^\alpha(\mathbf{k})-\mu$ and $\bm{\rho}=\mathbf{r}-\mathbf{r}'$.
In the presence of out-of-plane magnetic fields,  in the symmetric gauge $\mathbf{A}=\frac{H}{2}(-y,x,0)$ 
the one-particle Green's function can be represented as
\begin{align}
G^\alpha_{\omega_n,\pm}(\mathbf{r} j,\mathbf{r}' j')=&\exp\left(\mathrm{i}\frac{[\mathbf{r}\times\mathbf{r}']_z}{2l^2}\right)\nonumber\\
\times&\sum_{k_z}\!\mathrm{e}^{-\mathrm{i}k_z( j\!- j')}g^\alpha_{\omega_n,\pm}(\rho,k_z)
\label{GreenFunH}
\end{align}
where $l=\sqrt{c/(eH)}$ is the magnetic length and  $\rho=|\bm{\rho}|$. We use the quasiclassical approximation for the Green's function of the deep $e$ band 
\begin{equation}
g^e_{\omega_n,\pm}\approx \sum_{\mathbf{k}}\frac{\mathrm{e}^{-\mathrm{i}\mathbf{k}\cdot\bm{\rho}}}
{\mathrm{i}\omega_n\mp\mu_zH-\xi^e_\mathbf{k}+2t^e_z\cos k_z},\label{ge}
\end{equation}
and expand the Green's function of the shallow $h$ band over the exact Landau-level basis \cite{Rajagopal:PRB44.1991,Maniv:RMP73.2001}
\begin{equation}
g^h_{\omega_n,\pm}=\!\frac{1}{2\pi l^2}\sum_{\ell=0}\frac{L_ \ell(\frac{\rho^2}{2l^2})\exp(-\frac{\rho^2}{4l^2})}
{\mathrm{i}\omega_n\mp\mu_zH+\mathcal{E}_\ell-\mu_h\!+\!2t^h_z\cos k_z},\label{gh}
\end{equation}
where $\mu_h=\varepsilon_0-\mu$, $\mathcal{E}_\ell=\omega_c( \ell+\frac{1}{2})$, $\omega_c=eH/m_hc$ is the cyclotron frequency, and $L_\ell(x)$ are the Laguerre polynomials.

We note that in the single-band case the chemical potential of the shallow band may depend on the magnetic field \cite{Vagner:PRL51.1983}.  However, in our case the deep band serves as the particle reservoir that stabilizes the chemical potential. This allows us to solve the gap function assuming fixed chemical potential (see also Ref.\ \cite{Itskovsky:PRB61.2000}).

\section{Transition temperature in zero magnetic field ($T_C$)}\label{sec:TC}

For zero magnetic field, the superconducting order parameter is homogeneous, $\Delta^\alpha_{\mathbf{r} j}=\Delta^\alpha_{0}$.
This gives the linearized gap equation, which we represent as 
\begin{equation}
\Delta^\alpha_{0}=\sum_{ \beta}\Lambda_{\alpha\beta}\Lambda^{-1}_{0,\beta}\Delta^\beta_{0},
\end{equation}
where we have introduced the notations for the coupling matrix
$\Lambda_{\alpha\beta}=U_{\alpha\beta}N_\beta$ with $N_\alpha=m_\alpha/(2\pi)$ is the densities of states per layer (for shallow band, $N_h$ is the true density of states only in the regime of open Fermi surface, $\mu_h>2t_z^h$), and
\begin{equation}
\Lambda^{-1}_{0,\alpha}\equiv N^{-1}_\alpha T\sum^\infty_{\omega_n=-\infty}\sum_{j'}\int_{\mathbf{r}'}K^\alpha_{\omega_n}(\mathbf{r} j,\mathbf{r}' j').
\label{Lambda0}
\end{equation} 
The kernels are determined by the zero-field Green's function in Eq.\ \eqref{G0} giving
\[\Lambda^{-1}_{0,\alpha}\!=N^{-1}_\alpha T\sum_{\omega_n}\sum_{\mathbf{k}k_z}[\omega^2_n+(\xi^\alpha_\mathbf{k}-2t_z\cos k_z)^2 ]^{-1},\]
(see Appendix \ref{App:kernelsH0}).
We further integrate out the wave vectors in $\Lambda^{-1}_{0,\alpha}$ by using $\sum_\mathbf{k}=N_e\int^{\Omega}_{-\Omega}\mathrm{d}\xi^e$ for $e$ band ($\sum_\mathbf{k}=N_h\int^{\mu_h}_{-\Omega}\mathrm{d}\xi^h$ for $h$ band) and $\sum_{k_z}=\int^{\pi}_{-\pi}\frac{\mathrm{d}k_z}{2\pi}$. Here, $\Omega$ is the energy cutoff and we have assumed that $\Omega\gg t_z, T_C,\mu_h$. We therefore obtain the following gap equation near $T_C$:
\begin{equation}\label{eqn:H0GapEq}
\left[\hat{\Lambda}^{-1}-\hat{\Lambda}^{-1}_0\right]\begin{pmatrix}
\Delta^h_{0}\\
\Delta^e_{0}
\end{pmatrix}=0,
\end{equation}
where $\hat{\Lambda}^{-1}_0=\text{diag}[\Lambda^{-1}_{0,h},\Lambda^{-1}_{0,e}]$ with
\begin{subequations}
\begin{align}
\Lambda^{-1}_{0,e}&\approx\sum^\infty_{\omega_n>0}\frac{2 T_C}{\omega_n}\tan^{-1}\frac{\Omega}{\omega_n}\approx\ln\frac{\mathit{A}\Omega}{ T_C},\label{eqn:L0e}\\
\Lambda^{-1}_{0,h}& \approx
\frac{1}{2}\ln\frac{\mathit{A}\Omega}{T_C}+\Upsilon_{C},\label{eqn:L0h}
\end{align}
\end{subequations}
$\mathit{A}\!=\!2\mathrm{e}^{\gamma_{\mathrm E}}/\pi\!\approx\! 1.134$, and $\gamma_{\mathrm E}\!\approx\! 0.5772$ is the Euler-Mascheroni constant.
Here, the parameter $\Upsilon_{C}=\Upsilon_{T_C}$ is the value of the temperature-dependent function
\begin{equation}
\Upsilon_T=\sum^\infty_{\omega_n>0}\int^{\pi}_{-\pi}\frac{\mathrm{d}k_z}{2\pi}\frac{2 T}{\omega_n}\tan^{-1}\left[\frac{\mu_h-2t_z^h\cos k_z}{\omega_n}\right]
\end{equation}
at $T\!=\!T_{C}$. This parameter appears due to the cut off at the band edge for
the shallow band. We can carry out $k_z$ integration and the Matsubara-frequency
sum in $\Upsilon_T$ using the relation
$\tan^{-1}\frac{1}{a}=-\int^\infty_0\frac{dx}{x}\mathrm{e}^{-ax}\sin x$. This
gives
\begin{equation}
\Upsilon_T
=-\int^{\infty}_0\!\frac{\mathrm{d}s}{\pi s}\ln\tanh(\pi T s)\sin(2\mu_h s)J_0(4t_z^hs),
\label{UpsT}
\end{equation}
where $J_0(x)$ is the Bessel's function. 
In the limit of low temperatures, $T\ll \mu_h$, this function diverges logarithmically, 
$
\Upsilon_{T} \approx \tfrac{1}{2}\ln\left[\mathit{A}\left(\mu_{h}\!+\!\sqrt{\mu_{h}^{2}\!-\!\left(2t_{z}^{h}\right)^{2}}\right)/(2 T)\right].
$

The transition temperature  $T_C$ is determined by the condition of degeneracy of the matrix $\hat{W}\equiv\hat{\Lambda}^{-1}-\hat{\Lambda}^{-1}_0$, i.\ e., $\det[\hat{W}]=0$. This equation determines $\Lambda_{0,e}$ as
\begin{align}
\Lambda _{0,e}^{-1} &=\frac{\Lambda _{ee}+\frac{\Lambda _{hh}}{2}}
{\mathcal{D}_{\Lambda }}-\Upsilon _{C}\notag\\
+&\delta _{\Lambda }\sqrt{\left( \frac{\Lambda _{ee}\!-\!\frac{\Lambda _{hh}}{2}}
	{\mathcal{D}_{\Lambda }}-\Upsilon _{C}\right) ^{2}+2\frac{\Lambda _{eh}\Lambda _{he}}{\mathcal{D}_{\Lambda }^{2}}} 
\label{eqn:Lam0e}  
\end{align}
with $\delta_{\Lambda}\!=\!-\mathrm{sign}\left[ (1\!-\!\Upsilon_C
\Lambda_{hh}) \mathcal{D}_{\Lambda }\right]$ and $\mathcal{D}_{\Lambda}
\!=\!\Lambda _{ee}\Lambda _{hh}\!-\!\Lambda _{eh}\Lambda _{he}$. This parameter
directly determines  $T_C$ by Eq.\ \eqref{eqn:L0e}. In the following
consideration, we will use the zero-field equation, Eq.\  \eqref{eqn:H0GapEq}, to
eliminate the logarithmic divergences from the gap equation at finite magnetic
field.

\section{Superconducting instability in magnetic field}\label{HC2eqn}

In the presence of magnetic field, the superconducting order parameter is nonuniform. Near the onset of the superconducting instabilities in the magnetic field, the solutions for the linear problem in Eq.\ \eqref{eqn:gap-eqn} are given by the Landau-level eigenfunctions. Typically, the \textit{lowest Landau-level} eigenfunction yields the optimal $H_{C2}$, where the superconducting instability develops first
\footnote{For strong Zeeman spin splitting, the largest $H_{C2}$ may be realized for \textit{higher Landau-level} eigenfunction \cite{Buzdin:EPL35.1996,*Buzdin:PLettA218.1996}. 
}. 
This solution has the following shape\cite{Gruenberg:PRL16.1966}
\begin{equation}\label{eqn:ansatz}
\Delta^\alpha_{\mathbf{r}, j}=\Delta^\alpha_0\exp\left(-\frac{r^2}{2l^2}+\mathrm{i}Q_z j\right).
\end{equation}
We have assumed the possibility of the layer-to-layer modulation in the order parameters with the wave vector $Q_z$. %
\footnote{To be accurate, the complex order parameter in Eq.\ \eqref{eqn:ansatz} was suggested by Fulde and Ferrel \cite{Fulde:PRev135.1964}. As demonstrated by Larkin and Ovchinnikov\cite{Larkin:JETP20.1965}, for purely paramagnetic case the ground state below the transition is actually given by the order parameter with the amplitude modulation $\propto \cos(Q_z j)$.  
In the presence of orbital effects, both states may realize \cite{HouzetPhysRevB01}.
We only investigate the instability location here, which is the same for both states.} 
Such modulation is the realization of the nonuniform FFLO state \cite{Fulde:PRev135.1964,Larkin:JETP20.1965}.
Superconducting instability is determined by the condition that Eq.\ \eqref{eqn:ansatz} provides a solution of the gap equation at least for one value of $Q_z$ and then this optimal modulation wave vector realizes in the emerging superconducting state. 
The \textit{ansatz} in Eq.\ \eqref{eqn:ansatz} is the kernel eigenfunction in the magnetic field,
\begin{equation}\label{eqn:kernel-eig}
\sum_{ j'}\int_{\mathbf{r}'}K^\alpha_{\omega_n}(\mathbf{r} j,\mathbf{r}' j')\Delta^\alpha_{\mathbf{r}' j'}\!=\!\pi N_\alpha\lambda^\alpha_{\omega_n,Q_z}\Delta^\alpha_{\mathbf{r}, j}
\end{equation}
with
\begin{align}
&\lambda^{\alpha}_{\omega_n,Q_z}=\frac{2}{N_\alpha}\int^\infty_0\rho\mathrm{d}\rho \exp\left(-\frac{\rho^2}{2l^2}\right) \notag\\
&\times \left\langle g^\alpha_{\omega_n,+}\left(\rho,k_z\!-\tfrac{Q_z}{2}\right)
g^\alpha_{-\omega_n,-}\left(\rho,k_z\!+\tfrac{Q_z}{2}\right)\right\rangle_z.\label{lomega}
\end{align}
Here and below we use notation $\langle F(k_z)\rangle_z\!\equiv\!\int^\pi_{-\pi}\frac{\mathrm{d}k_z}{2\pi} F(k_z)$ for the averaging with respect to $k_z$. We omit the dependence of $\lambda^{\alpha}_{\omega_n,Q_z}$ on $T$, $H$, and the electronic parameters.
As follows from the definitions of the Green's functions, Eqs.\ \eqref{ge} and \eqref{gh}, $(\lambda^\alpha_{\omega_n,Q_z})^\ast=\lambda^\alpha_{-\omega_n,Q_z}$. 
Next, we first discuss the kernel eigenvalue for the deep $e$-band and then for the shallow $h$-band. 

For the deep $e$-band, the Landau quantization does not play a role. Using Eqs.\ \eqref{ge} and \eqref{lomega}, 
linearizing the band dispersion at Fermi level [$\varepsilon^e_{\mathbf{k}_F+\mathbf{q}}\approx\varepsilon^e_{\mathbf{k}_F}+\bm{v}_e\cdot\mathbf{q}$, where $\bm{v}_e=\nabla_{\mathbf{k}}\varepsilon^e_{\mathbf{k}}|_{\mathbf{k}^F}=\mathbf{k}^F/m_e$ with the Fermi wave vector $\mathbf{k}^F=(k^F_x, k^F_y)$], 
we obtain the following quasiclassical result for the kernel eigenvalue
\begin{align}
&\lambda_{\omega_{n},Q_{z}}^{e}\!=\!2\int\limits_{0}^{\infty}\!ds \Bigg\langle
\exp\!\Bigg[\!-\!\frac{s^{2}\left(\mu\!+\!2t_{z}^{e}\cos k_{z}\cos\frac{Q_{z}}{2}\right)}{m_el^{2}} \notag \\ 
&-\!2s\zeta_\omega\left(\omega_{n}\!+\!\mathrm{i}\mu_{z}H\!-\!2\mathrm{i}t_{z}^{e}\sin k_{z}\sin\frac{Q_{z}}{2}\right)\Bigg]\Bigg\rangle_z,\label{eqn:le}
\end{align}
where $\zeta_\omega\equiv\mathrm{sign}(\omega_n)$. The derivation details are described in the Appendix \ref{App:kernelsFinHDeep}.

For the shallow $h$ band, substituting  the Green's function, Eq.\ \eqref{gh}, into the general presentation in Eq.\ \eqref{lomega}, and 
using relation $\int^\infty_0 \mathrm{d} x L_\ell(x)L_{\ell'}(x)\mathrm{e}^{-2x}=\frac{(\ell+\ell')!}{2^{\ell+\ell'+1}\ell!\ell'!}$,
we derive
\begin{align}\label{lh}
&\lambda^h_{\omega_n,Q_z}\!=\! -\frac{1}{2\pi\omega_c}\\
\times\!&
\sum_{\ell\ell'}\!\left \langle\frac{(\ell+\ell')!/(2^{\ell+\ell'}\ell!\ell'!)}
{(\mathrm{i}\bar{\omega}_n\!+\!\ell\!+\frac{1}{2}-\!\tilde{\gamma}_{z}\!+\!\tilde{\mu}_{h})
	(\mathrm{i}\bar{\omega}_n\!-\!\ell'\!-\frac{1}{2}\!-\!\tilde{\gamma}_{z}\!+\!\tilde{\mu}_{h})}\right \rangle_z,\nonumber
\end{align}
where we introduced the following notations
\begin{subequations}
\begin{align}
\tilde{\mu}_{h}(k_z,Q_z)&=\bar{\mu}_h-2\bar{t}_z^h\cos k_z\cos (Q_z/2)\label{tmh},\\
\tilde{\gamma}_{z}(k_z,Q_z)&=\gamma_z-2\bar{t}_z^h\sin k_z\sin (Q_z/2)\label{tg}.
\end{align}
\end{subequations}
Here all ``barred'' normalized quantities are defined as $\bar{a}\equiv a/\omega_c$ (with $a=\omega_n,\mu_h,t_z$). Furthermore,
$\gamma_z\!=\!\mu_zm_hc/e\!=\!gm_h/4m_0$ is the reduced spin-splitting parameter, where $m_0$ is the free-electron mass and $g$ is the band-electrons $g$-factor. 

Therefore, the gap equation \eqref{eqn:gap-eqn}, in the presence of the magnetic
field, becomes
\begin{equation}
\hat{\Lambda}^{-1}\begin{bmatrix}
\Delta^h_{0}\\
\Delta^e_{0}
\end{bmatrix}
=2\pi T\, \text{Re}\sum^\Omega_{\omega_n>0}\begin{bmatrix}
\lambda^h_{\omega_n,Q_z}\Delta^h_{0}\\
\lambda^e_{\omega_n,Q_z}\Delta^e_{0}
\end{bmatrix}.
\end{equation}
The Matsubara-frequency sums, $\mathcal{S}_{\alpha}(H,T,Q_z)\!\equiv\!2\pi T\sum^\Omega_{\omega_n>0}\lambda^{\alpha}_{\omega_n,Q_z}$, are logarithmically divergent and have to be cut at $\omega_n\!\sim \! \Omega$. Similarly to the two-dimensional case \cite{Song:PRB95.2017}, we can regularize the gap equation using its zero-field counterpart, Eq.\ \eqref{eqn:H0GapEq}. Namely, we decompose $\mathcal{S}_{\alpha}(H,T,Q_z)$ as $\mathcal{S}_{\alpha}(H,T,Q_z)\!=\![\mathcal{S}_{\alpha}(H,T,Q_z)\!-\!\mathcal{S}_{\alpha}(0,T,0)]\!+\![\mathcal{S}_{\alpha}(0,T,0)\!-\!\mathcal{S}_{\alpha}(0,T_C,0)]
\!+\!\mathcal{S}_{\alpha}(0,T_C,0)$ with $\mathcal{S}_{\alpha}(0,T_C,0)\!\equiv \! \Lambda_{0,\alpha}^{-1}$ being the only log-diverging term. 
This leads to
\begin{equation}\label{eqn:gap-regularized}
\hat{W}\begin{bmatrix}
\Delta^h_{0}\\
\Delta^e_{0}
\end{bmatrix}
+\begin{bmatrix}
\mathcal{A}_1\Delta^h_{0}\\
\mathcal{A}_2\Delta^e_{0}
\end{bmatrix}
=\begin{bmatrix}
\mathcal{J}_1\Delta^h_{0}\\
\mathcal{J}_2\Delta^e_{0}
\end{bmatrix}.
\end{equation}
Here, $\mathcal{A}_\alpha(T)\!\equiv\! \mathcal{S}_{\alpha}(0,T_C,0)\!-\!\mathcal{S}_{\alpha}(0,T,0)$
are the temperature-dependent parts of the pairing eigenvalues,
$\mathcal{A}_2(T)\!=\!\ln(T/T_C)$ and  $\mathcal{A}_1(T)\!=\tfrac{1}{2}\ln(T/T_C)\!-\!\Upsilon_T\!+\!\Upsilon_{C}$,
where the function $\Upsilon_T(T/\mu_h, t_z^h/\mu_h)$ is defined in Eq.\ \eqref{UpsT}. 
The field-dependent parts, 
$\mathcal{J}_\alpha (H,T,Q_z)\!\equiv\! \mathcal{S}_{\alpha}(H,T,Q_z)\!-\!\mathcal{S}_{\alpha}(0,T,0)$, are
\begin{subequations}
\begin{align}
&\mathcal{J}_1(H,T,Q_z)\!=\!2\pi T\!\sum^\infty_{\omega_n>0}
\!\text{Re}\Big(\lambda^h_{\omega_n,Q_z}\!-\frac{1}{2\omega_n}\Big)\!-\!\Upsilon_T\label{Jh},\\
&\mathcal{J}_2(H,T,Q_z)\!=\!2\pi T\!\sum^\infty_{\omega_n>0}\!\text{Re}\Big(\lambda^e_{\omega_n,Q_z}\!-\frac{1}{\omega_n}\Big).\label{Je}
\end{align}
\end{subequations}
Note that, by definition, $\mathcal{A}_{\alpha}(T_C)=0$ and $\mathcal{J}_{\alpha}(0,T,0)\!=\!0$.
Therefore, the UV cutoffs are explicitly removed and the Matsubara-frequency sums in the right-hand side in Eqs.\ \eqref{Jh} and \eqref{Je} converge now in the limit of $\Omega\to\infty$. 
We can represent the functions $\mathcal{J}_\alpha$ in this limit as  
\begin{widetext}
\begin{subequations}
	\begin{align}
\mathcal{J}_{1}(H,T,Q_{z})\! & =\frac{1}{4}\!\sum_{m=0}^{\infty}\sum_{\ell=0}^{m}\frac{m!}{2^{m}\left(m\!-\!\ell\right)!\ell!}\left\langle \frac{\mathcal{T}(\ell+\tilde{\gamma}_{z}-\!\tilde{\mu}_{h})+\mathcal{T}(m\!-\!\ell-\tilde{\gamma}_{z}-\!\tilde{\mu}_{h})-2\mathcal{T}(\frac{m}{2}-\tilde{\mu}_{h})}{m+1-2\tilde{\mu}_{h}}\right\rangle _{\!z}\nonumber \\
& -\frac{1}{2}\left\langle \int\limits _{0}^{1/2}dx\frac{\mathcal{T}(\frac{x-1}{2}-\tilde{\mu}_{h0})}{x-2\tilde{\mu}_{h0}}+\sum_{m=0}^{\infty}\int\limits _{-1/2}^{1/2}dx\left[\frac{\mathcal{T}(\frac{m+x}{2}-\tilde{\mu}_{h0})}{m+1+x-2\tilde{\mu}_{h0}}-\frac{\mathcal{T}(\frac{m}{2}-\tilde{\mu}_{h})}{m+1-2\tilde{\mu}_{h}}\right]\right\rangle _{\!z},\label{J1llSumPres}		
	\\
	\mathcal{J}_{2}(H,T,Q_{z})\!= & 2\int_{0}^{\infty}\mathrm{d}s\ln\tanh\left(\frac{\pi T}{\omega_{c}^{e}}s\right)\left\langle \exp\left(-\tilde{\mu}_{e}s^{2}\right)\left[\tilde{\mu}_{e}s\cos\left(2\tilde{\gamma}_{z}^{e}s\right)+\tilde{\gamma}_{z}^{e}\sin\left(2\tilde{\gamma}_{z}^{e}s\right)\right]\right\rangle _{\!z},\label{eqn:J2}
	\end{align}
\end{subequations}
\end{widetext}
where we introduced notations $\mathcal{T}(x)\!\equiv\!\tanh[\omega_c(x\!+\!\tfrac{1}{2})/2T]$, $\tilde{\mu}_{h0}(k_{z}) \!=\!\bar{\mu}_{h}\!-2\bar{t}_{z}^{h}\cos k_{z}$. Furthermore,   $\omega_{c}^{e}\!=\!eH/cm_e$ is the $e$-band cyclotron frequency, $\tilde{\mu}_{e}\!\equiv\!\bar{\mu}\!+\!2\bar{t}_{z}^{e}\cos k_{z}\cos\frac{Q_{z}}{2}$,  $\tilde{\gamma}_{z}^{e}\!\equiv\!\gamma_{z}\!-\!2\bar{t}_{z}^{e}\sin k_{z}\sin\frac{Q_{z}}{2}$, $\bar{\mu}\!=\!\mu/\omega_{c}^{e}$, and $\bar{t}_{z}^{e}=t_{z}^{e}/\omega_{c}^{e}$.
We remind that the parameters $\tilde{\mu}_{h}$ and $\tilde{\gamma}_{z}$ also depend on $k_z$ [see Eqs.\ \eqref{tmh} and \eqref{tg}]. We describe derivation of $\mathcal{J}_{1}(H,T,Q_{z})$ in Appendix \ref{App:kernelsFinHSh}.  The double sum in Eq.\ \eqref{J1llSumPres} collects the contributions to pairing coming from the quasiparticles located at the Landau levels $\ell$ and $m\!-\!\ell$. The quantum kernel eigenvalue $\mathcal{J}_{1}$ depends on five independent dimensionless parameters: the reduced magnetic field $\omega_c/\mu_h$, the reduced temperature $T/\mu_h$, the modulation wave vector $Q_z$, the ratio $t_z^h/\mu_h$, and the spin-splitting factor $\gamma_z$.

The $H_{C2}$ problem is reduced to the solution of equation
\begin{equation}
\det\begin{bmatrix}
W_{11}+\mathcal{A}_1-\mathcal{J}_1 & W_{12}\\
W_{21}&W_{22}+\mathcal{A}_2-\mathcal{J}_2
\end{bmatrix}=0
\end{equation}
for a given $T$ and $Q_z$, and we need to find the optimized $Q_z$ in $H_{C2}(T,Q_z)$ 
for which the instability develops first.
As the matrix $\hat{W}$ is degenerate, this equation can be rewritten as
\begin{equation}
	\prod_{\alpha=1,2}\left( 1+\frac{\mathcal{A}_\alpha(T)-\mathcal{J}_\alpha(H,T,Q_z)}{W_{\alpha\alpha}}\right) =1.
	\label{eqn:HC2}
\end{equation}
This is our main equation for determination of the upper critical field in two-band layered superconductors. 
All information about the coupling matrix is contained in the two parameters, $W_{11}$ and $W_{22}$.
The analytical expressions for these parameters can be derived from Eq.\ \eqref{eqn:H0GapEq} (see also Ref.\ \cite{Song:PRB95.2017}),
\begin{subequations}
\begin{align}
W_{11}&=\frac{\Lambda_{ee}-\frac{1}{2}\Lambda_{hh}}{2\mathcal{D}_\Lambda}-\frac{\Upsilon_{C}}{2}+\delta_{W}\frac{R}{2}\\
W_{22}&=-\frac{\Lambda_{ee}-\frac{1}{2}\Lambda_{hh}}{\mathcal{D}_\Lambda}-\Upsilon_{C}+\delta_{W}R
\end{align}
\end{subequations}
with $\mathcal{D}_\Lambda\!=\!\Lambda_{ee}\Lambda_{hh}\!-\!\Lambda_{he}\Lambda_{eh}$, $\delta_{W}\!=\!\text{sign}[\mathcal{D}_\Lambda(1\!-\!\Upsilon_{C}\Lambda_{hh})]$, and
\begin{equation}
R=\sqrt{\Big(\frac{\Lambda_{ee}-\frac{1}{2}\Lambda_{hh}}{\mathcal{D}_\Lambda}-\Upsilon_{C}\Big)^2+2\frac{\Lambda_{he}\Lambda_{eh}}{\mathcal{D}_\Lambda^2}}.
\end{equation}
As follows from Eq.\ \eqref{eqn:HC2} the weights with which the bands contribute to the pairing near the upper critical field scale as $1/|W_{\alpha\alpha}|$. 

The behavior of the upper critical field is mostly determined by the shape of the kernel
eigenvalues $\mathcal{J}_\alpha(H,T,Q_z)$. \emph{In general, larger values of $\mathcal{J}_\alpha$ correspond to stronger pairing strength}. 
In the next section, we will explore in detail the quantum kernel eigenvalue
$\mathcal{J}_1$. 

\section{Behavior of  the quantum pairing kernel eigenvalue} 
\label{sec:J1}
\begin{figure}[htbp]
	\centering
	\includegraphics[width=3.4in]{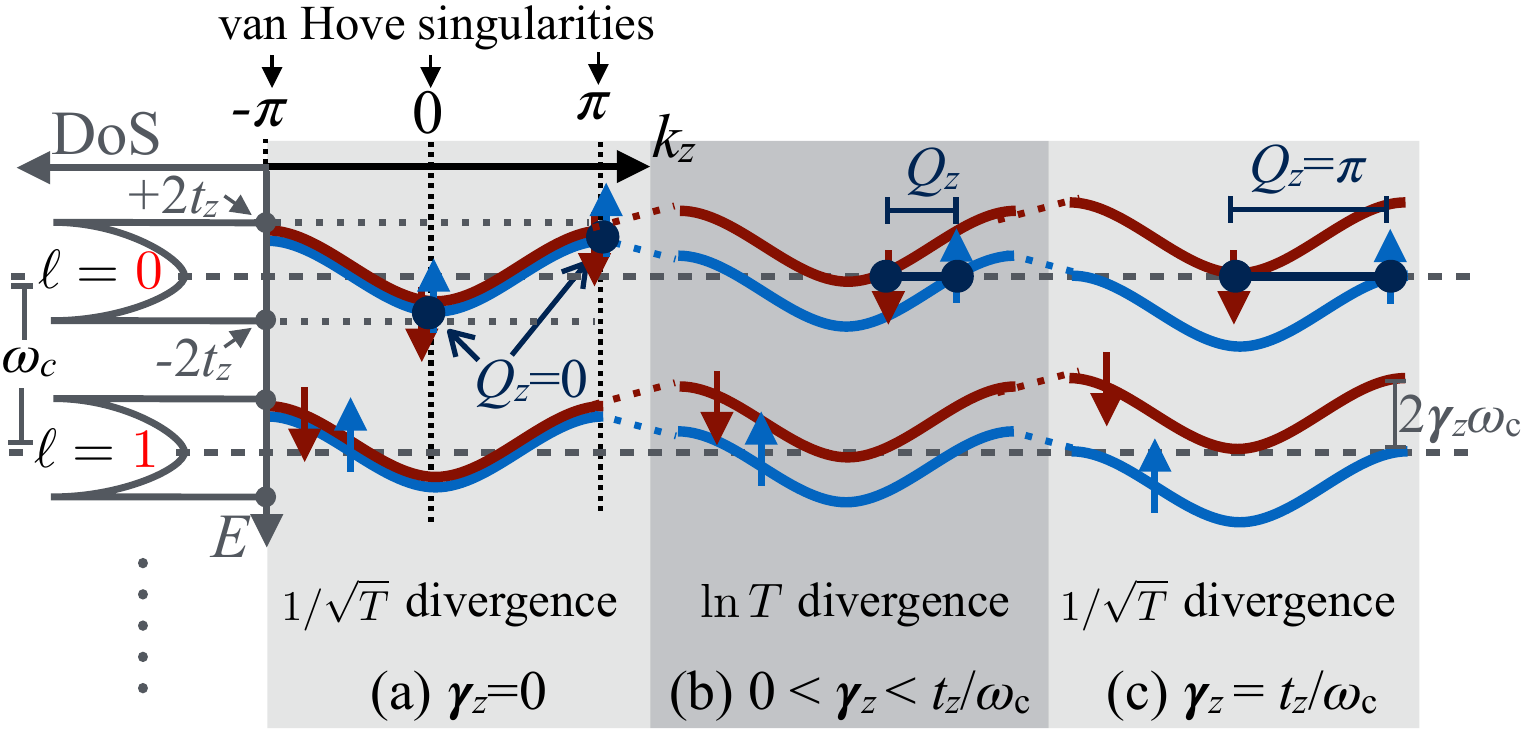} 
	\caption{The schematic diagram illustrating the $k_z$ spectrum of the shallow $h$-band in the magnetic field in different situations. The dispersion along $k_z$-direction has two van Hove singular points at $k_z\!=\!0,\pi$. If the spin-up and spin-down van Hove singular points match the Fermi level, the superconducting pairings lead to $1/\sqrt{T}$-divergence in the pairing kernel eigenvalue $\mathcal{J}_1$ [cases (a) and (c)]. The matching condition for (a) is $\omega_c\!=\!(\mu_h\!\pm\!2t_z)/(\ell\!+\!\frac{1}{2})$ and for (c) is $\omega_c\!=\!\mu_h/(\ell\!+\!\frac{1}{2})$. The pairing instabilities for cases (b) and (c) favor the formation of the FFLO states with non-zero  $Q_z$.}
	\label{fig:zeeman}
\end{figure}

The shape of magnetic field--temperature phase diagrams is determined by the behavior of the pairing kernel eigenvalues $\mathcal{J}_\alpha$. While the quasiclassical kernel eigenvalue $\mathcal{J}_2$ is well studied and has monotonic dependence on the magnetic field for all temperatures, the quantum kernel eigenvalue $\mathcal{J}_1$ has rather complicated nonmonotonic dependence on the magnetic field which is very sensitive to the electronic-spectrum parameters $\mu_h$ and $t_z^h$ as well as the spin-splitting factor $\gamma_z$. 
The electronic spectrum of the shallow band in the magnetic field is composed of the Landau-level minibands with width $4t_z^h$ [see Fig. \ref{fig:zeeman}(a)]. The system has series of Lifshitz transitions with increasing magnetic field when the chemical potential enters or exits a particular miniband. 
At low temperatures, the magnetic-field dependences of $\mathcal{J}_1$ have features at these transitions whenever the chemical potential crosses the van Hove points at the miniband edges. There are two such points for every miniband corresponding to two values of $k_z$, $0$ and $\pi$. In addition, there are two minibands for every Landau level for two spin orientations.
This gives \emph{four} miniband-edge magnetic fields per Landau level, $H_{\ell, \sigma,\delta_{t}}=(cm_h/e)\omega_{\ell, \sigma,\delta_{t}}$, corresponding to cyclotron frequencies
\begin{equation}
\omega_{\ell, \sigma,\delta_{t}}
=\frac{\mu_h+2\delta_{t} t_h^h}{\ell+\tfrac{1}{2}+\sigma \gamma_z},
\label{wc-vH}
\end{equation}   
where $\sigma=\pm 1$ ($\uparrow$/$\downarrow$) describes spin orientation and $\delta_{t}=1$ and $-1$ corresponds to $k_z\!=\!\pi$ and $0$, respectively. 
In addition, we have to consider the behavior of $\mathcal{J}_1$  for different modulation wave vectors $Q_z$. In general, as the larger $\mathcal{J}_1$ corresponds to stronger pairing strength, the shallow band favors the states which maximize $\mathcal{J}_1$.

Before proceeding to the investigation of the magnetic field--temperature phase diagrams for two-band systems, it is very instructive to study the analytical properties of the function $\mathcal{J}_1(H,T,Q_z)$, particularly, in the low-temperature limit. We first focus on its singularities when the magnetic field crosses the typical values in Eq.\ \eqref{wc-vH} and on
identifying the possible divergences for $T\to0$, since these features have important implications on the superconducting instabilities. Next, we consider the magnetic field dependences of $\mathcal{J}_1$ for representative cases. Furthermore, by studying the $Q_z$ dependences of $\mathcal{J}_1$ for different Zeeman spin-splitting parameters, we identify possible parameter ranges for FFLO instabilities. 

\subsection{Low-temperature limit and its divergences}
\label{sec:J1lowT}

In this subsection, we investigate the leading divergences of the quantum kernel eigenvalues $\mathcal{J}_\alpha$ as $T\to 0$ for different cases.
We review first the behavior of the quasiclassical kernel eigenvalue $\mathcal{J}_2$ [Eq.\ \eqref{eqn:J2}]. For all magnetic fields, $\mathcal{J}_2$ has the same logarithmic divergence as $\mathcal{A}_2$ so that $\mathcal{J}_2-\mathcal{A}_2$ approaches a finite value. In the typical situation of a moderate spin-splitting factor, $\gamma_z\ll \sqrt{\mu/\omega_c^e}$, it can be treated perturbatively. For the uniform case, $Q_z=0$, we derive from Eq.\ \eqref{eqn:J2} the result for the 
zero-temperature limit of the full quasiclassical kernel eigenvalue
\begin{align}\label{A2J2LowT}
&\mathcal{J}_2\left(H,T\!\to \!0,0 \right) -\mathcal{A}_2\left(T\!\to \!0 \right)
=-\frac{1}{2}\ln r_C^{(0)},\\
&r_C^{(0)}\!\approx\!\frac{\mathrm{e}^{\gamma_{E}}\omega_{c}^{e}}{\pi^{2}T_{C}^{2}}
\frac{\mu\!+\!\sqrt{\mu^{2}\!-\!(2t_{z}^{e})^{2}}}{2}\left(1\!+\!\frac{2\gamma_{z}^{2}\omega_{c}^{e}}{\sqrt{\mu^{2}\!-\!\left(2t_{z}^{e}\right)^{2}}}\right).\nonumber
\end{align}
The parameter $r_C^{(0)}$ is just the ratio $H/H_{c2}^e(0)$, where $H_{c2}^e(0)$ is the upper critical field of the deep band at zero temperature.
In the case of finite $Q_z$, the parameter $r_C^{(0)}$ has to be replaced by the function $r_C(Q_z)$. The closed analytical result for  $r_C(Q_z)$ is not available even for $\gamma_z=0$. One can only derive an approximate result in the limits,  $\gamma_z\ll \sqrt{\mu/\omega_c^e}$ and $t_{z}^{e}/\mu, t_{z}^{e}\sin(Q_z/2)/\sqrt{\omega_c^e\mu}\ll 1$:
\begin{align}
r_C(Q_z)\!\approx &\frac{\mathrm{e}^{\gamma_{E}}\omega_{c}^{e}\mu}{\pi^{2}T_{C}^{2}}\Big[1\!+\!2\frac{\omega_{c}^{e}\gamma_{z}^{2}}{\mu}\!+\!4\frac{\left(t_{z}^{e}\right)^{2}}{\mu\omega_{c}^{e}}\sin^{2}\frac{Q_{z}}{2}\nonumber\\
&\!-\!\frac{(t_{z}^{e})^2}{\mu^{2}}\cos^{2}\frac{Q_{z}}{2}\!\Big].
\label{rCQz}
\end{align}
Generally, the quasiclassical kernel eigenvalue $\mathcal{J}_2\!-\!\mathcal{A}_2$ is a monotonically decreasing function of $H$ at all temperatures and spin-splitting parameters  meaning that the magnetic field \textit{always} suppresses  superconductivity. As expected, it has maximum at $Q_z\!=\!0$ in the limit  $\gamma_z\ll \sqrt{\mu/\omega_c^e}$ meaning that the deep band favors the uniform state. 

The quantum kernel eigenvalue $\mathcal{J}_1$ typically behaves similarly to $\mathcal{J}_2$, i.\ e., it  has the same logarithmic divergence as $\mathcal{A}_1$, $\mathcal{J}_1 \propto \ln (T/T_C)$  so that the total kernel eigenvalue $\mathcal{J}_1-\mathcal{A}_1$ approaches a finite value in the zero-temperature limit.
This zero-temperature value, however, has singular contributions when the magnetic field crosses the typical values given by Eq.\ \eqref{wc-vH}, corresponding to Lifshitz transitions for the Landau-level minibands. We discuss these singularities in the next subsection.  
In several exceptional resonant cases, when two miniband-edge fields with opposite spin orientations are identical, the Landau quantization leads to faster divergence $\mathcal{J}_1\propto 1/\sqrt{T}$.  In the case when these two fields originate from  the van Hove singular points of the same type (either $k_z=0$ or $\pi$), the divergence occurs in the uniform channel $Q_z\!=\!0$. On the other hand, if the two fields correspond to the opposite  van Hove points, the divergence takes place in the alternating channel $Q_z\!=\!\pi$. We discuss both these cases below.
Another divergence appears within the field ranges where the chemical potential simultaneously crosses two minibands with opposite spin orientations. In this case, the total kernel eigenvalue $\mathcal{J}_1-\mathcal{A}_1$ diverges logarithmically for $T\to 0$ at the optimal wave vector $Q_z=Q_\text{op}$ connecting  the minibands' Fermi momenta [see Fig.\ \ref{fig:zeeman}(b)].
All these low-$T$ divergences may lead to the high-field superconducting states.

\subsubsection{Square-root singularity of the pairing kernel at the miniband-edge fields }

As discussed above, the system has series of the miniband Lifshitz transitions at the magnetic fields given by Eq.\ \eqref{wc-vH}. In this subsection we discuss singularity of the pairing kernel eigenvalue $\mathcal{J}_{1}(H,T\to0,Q_{z})$ at these transitions, when the cyclotron frequency $\omega_{c}$ crosses the miniband-edge value $\omega_{\ell_{0},\sigma,\delta_{t}}$. We consider here only a general nondegenerate situation when the corresponding transition magnetic field, $H_{\ell_{0},\sigma,\delta_{t}}$, is separated from other typical fields. The derivation in Appendix \ref{app:J1sqrtCross} gives the result for the singular contribution at zero temperature
\begin{align}
&\mathcal{J}_{1}(\omega_{c})\!-\mathcal{J}_{1}(\omega_{\ell_{0},\sigma,\delta_{t}})\nonumber\\
&\approx\!-  \frac{\delta_{t}}{2\pi}\sqrt{\frac{\mu_{h}\!+\!2\delta_{t}t_{z}^{h}}{t_{z}^{h}}}
\sqrt{\left|1\!-\!\frac{\omega_{c}}{\omega_{\ell_{0},\sigma,\delta_{t}}}\right|}\,
\theta\!\left[\delta_{t}\!\left(\!1\!-\!\frac{\omega_{c}}{\omega_{\ell_{0},\sigma,\delta_{t}}}\right)\!\right]\nonumber\\
&\times\mathcal{G}_{\ell_{0}}\left[\frac{\mu_{h}\!+\!2\delta_{t}t_{z}^{h}\cos^{2}\tfrac{Q_{z}}{2}}{\omega_{c}}\right]\label{eq:SqRJ1}
\end{align}
with $\theta(x)$ being the step function and
\[
\mathcal{G}_{\ell}(x)\equiv\sum_{m=\ell}^{\infty}\frac{m!}{2^{m}\left(m-\ell\right)!\ell!}\frac{1}{m+1-2x}.
\]
We see that the square-root singularity $\mathcal{J}_{1}(H)\propto\sqrt{\left|H-H_{\ell_{0},\sigma,\delta_{t}}\right|}$ appears near the transition point when the chemical potential is inside the Landau-level miniband, i.e., $H\gtrsim H_{\ell_{0},\sigma,-}$ or $H\lesssim H_{\ell_{0},\sigma,+}$. It reflects the pairing enhancement caused by the square-root divergence of the density of state at the edge of one-dimensional miniband. Finite temperature smears this singularity.

\subsubsection{Resonant cases for the uniform state ($Q_z=0$)}

In the uniform state, $Q_z=0$, the resonant condition corresponds to the matching of the Zeeman
spin-splitting energy, $2\gamma_z\omega_c$, and the Landau-level energy spacing, $j_z\omega_c$,
giving $2\gamma_z\!=\!j_z$ with integer $j_z\geq 0$. 
The divergence occurs when the Fermi level matches the van Hove singular point at $k_z\!=\!0$
or $\pi$ corresponding to the magnetic field
\begin{equation}\label{eqn:ResCond}
\omega_c=\frac{\mu_h\mp2t_z}{\ell_0+(j_z\!+\!1)/2}.
\end{equation}
Figure \ref{fig:zeeman}(a) illustrates the simplest case with $j_z\!=\!0$.
We derive in Appendix \ref{App:J1Q0smallT} the following asymptotic behavior of $\mathcal{J}_{1}(H,T,0)$ for $T\ll \omega_c, t_z^h$ and arbitrary $\ell_0$ and $j_z$:
\begin{equation}\label{eqn:J1Resonant}
\mathcal{J}_{1}(H,T,0)\sim \mathit{C}\frac{\left(2\ell_{0}\!+\!j_{z}\right)!}{2^{2\ell_{0}+j_{z}}\left(\ell_{0}\!+\!j_{z}\right)!\ell_{0}!}\frac{\omega_{c}}{\sqrt{t_{z}^hT}},
\end{equation}
where $\mathit{C}\!=\!(2\sqrt{2}\!-\!1)\zeta(\tfrac{3}{2})/(8\pi^{3/2})\!\approx\! 0.1072$ and $\zeta(x)$ is the Riemann zeta function [$\zeta(\frac{3}{2})\!\approx\! 2.6124$].
We can see
that the $\mathcal{J}_1$ diverges as $\sim1/\sqrt{T}$ manifesting the enhancement of pairing. The
coefficient decreases with increasing interlayer hoping $t_z^h$ and with increasing of the
resonance order described by the integers $\ell_0$ and $j_z$. We note that for any small deviations from the condition $(\ell_0+\frac{1}{2}j_z+\frac{1}{2})\omega_c=\mu_h\pm2t_z^h$, the total eigenvalue  $\mathcal{J}_1-\mathcal{A}_1$ approaches a finite limit at $T\to 0$. 
We can also note that this divergence is somewhat weaker than the $1/T$ divergence for the 2D case \cite{Song:PRB95.2017}, due to the smearing of $\delta$-function singularity in the density of states at the Landau levels by the interlayer hopping. Such $1/\sqrt{T}$ divergence in the uniform state for zero or resonant spin-splitting factor has also been reported in Refs. \onlinecite{Gruenberg:PRev176.1968,Maniv:RMP73.2001}. 

For the strongest resonance at $\ell_0\!=0$, $j_z\!=0$, and $\omega_c\!=2(\mu_h\!\mp 2t_z^h)$ for $k_z\!=0$/$\pi$, we derived in Appendix \ref{App:J1Q0smallT} the more accurate asymptotic for the total kernel eigenvalue
\begin{align}
	\mathcal{J}_{1}\!-\!\mathcal{A}_{1} & \!\approx\!\mathit{C}\frac{\omega_{c}}{\sqrt{t_{z}^hT}}\!+\!\mathcal{R}_{\mp}(\bar{t}_{z}^h)\!-\frac{1}{2}\ln\frac{\omega_{c}}{\pi T_{C}}-\Upsilon_{C},
	\label{eqn:J1T0R}\\
	\mathcal{R}_{\mp}(\bar{t}_{z}^h) & =\frac{1}{2}\!\sum_{m=1}^{\infty}\left(\frac{1}{\sqrt{m\left(m\mp8\bar{t}_{z}^h\right)}}-\frac{1}{m}\right).\nonumber
\end{align}
For $k_z=0$, the function $\mathcal{R}_{-}(\bar{t}_{z}^h)$ diverges for $\omega_{c}\!=8t_{z}^h$. As $\omega_{c}\!=2(\mu_{h}\!-2t_{z}^h)$, meaning that this result is only valid for $\mu_h>6t_z^h$.

We mention that the spin splitting is a fixed material's parameter and therefore the resonance cases, $2\gamma_z=j_z$, are exceptional. To some extent, the effective spin splitting can be tuned by tilting the magnetic field \cite{Wosnitza:FSofLowDSC.1996}. We note, however, that in contrast to the two-dimensional case \cite{Song:PRB95.2017}, in layered superconductors the in-plane magnetic field also influences the orbital motion of quasiparticles, meaning that the problem of the upper critical field for this case requires separate consideration.

\subsubsection{Resonant cases for alternating FFLO state ($Q_z=\pi$)}
\label{sec:J1LowTQzpi}

The resonance enhancement of pairing may also take place in the \textit{alternating} FFLO state because in this case the modulation wave vector $Q_z\!=\!\pi$ couples the van Hove singular points at $k_z\!=\!0$ and $\pi$. Such enhancement is a unique property of a layered superconductor with open Fermi surface and it only exists for specific relations between the band parameters $\mu_h$ and $t_z^h$. %
Writing the full quasiparticle energy in the magnetic field as  
$\xi_{\pm}^{h}(\ell,k_{z})=\mu_h-\omega_{c}\left(\ell+\frac{1}{2}\pm\gamma_{z}\right)-2t_{z}^{h}\cos k_{z}$, 
we can identify that the resonance conditions are realized when the spin-down (spin-up) energy at $k_z\!=\!0$ at the Landau level $\ell_0$ simultaneously matches with the spin-up (spin-down) energy at $k_z\!=\!\pi$ at the Landau level $\ell_\pi$ and with the chemical potential, i.\ e.,
$\xi_{\mp}^{h}(\ell_{0},0) =\xi_{\pm}^{h}(\ell_{\pi},\pi)=0$ [see the first-case example with $\ell_0\!=\!\ell_\pi\!=\!0$ in Fig.\ \ref{fig:zeeman} (c)].
This gives the conditions
\begin{subequations}
\begin{align}
\omega_{c} &  =\frac{4t_{z}^{h}}{\ell_{\pi}-\ell_{0}\pm2\gamma_{z}}=\frac{2\mu_h}{\ell_{0}+\ell_{\pi}+1},\label{eqn:pi-wc}\\
\mu_{h} & =\frac{\omega_{c}}{2}\left(\ell_{0}+\ell_{\pi}+1\right)
 =\frac{2t_{z}^{h}\left(\ell_{0}+\ell_{\pi}+1\right)}{\ell_{\pi}-\ell_{0}\pm 2\gamma_{z}}.\label{eqn:pi-cond}
\end{align}
\end{subequations}

To derive the low-temperature behavior of $\mathcal{J}_1$,  we substitute the conditions in Eqs.\ \eqref{eqn:pi-wc} and \eqref{eqn:pi-cond} into Eq.\ \eqref{J1llSumPres} with  $Q_z=\pi$. The derivation described in Appendix \ref{App-J1pi}
yields the low-temperature asymptotic for generic case of noninteger $2\gamma_z$:
\begin{equation}\label{eqn:J1piFFLO}
\mathcal{J}_{1}(H,T,\pi)\simeq\frac{\mathit{C}}{2}\!\frac{\left(\ell_{0}+\ell_{\pi}\right)!}{2^{\ell_{0}+\ell_{\pi}}\ell_{0}!\ell_{\pi}!}\frac{\omega_{c}}{\sqrt{Tt_{z}^{h}}}.
\end{equation}
where the numerical constant $\mathit{C}$ is defined above, after Eq.\ \eqref{eqn:J1Resonant}.
We note that this additional divergence is purely a consequence of the interplay between the Landau quantization and interlayer tunneling. In particular, such FFLO instability is absent in the 3D case with $k^2_z$ dispersion. 

For the strongest resonance, $\ell_0=\ell_\pi=0$,  $\mu_h=t_z^h/\gamma_z$ ($\gamma_z<0.5 $), and $\omega_c=2\mu_h$, we derived the more accurate asymptotic
\begin{equation}
\mathcal{J}_{1}-\mathcal{A}_{1}\simeq\frac{\mathit{C}}{2}\frac{\omega_{c}}{\sqrt{Tt_{z}^{h}}}-\frac{1}{2}\ln\frac{2\omega_{c}}{\pi T_{c}}-\Upsilon_{C}.\label{eqn:J1T0pi}
\end{equation}
In the real systems such instability can be obtained by tuning the chemical potential $\mu_h$  by doping or pressure. 

\subsubsection{Logarithmic divergence at the optimal modulation wave vector}

In the case when the chemical potential crosses two Landau-level branches with opposite spin orientations, the kernel eigenvalue also has logarithmic divergence at the wave vector connecting the crossing points [see Fig.\ \ref{fig:zeeman}(b)]. This divergence originates from the one-dimensional character of the electronic spectrum. The similar result was also found in the study of a three-dimensional model with parabolic Landau minibands \cite{Maniv:RMP73.2001}. For illustration, we consider such divergence for such branches belonging to the same Landau level with index $\mathcal{\ell}_{0}$. In this case, the diverging term in Eq.\ \eqref{J1llSumPres} is the one with $m=2\ell_{0}$ and $\ell=\ell_{0}$:
\begin{align*}
\mathcal{J}_{1,\ell_{0}}\! & =\!\frac{\omega_{c}}{4}\!\frac{(2\ell_0)!}{2^{2\ell_0}\left(\ell_0!\right)^{2}}\\
\times&\int\limits _{-\pi}^{\pi}\!\frac{\mathrm{d}k_{z}}{2\pi}\!\sum_{\delta=\pm1}\!\frac{\tanh[\frac{\omega_{c}(\ell_{0}+\frac{1}{2}+\delta\gamma_{z})-\mu_{h}+2t_{z}^{h}\cos(k_{z}\!+\delta\tfrac{Q_{z}}{2})}{2T}]}{\omega_{c}\left(2\ell_{0}\!+\!1\right)-2\left(\mu_{h}\!-\!2t_{z}^{h}\cos k_{z}\cos\tfrac{Q_{z}}{2}\right)}.
\end{align*}
For the optimal modulation wave vector $Q_{z}\!=\!Q_{\mathrm{op}}$ shown
in Fig. 2(b), we have the relations 
\[
\omega_{c}(\ell_{0}+\tfrac{1}{2}\!+\!\delta\gamma_{z})-\!\mu_{h}\!+2t_{z}^{h}\cos\left(k_{z0}\!+\delta\tfrac{Q_{\mathrm{op}}}{2}\right)=0,
\]
for $\delta=\pm1$, where $k_{z0}\pm Q_{\mathrm{op}}/2$ are the Fermi wave vectors for the two considered branches. These relations determine 
$Q_{\mathrm{op}}$ as
\begin{align}
Q_{\mathrm{op}}=&\sum_{\delta=\pm 1}\delta\arccos\!\left[\frac{\mu_{h}
	\!-\!\omega_{c}(\ell_{0}+\tfrac{1}{2}+\!\delta\gamma_{z})}{2t_{z}^{h}}\right].
\label{eq:Qop}
\end{align}
\begin{figure*}[htbp] 
	\centering
	\includegraphics[width=6.8in]{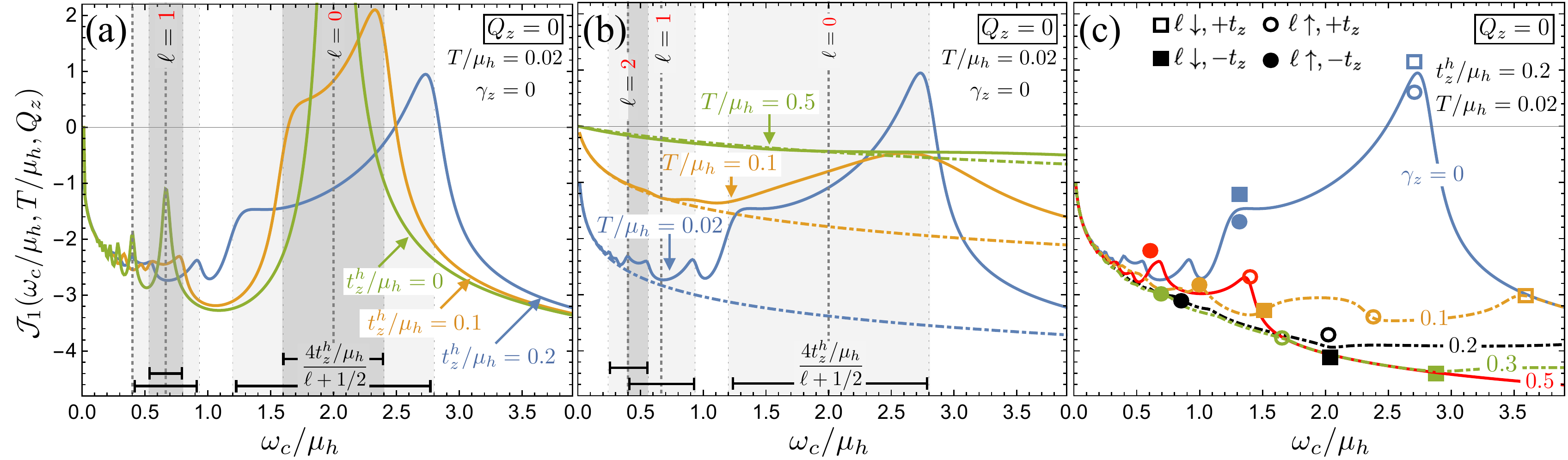}
	\vspace{-0.1in}  
	\caption{
The behavior of the kernel eigenvalue $\mathcal{J}_1$ in the uniform state,  $Q_z=0$, for
different interlayer tunnelings $t_z^h$ (a), temperatures (b), and spin-splitting factors
$\gamma_z$ (c). We can see that at low temperatures the function $\mathcal{J}_1(H)$ has pronounced
peaks when the chemical potential matches Landau levels broadened by the interlayer tunneling. From plots in
(a) we can see that the interlayer tunneling splits a single 2D peak into two smaller peaks
corresponding to the locations of the van Hove singularities in the spectrum at $k_z=0$ and $\pi$
[Eq.\ \eqref{eqn:ResCond}]. Plots in (b) show that thermal fluctuations smear the Landau levels
and eventually wash out the quantum effects near $T\sim \omega_c/2$. The dotted-dashed lines are the
quasiclassical results. Plots in (c) indicate that the uniform state is very sensitive to the
Zeeman spin-splitting effects: the Landau peaks are rapidly suppressed at very small $\gamma_z$.
Smaller peaks reappear when the resonant conditions are met ($2\gamma_z=j_z$). The symbols mark the miniband-edge magnetic fields, at which the chemical potential matches the van Hove points [Eq.\ \eqref{wc-vH}].
}	
	\label{fig:J1Qz0}
	\vspace{-0.1in}
\end{figure*} 

In the limit $T\ll t_{z}^{h}$ the dominating contribution to $\mathcal{J}_{1,0}$
is coming from the regions near the Fermi momenta meaning that we
can use expansion $k_{z}=k_{z0}+\tilde{k}$ with $\tilde{k}\ll k_{z0}$ and approximate
\begin{align*}
\mathcal{J}_{1,\ell_{0}} & \approx\!\frac{(2\ell_{0})!}{32\pi2^{2\ell_{0}}\left(\ell_{0}!\right)^{2}}\frac{\omega_{c}}{t_{z}^{h}\sin k_{z0}\cos\tfrac{Q_{\mathrm{op}}}{2}}\\
\times & \sum_{\delta=\pm1}\int\limits _{-\pi}^{\pi}\!\frac{\mathrm{d}\tilde{k}}{\tilde{k}}\tanh\left[\frac{t_{z}^{h}}{T}\sin\left(k_{z0}\!+\!\delta\tfrac{Q_{\mathrm{op}}}{2}\right)\tilde{k}\right].
\end{align*}
Evaluating the integral, we obtain 
\begin{align}
\mathcal{J}_{1,\ell_{0}} & \approx\frac{(2\ell_{0})!}{8\pi2^{2\ell_{0}}\left(\ell_{0}!\right)^{2}}\frac{\omega_{c}}{t_{z}^{h}\sin k_{z0}\cos\tfrac{Q_{\mathrm{op}}}{2}}\nonumber \\
\times &\ln\left[
\frac{t_{z}^{h}}{T}\sqrt{\left|\sin\left(k_{z0}\!-\!\tfrac{Q_{\mathrm{op}}}{2}\right)\sin\left(k_{z0}\!+\!\tfrac{Q_{\mathrm{op}}}{2}\right)\right|}
\right].
\label{eq:J1QzLowT}
\end{align}
This logarithmic divergence for $T\to 0$ is similar to the well-known Peierls divergence of the
electronic susceptibility for one-dimensional metals at the nesting wave vector which, in
particular, leads to the charge-density-wave transition. In the simplest scenario when the chemical potential crosses only two minibands,  the kernel eigenvalue diverges only at one wave vector. In more complicated cases with many minibands at the Fermi level, the divergence takes place at several wave vectors and one can expect competition between multiple ground states, similar to the situation considered in Ref.\ \cite{Takeshi:JPSJ83.2014,*Takahashi:PRB89.2014}.

\subsection{Magnetic-field dependences of quantum kernel eigenvalue $\mathcal{J}_1$ for different cases}
\label{sec:J1-H}

\begin{figure}[htbp] 
	\centering
	\includegraphics[width=3.3in]{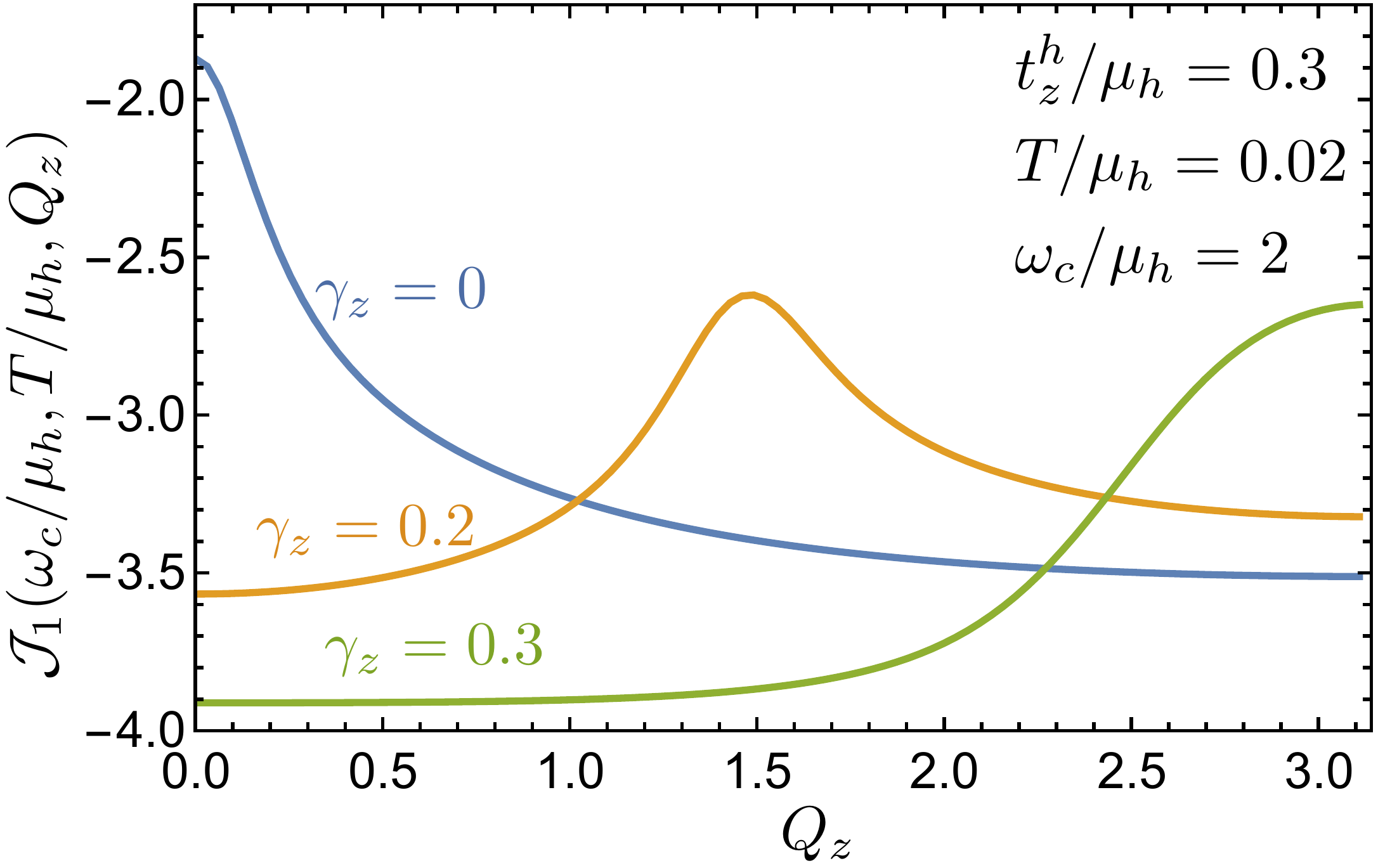}
	\caption{			
		The representative dependences of the pairing kernel eigenvalue on the modulation wave vector $Q_z$. In the resonance cases with integer $2\gamma_z$, the maximum of $\mathcal{J}_{1}$ is always realized at  $Q_z\!=\!0$ corresponding to the uniform state. The curve with $\gamma_z=0.2$ illustrates the nonresonant case, when the chemical potential crosses both spin-up and -down zero Landau-level minibands, as illustrated in Fig.\  \ref{fig:zeeman}(b). The maximum is realized at the modulation vector connecting the Fermi momenta of these minibands. Finally, the curve with $\gamma_z\!=\!0.3$ illustrates a special situation when the chemical potential matches two opposite van Hove points at $k_z\!=\!0$ and $\pi$ [see Fig.\  \ref{fig:zeeman}(c)]. In this case, the maximum is realized at $Q_z\!=\!\pi$, favoring the alternating FFLO state.
	}
	\label{fig:J1Qz}
\end{figure}

Now, we discuss the general behavior of the kernel eigenvalue $\mathcal{J}_1$. In contrast to the monotonic behavior of the quasiclassical result, the low-temperature divergences discussed in the Sec.\ \ref{sec:J1lowT} may lead to the emergence of strong peaks in $\mathcal{J}_1$ at the specific magnetic fields and modulating wave vectors $Q_z$ which indicate the enhancement of pairing. The magnitude and location of these peaks are sensitive to the details of electronic band properties. 

\subsubsection{Uniform state: $Q_z=0$}

Figure \ref{fig:J1Qz0} illustrates the effects of interlayer tunneling, temperature, and Zeeman spin-splitting in the \emph{uniform state} ($Q_z=0$). Without spin splitting, the interlayer tunneling broadens the Landau levels and splits a single divergent peak into two smaller peaks at low temperatures [see Fig. \ref{fig:J1Qz0}(a)]. The smaller peaks correspond to the matching between the chemical potential and the van Hove singularities in the spectrum at $k_z=0$ and $\pi$ [Eq.\ \eqref{eqn:ResCond}]. Their values are estimated by Eq.\ \eqref{eqn:J1Resonant}. The shaded regions mark the field ranges for which the chemical potential is located within the broadened Landau levels, and their boundaries correspond to the crossing of the van Hove singularities.

Figure \ref{fig:J1Qz0}(b) shows that the peaks are rapidly smeared by the thermal effects and disappear at $T\sim\omega_c/2$. In this figure, we compare the exact kernel eigenvalue with the quasiclassical results (dotted-dashed lines), which are calculated by using
\begin{align}
\mathcal{J}^{qc}_1(H,T,Q_z)&=2\int^\infty_0\mathrm{d}s\Big\langle\ln\tanh\left(\frac{\pi T}{\omega_c}s\right)\exp\left(-\tilde{\mu}_{h}s^2\right)\notag\\
\times&[\tilde{\mu}_{h}s\cos(2\tilde{\gamma}_{z}s)\!+\!\tilde{\gamma}_{z}\sin(2\tilde{\gamma}_{z}s)]\Big\rangle_z.\label{eqn:J1qc}
\end{align}
This quasiclassical approximation of $\mathcal{J}_1$ can be obtained from the quantum kernel in Eq.\ \eqref{J1llSumPres} by assuming small-field limit\cite{Champel:PhilMagB81.2001}, $\omega_c\ll \mu_h$ with the temperature range, $\omega_c\ll T\ll \mu_h$. One can see that this approximation yields a monotonic behavior for the kernel eigenvalue in the magnetic field and gives a good approximation at low fields. 

We can see in Fig.\ \ref{fig:J1Qz0}(c) that the divergent peaks are also rapidly suppressed by finite spin splitting $\gamma_z$ meaning that the uniform state is highly susceptible to the Zeeman effect. The smaller peaks reappear if the resonant conditions are met (the solid red line for $\gamma_z\!=0\!.5$). 
We also observe a noticeable enhancement of $\mathcal{J}_1$ in the field ranges where the chemical potential crosses both spin up and spin down minibands for the zero Landau level (e. g., in the range $1.5<\omega_c/\mu_h<2.4$ for $\gamma_z\!=\!0.1$). We point, however, that in this range the maximal  $\mathcal{J}_1$ is not at $Q_z\!=\!0$. We discuss the $Q_z$ dependences in the next subsection.

\subsubsection{FFLO modulated state: $Q_z\neq0$}

We now consider the behavior of the quantum kernel eigenvalue at finite modulation wave vectors $Q_z$.
Figure \ref{fig:J1Qz} shows the dependences of $\mathcal{J}_1$ on $Q_z$ for fixed electronic-spectrum parameters, fixed magnetic field, and for three spin-splitting factors representing different cases.
In the resonant cases with half-integer spin-splitting factors the kernel eigenvalue always has the maximum at $Q_z\!=\!0$, as illustrated by the curve with $\gamma_z\!=\!0$. The finite-$Q_z$ modulated states are not favorable in these cases, due to the absence of strong Zeeman pair breaking at the Landau levels [see Fig.\ \ref{fig:zeeman}(a)], and \emph{the uniform state always dominates}. Away from the uniform-state resonances, $2\gamma_z\neq j_z$, the Zeeman pair breaking favors developing of the finite-$Q_z$ states. The second curve with  $\gamma_z=0.2$ illustrates the situation when the chemical potential crosses both spin-up and -down zero Landau-level minibands. In this case, the maximum is realized at the modulation wave vector connecting the minibands Fermi momenta [see Fig.\  \ref{fig:zeeman}(b)]. This modulation vector varies with the magnetic field. The third curve with $\gamma_z=0.3$ illustrates a special situation when the chemical potential matches two opposite van Hove points at $k_z=0$ and $\pi$ [see Fig.\  \ref{fig:zeeman}(c)]. Such matching is realized when the parameters satisfy the relation given by Eqs.\  \eqref{eqn:pi-wc} and \eqref{eqn:pi-cond} (the plot is made for $\ell_0=\ell_\pi=0$). In this case, the maximum is realized at $Q_z=\pi$, favoring the alternating FFLO state.

\begin{figure}[htbp] 
	\centering \includegraphics[width=3.4in]{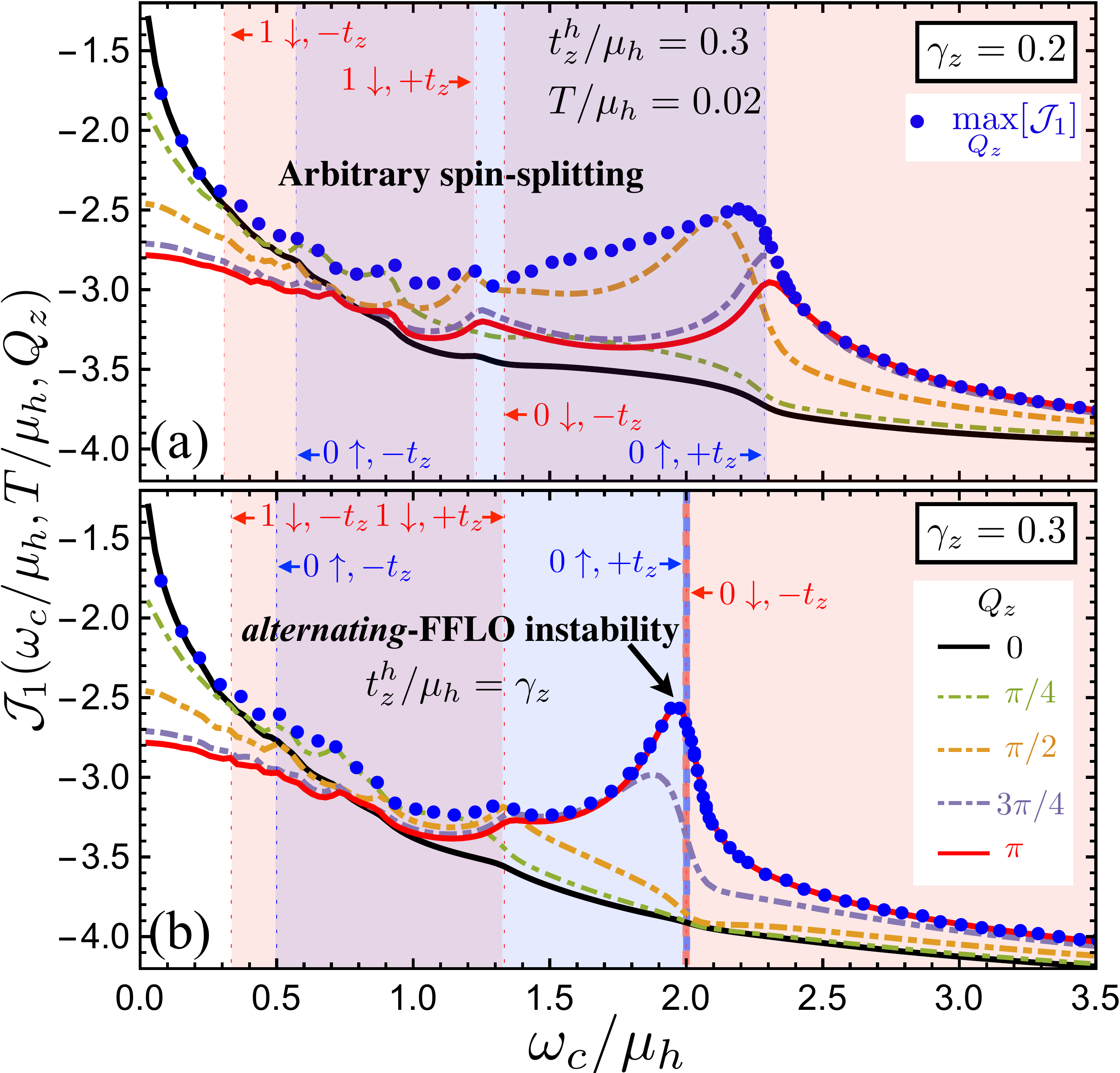}
	\caption{	
		The magnetic field dependences of $\mathcal{J}_1$ for different
		modulation wave vectors $Q_z$ for noninteger $2\gamma_z$. 
		The parameters in the panel (a),
		$t_z/\mu_h=0.3,\gamma_z=0.2$, correspond to a general case without any resonances. In this
		case, states with different $Q_z$ are favorable for different magnetic fields.
		The parameters in the panel (b),
		$t_z/\mu_h\!=\!\gamma_z\!=\!0.3$, favor $\pi$-FFLO instability. As a consequence, a new diverging
		peak appears for $Q_z=\pi$ state at $\omega_c=2\mu_h$. 
		In both panels, we also show the magnetic-field dependence of the maximum value of $\mathcal{J}_1$ with respect to $Q_z$ (blue circles).
	}
	\label{fig:J1QzNoRes}
\end{figure}
Figure \ref{fig:J1QzNoRes} illustrates the magnetic field dependences of  $\mathcal{J}_1$ for different $Q_z$ in two situations with noninteger $2\gamma_z$. We also show the maximal $\mathcal{J}_1$ with respect to $Q_z$. The light blue and pink regions mark the crossing of the chemical potential with the mismatching spin-up and -down Landau-level minibands. This regions are limited by the miniband-edge magnetic fields given by Eq.\ \eqref{wc-vH} with $\delta_{t}=\pm 1$ while $\ell$ and $\sigma$ are fixed. The \emph{simultaneous} crossing of both such bands at the chemical potential only occurs in the overlapping regions in which the pairing favors finite-$Q_z$ states as illustrated in Fig. \ref{fig:zeeman}(b). Figure \ref{fig:J1QzNoRes}(a) corresponds to a general situation without any resonances. In this case, the states with different $Q_z$ become favorable in different fields. In particular, for $Q_z=\pi/2$ the maximum at $\omega_c/\mu_h\approx 2.1$ corresponds to the condition for optimum modulation vector $Q_{op}(\omega_c)=Q_z$ [Eq.\ \eqref{eq:Qop}].  We can also see that the pairing strength is noticeably enhanced in the region $1.3<\omega_c/\mu_h<2.3$ when the chemical potential crosses zero-Landau level minibands for both spin orientations. The choice of parameters in Fig.\ \ref{fig:J1QzNoRes}(b) allows for $Q_z=\pi$ singularity [see Eq.\ \eqref{eqn:pi-cond} and the illustration in Fig.\ \ref{fig:zeeman}(c)]. At $\omega_c/\mu_h=2$, the chemical potential matches simultaneously two van Hove singular points: the spin-up energy at $k_z=0$ and spin-down energy at $k_z=\pi$. This matching leads to the additional divergent peak at this point strongly favoring $Q_z=\pi$ instability and may lead the formation of the superconducting state with the alternating sign of the order parameter between the layers. In both cases, the optimal kernel eigenvalue (blue circles) significantly exceeds the one for the uniform state (black line) almost in the whole field range. 

\begin{figure*}[hptb] 
	\centering
	\includegraphics[width=2.42in]{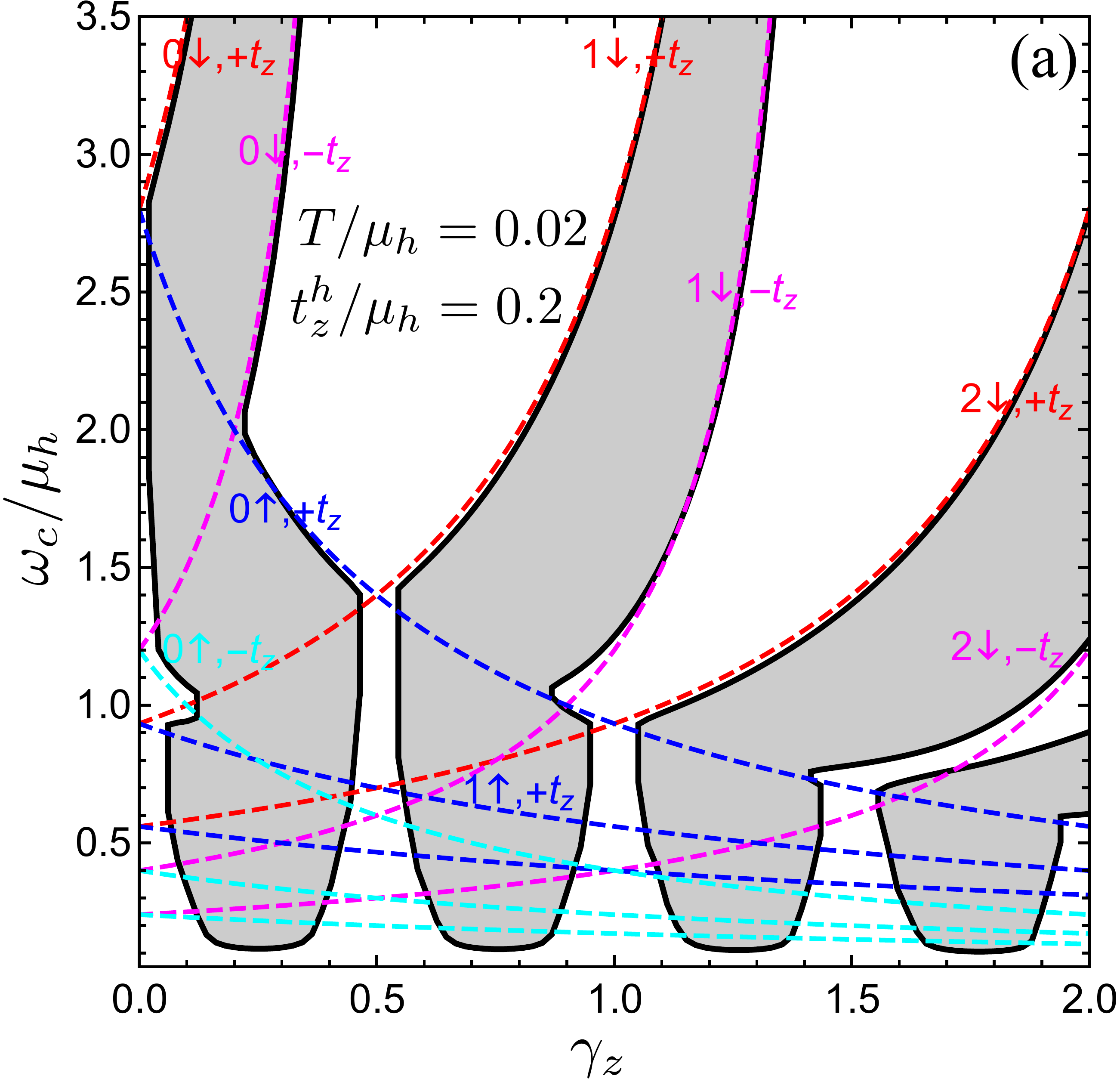} 
	\includegraphics[width=2.27in]{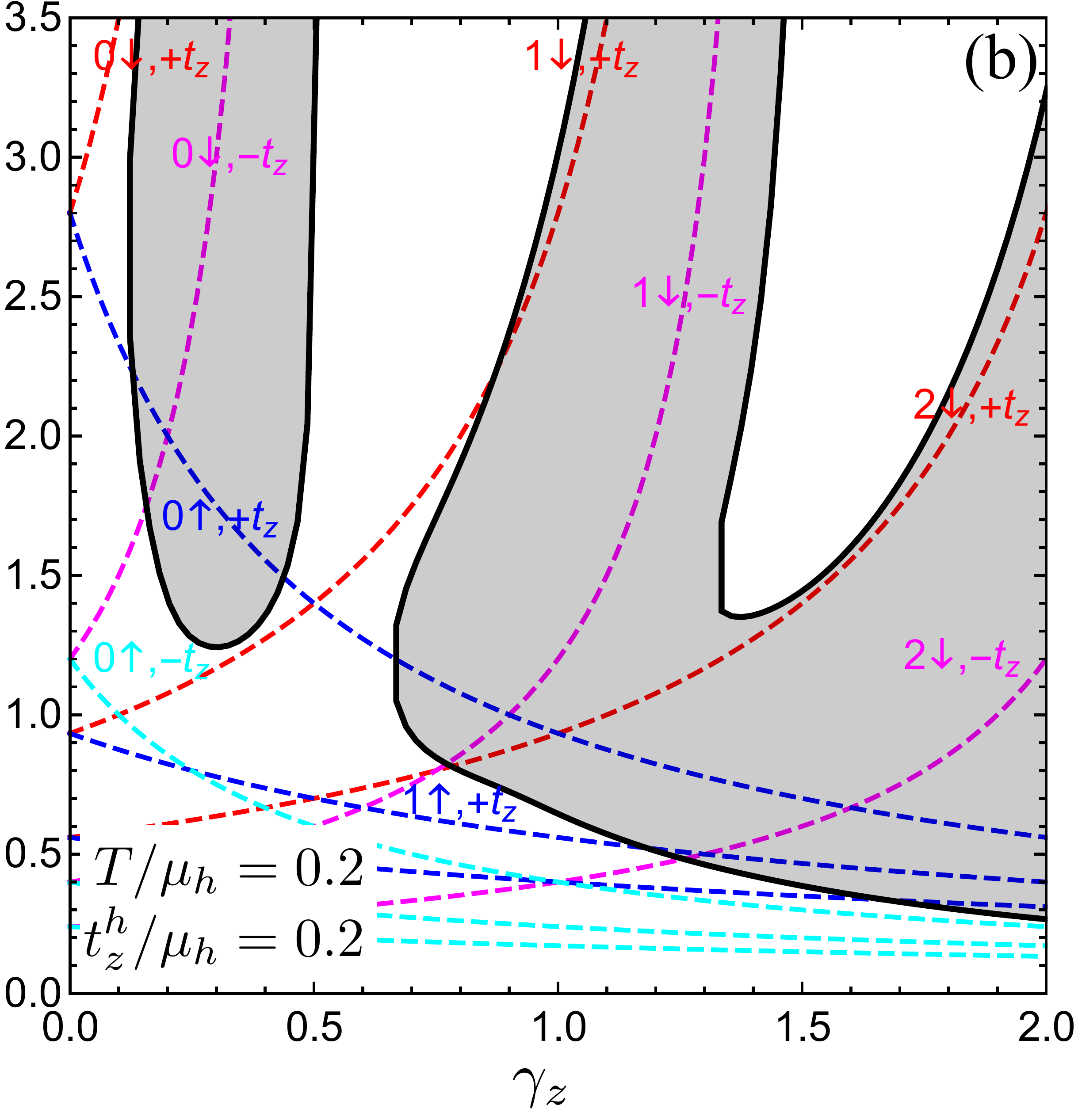} 
	\includegraphics[width=2.27in]{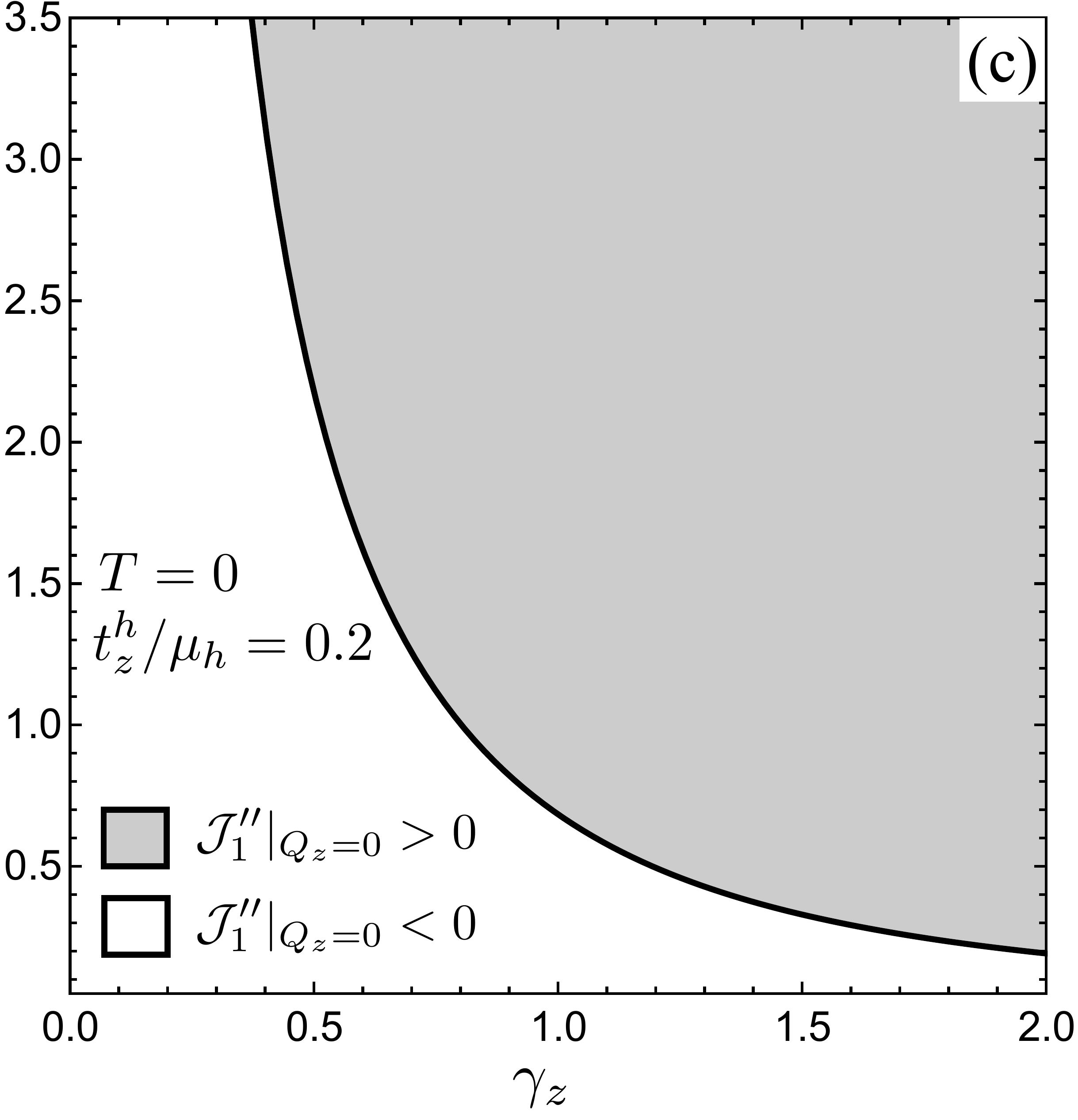} 
	\caption{These plots show positive and negative regions of the  kernel-eigenvalue second derivative with respect to the modulation wave vector $Q_z$, $\mathcal{J}''_1(H,T,0)$, in the spin-splitting--magnetic field plane. In the shaded regions with $\mathcal{J}''_1(H,T,0)\!<\!0$ the onset of the finite-$Q_z$ FFLO states may be expected. The (a) and (b) diagrams are made using the exact kernel eigenvalue at $T\!=\!0.02\mu_h$ and $0.2\mu_h$. The dashed lines trace the magnetic fields for which the  van Hove singularities of the Landau-level minibands match the chemical potential exactly, Eq.\ \eqref{wc-vH}. The label `$+t_z$' (`$-t_z$') in the diagram marks the $k_z=\pi$ ($k_z=0$) van Hove singular point. (c) The diagram for the quasiclassical pairing kernel at zero temperature, Eq.\ \eqref{eqn:J1qc}.}
	\label{fig:FFLOcond}
\end{figure*}
To obtain a better presentation for the range of parameters where the shallow band favors nonuniform states,
we expand $\mathcal{J}_1$ with respect to small $Q_z$:
\begin{equation}
\mathcal{J}_1(H,T,Q_z)\approx\mathcal{J}_1(H,T,0)+\frac{1}{2}\mathcal{J}''_1(H,T,0)Q_z^2.
\end{equation}
If $\mathcal{J}''_1(H,T,0)$ is positive, then $Q_z\!=\!0$ does not maximize $\mathcal{J}_1$ and the finite non-zero $Q_z$ state is favorable. For better visual impression of the parameter range where the shallow band favors nonuniform states, we present in Fig.\ \ref{fig:FFLOcond} the regions of positive  (shaded) and negative (unshaded) $\mathcal{J}''_1(H,T,0)$ in the magnetic field--spin splitting plane for representative parameter $t_z^h/\mu_h=0.2$ and two values  of temperature $T=0.02\mu_h$(a) and $0.2\mu_h$(b).  For comparison, we also show in Fig.\ \ref{fig:FFLOcond}(c) the same diagram for the quasiclassical kernel eigenvalue, Eq.\ \eqref{eqn:J1qc}. In the latter case an analytical analysis \cite{Gruenberg:PRL16.1966} suggests that the boundary should behave as $\omega_{c}/\mu_{h}\propto 1/\gamma_z^2$ while the numerical fitting gives $\omega_{c}/\mu_{h}\approx 1/(0.14+1.07\gamma_z)^2$. If the upper critical field with $Q_z=0$ state falls into the shaded region, this means that the shallow band favors the nonuniform state, and the order parameter has a tendency to develop nonzero $Q_z$ modulation in real space. We see that at low temperatures the shaded regions  cover substantial part of the phase diagram except locations around the integer $2\gamma_z$. Especially surprising is that they extend to very low fields, down to $\omega_c/\mu_h\!\sim\! 0.1$. These low-field regions are eliminated at higher temperatures. 

To avoid misunderstanding, we note that for two-band systems the nonuniform states do not automatically appear in the shaded regions, because the ground state is determined by both bands and the deep band typically favors the uniform state, $\mathcal{J}_{2}^{\prime\prime}(H,T,0)<0$. The FFLO modulation may appear below certain temperature $T_{\mathrm{FFLO}}$, at which the second derivative $\partial^2 H_{C2}/\partial Q_z^2$ changes sign. 
From the general equation for the upper critical field [Eq.\ \eqref{eqn:HC2}] we can derive the equation for  this temperature, $T=T_{\mathrm{FFLO}}$,
\begin{equation}
\frac{\mathcal{J}_{1}^{\prime\prime}(H,T,0)}{W_{11}\!+\!\mathcal{A}_{1}(T)\!-\!\mathcal{J}_{1,0}(H,T)}
+\frac{\mathcal{J}_{2}^{\prime\prime}(H,T,0)}{W_{22}\!+\!\mathcal{A}_{2}(T)\!-\!\mathcal{J}_{2,0}(H,T)}\!=\!0
\label{eq:TFFLO}
\end{equation}
with $\mathcal{J}_{\alpha,0}(H,T)\equiv\mathcal{J}_{\alpha}(H,T,0)$, 
in which we have to substitute $H\to H_{C2}(T)$ at $Q_z\!=\!0$. This equation may have solution only if (i) $\mathcal{J}_{1}^{\prime\prime}(H,T,0)>0$ at low temperatures [i.e., the parameters are in the shaded region in Fig.\ \ref{fig:FFLOcond}(a)] and (ii) the ratio $W_{11}/W_{22}$  characterizing the relative weight of the shallow band is not too small.

\section{Superconducting instabilities and phase diagrams}\label{sec:HT}

The exact shape of the $H$-$T$ phase diagram of a layered multiple-band superconductor depends on many parameters: electronic spectra properties of the bands, spin-splitting factors, and structure of the coupling matrix. The most crucial factor is the quantum pairing kernel eigenvalue for the shallow band, $\mathcal{J}_1$, whose shape at low temperatures is extremely sensitive to the electronic parameters of the shallow band, $\mu_h$, $t_z^h$, and $\gamma_z$, as we discussed in the previous section. With the finite interlayer tunneling, the $k_z$ dispersion in the Landau-quantization spectrum can lead to complex behavior of the superconducting state in the magnetic field.

In addition, the Zeeman spin splitting in the shallow band favors the formation of the FFLO states (see Figs.\ \ref{fig:zeeman} and \ref{fig:J1QzNoRes}). As the Landau quantization strongly influences the Cooper pairing in the bands with relatively low Fermi energies, this brings a natural general question about the role of quantum effects in the FFLO instability, even in the common case of a single-band superconductor in the quasiclassical limit. Surprisingly, this important topic has been never addressed before and  we leave it to a separate paper. Here, we focus on the general behavior of the $H$-$T$ phase diagrams for the two-band system where the shallow band is close to the extreme quantum limit. We illustrate this behavior for several representative cases.

The most spectacular consequence of the Landau quantization is the emergence of the high-field superconducting states at the magnetic fields corresponding to matching of the chemical potential with the Landau levels. These states are most pronounced in the two-dimensional case and for the lowest Landau level \cite{Song:PRB95.2017}.  The interlayer tunneling smears the delta-function singularity in the density of states, leading to suppression of the high-field superconductivity. 
To quantify these effects, we first calculate the transition temperatures in magnetic field, $T_{C2}$, in the resonance cases for which the kernel eigenvalue diverges as $\sim 1/\sqrt{T}$, see Sec.\ \ref{sec:J1lowT}. For this we find solution of Eq.\ \eqref{eqn:HC2} assuming that $T_{C2}\ll T_C, t_z$. 
\begin{figure}[htbp] 
	\centering
	\includegraphics[width=3.4in]{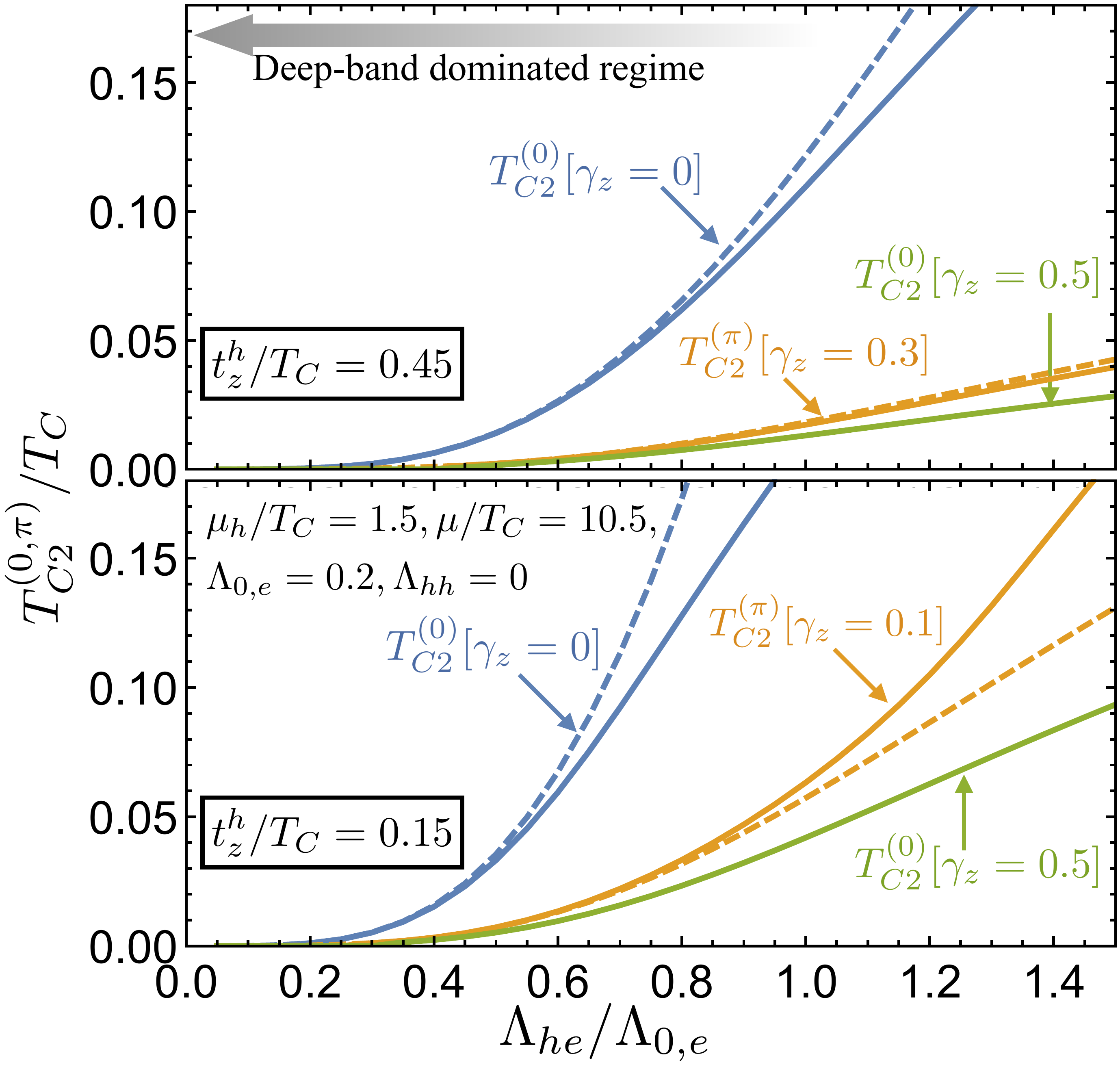} 
	\caption{The dependences of the transition temperatures $T_{C2}$ on the interband coupling constant $\Lambda_{he}$ for the superconducting states that are induced by the strongest $1/\sqrt{T}$ divergences at high magnetic field. The plots are made assuming fixed effective coupling constant $\Lambda_{0,e}\!=\!0.2$. The blue (green) curves show the dependences of uniform state $T_{C2}$ on the interband coupling $\Lambda_{he}\!=\!\Lambda_{eh}$ at the magnetic field $\omega_c\!=\!2\mu_h\!+\!4t_z$ ($\omega_c\!=\!\mu_h\!+\!2t_z$) with $\gamma_z\!=\!0$ ($\gamma_z\!=\!0.5$), while the orange curves are for the alternating-state case at the magnetic field $\omega_c\!=\!2\mu_h$ with $\gamma_z\!=\!0.3$ (upper plot) and $0.1$ (lower plot). The dashed lines are the approximate results  given by Eqs.\ \eqref{eqn:TC20p} and \eqref{eqn:TC2pi} (which are accurate only if $t_z\ll T_{C2}$) and the solid lines are calculated using the exact formula in Eq.\ \eqref{eqn:HC2}. The upper and lower figures compare two different values of the interlayer hopping energy $t_z^h$.}
	\label{fig:TC2}
\end{figure}

First, we evaluate the transition temperature for the strongest resonance in the uniform state without Zeeman spin-splitting ($\gamma_z\!=\!0$), $T^{(0)}_{C2}$, which is realized when the chemical potential matches the van Hove singular points at $k_z\!=\!0,\pi$ at the magnetic fields $\omega_c\!=\!2(\mu_h\! \mp \! 2t^h_z)$. 
In this case, substituting the low-$T$ asymptotics of $\mathcal{J}_\alpha\!-\!\mathcal{A}_\alpha$ given by Eqs.\ \eqref{A2J2LowT} and \eqref{eqn:J1T0R} into Eq.\ \eqref{eqn:HC2}, we derive
\begin{equation}\label{eqn:TC20p}
T^{(0)}_{C2}\!=\!\frac{\mathit{C}^2\omega_c^2}{t^h_z}\!\Bigg[\frac{W_{11}\ln r_C^{(0)}}{2W_{22}\!+\!\ln r_C^{(0)}}
\!-\!\mathcal{R}_{\mp}(\bar{t}_z^h)
\!+\!\frac{1}{2}\ln\frac{\omega_c}{\pi T_C}\!+\!\Upsilon_{C}\!\Bigg]^{\!-2}\!.
\end{equation}
This result allows us to understand better the typical sizes of the high-field reentrant regions.
A similar result can be derived for the strongest alternating-state resonance with $Q_z\!=\!\pi$ realized when the parameters satisfy the relation  $\mu_h=t_z^h/\gamma_z$.  In this case, the low-temperature asymptotics of $\mathcal{J}_1-\mathcal{A}_1$ is given by Eq.\ \eqref{eqn:J1T0pi}, leading to the $\pi$-state transition temperature at $\omega_c=2\mu_h$:
\begin{equation}\label{eqn:TC2pi}
T^{(\pi)}_{C2}\!=\!\frac{\mathit{C}^2\omega_c^2}{4t^h_z}\Bigg[
\frac{W_{11}\ln r_C^{(\pi)}}{2W_{22}\!+\!\ln r_C^{(\pi)}}
\!+\!\frac{1}{2}\ln\frac{2\omega_c}{\pi T_C}\!+\!\Upsilon_{C}\Bigg]^{-2}
\end{equation}
with $r_C^{(\pi)}\equiv r_C(Q_z\!=\!\pi)$. 
Even though the overall scale of the above transition temperatures is given by $\omega_c^2/t^h_z$ with the small numerical factor $\mathit{C}^2\approx 0.0115$, their absolute values are very sensitive to the structure of the coupling matrix. The latter dependence is given by the first term in the square brackets in Eqs.\ \eqref{eqn:TC20p} and \eqref{eqn:TC2pi}. In general, these transition temperatures are not vanishingly small only if the shallow band gives substantial contribution to pairing. 

In particular, in the case when the deep band dominates pairing, $\Lambda_{ee}\!>\!
\Lambda_{hh},|\Lambda_{eh}|,|\Lambda_{he}|$, the constants
$W_{\alpha\alpha}$ can be estimated as $W_{11}\!\approx \!\Lambda_{ee}/\mathcal{D}_{\Lambda}$
and  $W_{22}\!\approx \!\Lambda_{eh}\Lambda_{he}/(\Lambda_{ee}\mathcal{D}_{\Lambda})$, meaning that
$|W_{22}|\ll |W_{11}|$. In addition, inequalities $\ln r_C \ll |W_{22}|$ and $W_{11}\ln r_C/W_{22}\!\approx\!
\Lambda_{ee}^2\ln r_C/\Lambda_{eh}\Lambda_{he}\!\gg\! \Upsilon_{C},\ln(\mu_h/T_c)$ are typically satisfied, leading to the estimate 
$T^{(0)}_{C2}\!\approx \!\left(\mathit{C}^2\omega_c^2/t^h_z\right) \left[2\Lambda_{eh}\Lambda_{he}/(\Lambda_{ee}^2\ln r_{C}^{(0)}) \right]^2$. In this case, the transition temperature has the additional small factor which rapidly decreases with decreasing of the interband couplings constants, $\Lambda_{eh}$ and $\Lambda_{he}$, and with increasing the deep-band coupling constant, $\Lambda_{ee}$. On the other hand, we see that $T^{(0)}_{C2}$ increases when the resonance field approaches the orbital field of the shallow band, $r_C^{(0)}\!\to \!1$. In the interband-coupling scenario, $|\Lambda_{eh}|,|\Lambda_{he}|\!>\!
\Lambda_{ee},\Lambda_{hh}$, the ratio $W_{11}/W_{22}\! \approx 1/2$, meaning that there is no additional smallness caused by the coupling factor. 

We illustrate the typical behavior of $T_{C2}$ in Fig.\ \ref{fig:TC2}, by plotting the transition temperatures for uniform and alternating states as function of the ratio $\Lambda_{h,e}/\Lambda_{0,e}$ which controls the shallow-band weight. The plots are made for fixed effective coupling constant $\Lambda_{0,e}=0.2$ which fixes the zero-field transition temperature $T_C$. As expected, in all cases the transition temperatures rapidly decrease with decreasing of the shallow-band contribution to pairing. For comparison, we present the $T_{C2}$-dependences for two values of the interband hopping energy $t_z^h=0.45T_C$ and  $0.15T_C$. We see that $T_{C2}$'s are much higher for smaller  $t_z^h$.

We also note that, in spite of the kernel low-temperature divergences in the resonance cases, the
two-band system does not always become superconducting at the corresponding magnetic fields. Indeed, the
above solutions for $T^{(0)}_{C2}$ and $T^{(\pi)}_{C2}$ are only valid if the arguments in the
square brackets of Eqs.\ \eqref{eqn:TC20p} and \eqref{eqn:TC2pi} are positive. When the interband
coupling dominates, $\mathcal{D}_{\Lambda}\!<\!0$, the parameters $W_{\alpha \alpha}$ are
negative. In this case, $T^{(\alpha)}_{C2}$ may vanish if $\ln r_C^{(\alpha)}=2|W_{22}|$. 
As $|W_{22}|\gg 1 $, this may only realize if the resonance magnetic field is much larger than the
deep-band orbital field,  $r_C^{(\alpha)}\gg 1$, corresponding to the very large Fermi energy, $\mu >
100 T_C$. 
\begin{figure}[htbp] 
	\centering
	\includegraphics[width=3.4in]{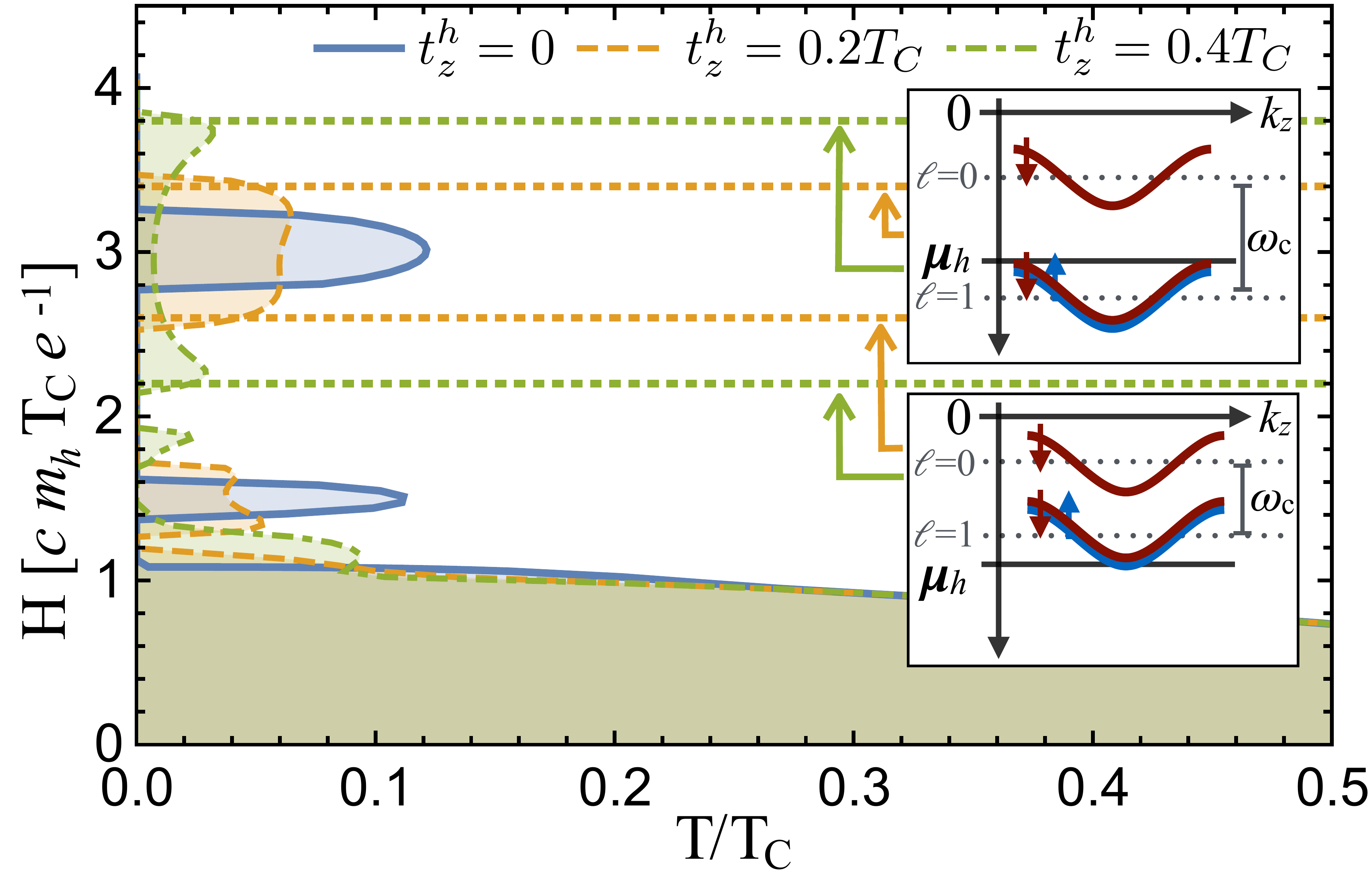} 
	\caption{The influence of the interlayer tunneling on the typical $H$-$T$ diagram for the resonant case ($2\gamma_z=1$) in the low-$T$ regime. We used the following parameters: the coupling constants  $\Lambda_{hh}\!=\!\Lambda_{ee}\!=\!0$, $\Lambda_{he}\!=\!\Lambda_{eh}\!=0.3$, the mass ratio is $m_e/m_h\!=\!1$,  $\varepsilon_0\!=\!12T_C$, and $\mu_h/T_C\!=\!3$. The insets demonstrate the crossing of the van Hove singular points at the chemical potential.}
	\label{fig:HC2R}
\end{figure}
\begin{figure*}[htbp] 
	\centering 
	\includegraphics[width=7in]{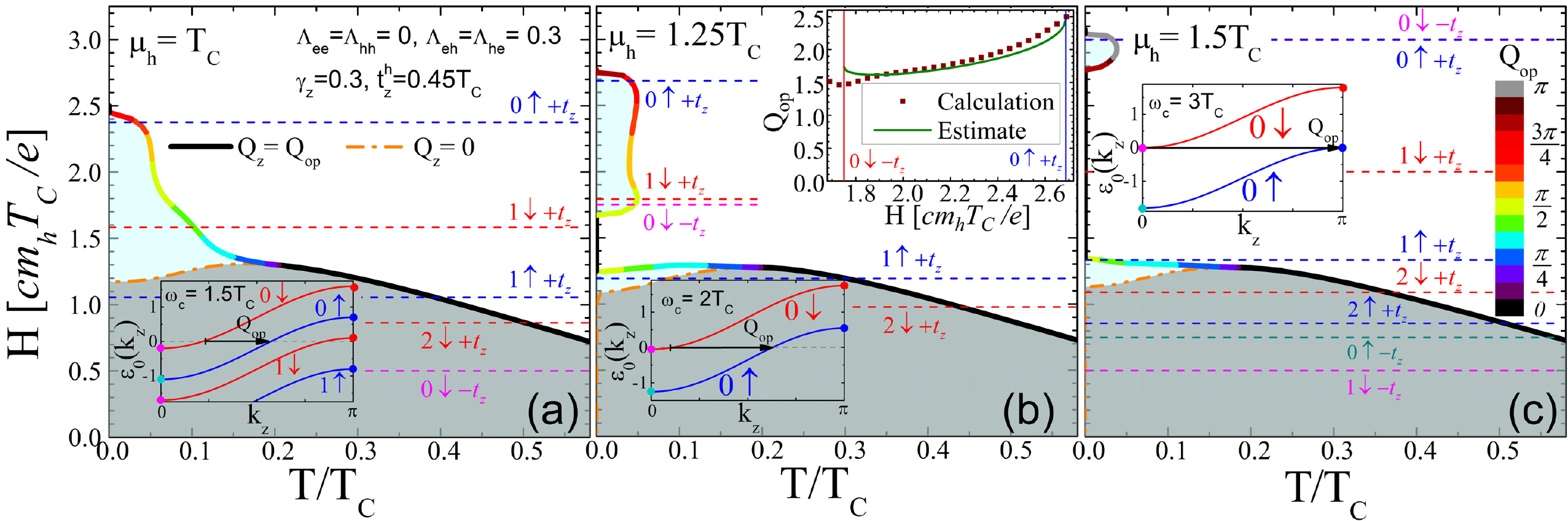} 		
	\caption{The representative $H$-$T$ phase diagrams for nonresonant spin-splitting factor $\gamma_z=0.3$ and interband-coupling scenario (the coupling constants $\Lambda_{hh}\!=\!\Lambda_{ee}\!=\!0$, $\Lambda_{he}\!=\!\Lambda_{eh}\!=\!0.3$) computed for three Fermi energies. Other used parameters are:  the interlayer hopping energies $t_z^h\!=\!t_z^e\!=\!0.45T_C$,  the mass ratio $m_e/m_h\!=\!1$, and $\varepsilon_0\!=\!12T_C$. The boundaries are color coded by the optimal-modulation wave vector $Q_{\mathrm{op}}$. The orange dotted-dashed line shows instability boundary for the uniform state $Q_z=0$. The horizontal dashed lines are the magnetic fields that match the Fermi level of spin-up/-down electrons in the $h$ band with $k_z=0,\pi$ [see Eq.\ \eqref{wc-vH}].  (a) For $\mu_h\!=\!T_C$ we can see that the formation of the nonuniform FFLO state below $0.18T_C$ leads to significant enhancement of $H_{C2}$ at low temperatures caused by the proximity to the lines $0\uparrow,+t_z$ and $1\downarrow,+t_z$. The inset in this and other plots demonstrate the Landau-level minibands  near the chemical potential for selected magnetic fields and corresponding optimal modulation vectors. (b) For the larger Fermi energy $\mu_h=1.25T_C$, the reentrant region appears located between the lines $0\uparrow,+t_z$ and $0\downarrow,-t_z$, where the chemical potential crosses the zero-Landau-level minibands for both spin orientations.	The upper right inset in this plot shows the optimal modulated wave vector $Q_{\mathrm{op}}$ evaluated from Eq.\ \eqref{eq:Qop} (green line) and the realized one (solid brown points). (c) The value of the Fermi energy $\mu_h\!=\!1.5T_C$ in this plot satisfies the relation $\mu_h=t_z^h/\gamma_z$, which is required for realization of the $\pi$-FFLO state around the field $\omega_c\!=\!2\mu_h$. Indeed, matching of the levels $0\uparrow,+t_z$ and $0\downarrow,-t_z$ at this field leads to the reentrant high-field state with $Q_{\mathrm{op}}=\pi$.  }	
	\label{fig:HC2q}
\end{figure*}

We now present numerically computed magnetic field--temperature phase diagrams for different typical situations. We first illustrate the influence of the interlayer tunneling on the reentrant high-field states in the resonant cases with integer $2\gamma_z$. Figure \ref{fig:HC2R} shows the $H$-$T$ diagrams for  $\gamma_z\!=\!0.5$, $\mu_h\!=\!3T_C$, and different hopping energies $t_z^h$. The diagrams are made for the case of dominating interband coupling scenario ($\Lambda_{hh}\!=\!\Lambda_{ee}\!=\!0$, $\Lambda_{he}\!=\!\Lambda_{eh}\!=\!0.3$), for which the discussed reentrant effects are very pronounced. 
The vertical scale of the magnetic field $cm_hT_C/e$ in the plot can be rewritten as $7.44 \text{T}(m_h/m_0)(T_C/10\text{K})$. Two reentrant states are realized for selected parameters, around the zeroth and first Landau levels. We can see that the interlayer tunneling smears these states. It splits the single peak in $T_{C2}(H)$ around the field $\omega_c\!=\!\mu_h(\ell+\frac{1}{2}+\frac{1}{2}j_z)^{-1}$ into the two peaks with maximums approximately located at $\omega_c\!=\!(\mu_h\pm2t_z^h)(\ell+\frac{1}{2}+\frac{1}{2}j_z)^{-1}$ (in the plot $\ell\!=\!0,1$ and $j_z\!=\!1$). The latter fields correspond to the crossing of the Fermi level and the van Hove singular points of the shallow $h$-band given by Eq.\ \eqref{wc-vH}, as illustrated by the insets. Since the increase of $t_z^h$ reduces the low-$T$ divergent peaks in $\mathcal{J}_1(H)$ dependence [Eq.\ \eqref{eqn:J1Resonant}] the reentrant $T_{C2}$ are always smaller than in the two-dimensional case. We emphasize again that for the resonant cases, $2\gamma_z\!=\!j_z$, the FFLO state is not favorable and $Q_z=0$ always gives the largest $T_{C2}$.

We now discuss the formation of the interlayer FFLO states due to spin splitting in the shallow band for a more general case of  noninteger  $2\gamma_z$. Figure \ref{fig:HC2q} shows the representative $H$-$T$ phase diagram for the spin-splitting factor $\gamma_z\!=\!0.3$, the interlayer hopping energy $t_z\!=\!0.45T_C$, and three values of the Fermi energy, $\mu_h\!=\!T_C$ (a), $1.25T_C$ (b), and $1.5T_C$ (c).  We again consider the interband-coupling scenario with the same coupling constants as in the previous figure. We plot the $H_{C2}$ lines with $Q_z$ at the optimal values ($Q_{\mathrm{op}}$) and the lines are color coded by $Q_{\mathrm{op}}$. For comparison, we also show the $Q_z\!=\!0$ transition lines. 

In the case  $\mu_h=T_C$ [Fig. \ref{fig:HC2q}(a)], the Zeeman spin splitting in the Landau levels of the shallow $h$ band leads to the pairings that favor the FFLO modulation along the out-of-plane direction for $T<0.18T_C$. The formation of this state leads to substantial enhancement of the upper critical field at low temperatures. The modulation wave vector at the transition  $Q_{\mathrm{op}}$ rapidly increases with decreasing temperature, reaching 2.17 at $T\to 0$. At somewhat higher Fermi energy $\mu_h\!=\!1.25T_C$ [Fig. \ref{fig:HC2q}(b)] the pairing at overlapping spin-up and spin-down zero-Landau-level minibands leads to the formation of the separated reentrant high-field state. This state also has $z$-axis modulation with the wave vector $Q_{\mathrm{op}}$ shown in the inset and its value is very close to the distance between spin-up and -down Fermi momenta given by Eq.\ \eqref{eq:Qop}. Note that the reentrant FFLO state occupies a quite extended field range corresponding to $1.5\!<\!\omega_c/T_C\!<2.75$, where the spin-up and -down zero-Landau-level minibands overlap.

As demonstrated in Sec.\ \ref{sec:J1LowTQzpi}, a layered superconductor with shallow band may have special resonance situations leading to appearance of the \emph{alternating} FFLO state. Namely, if $\mu_h$ satisfies the condition in Eq.\ \eqref{eqn:pi-cond}, $\mathcal{J}_1$ diverges for $T\to 0$ at the magnetic field that is given by Eq.\ \eqref{eqn:pi-wc}. This divergence typically gives rise to the FFLO modulation with $Q_z\!=\!\pi$. Figure \ref{fig:HC2q}(c) shows an example of the $H$-$T$ diagram for such situation with $\ell_0\!=\!\ell_{\pi}\!=\!0$ in Eqs.\ \eqref{eqn:pi-wc} and \eqref{eqn:pi-cond} corresponding to the strongest resonance. We can see that the small reentrant region indeed appears at the high magnetic field. However, only part of this region is occupied by the alternating state. In contrast to the uniform state, $T_{C2}(H)$ has only single maximum, as in the two-dimensional case. We see that, depending on parameters, the reentrant states can be either well separated from or very close to the main superconducting region.

\begin{figure}[htbp] 
	\centering
	\includegraphics[width=3.4in]{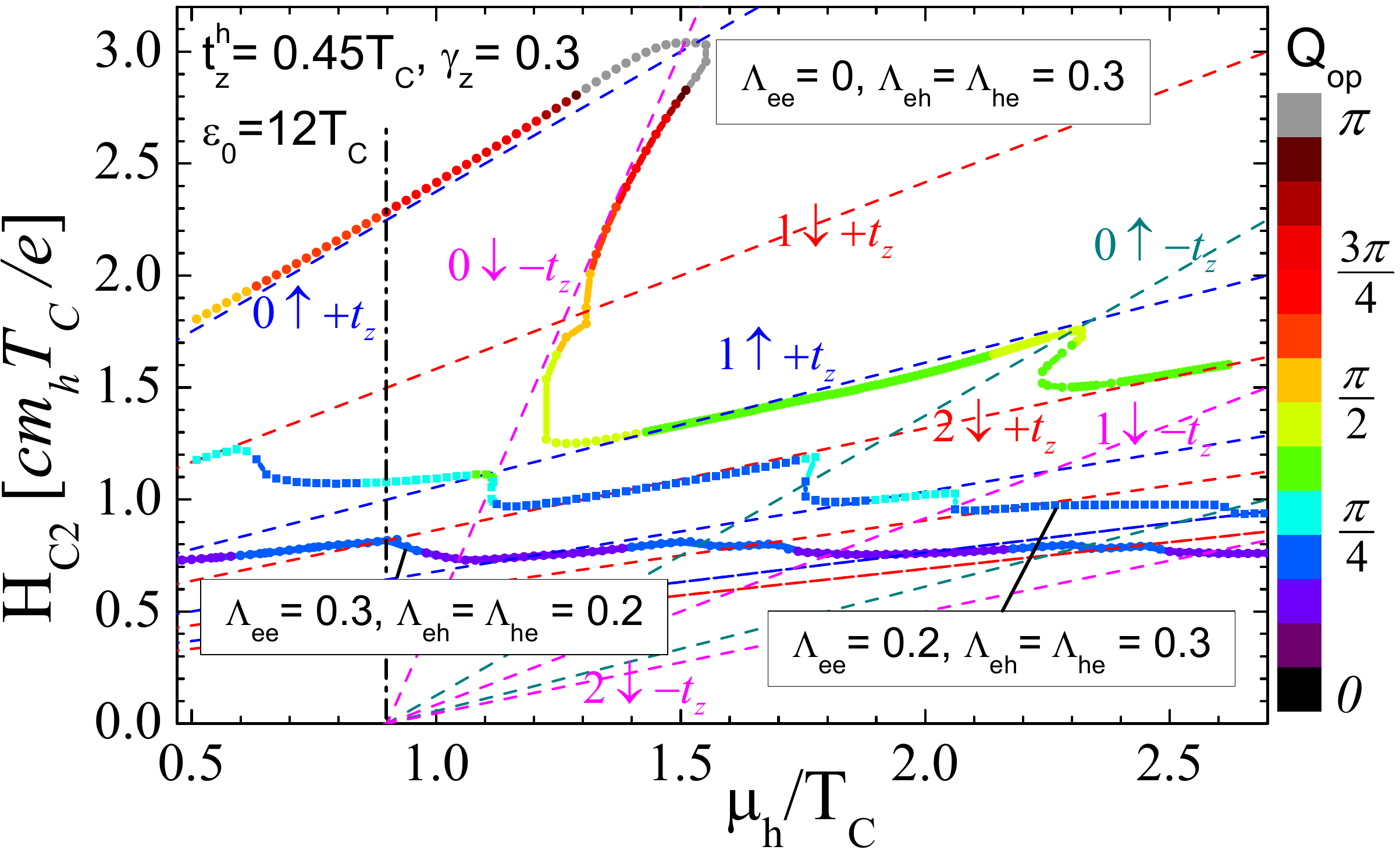}
	\caption{The dependences of the upper critical field on the shallow-band Fermi energy $\mu_h$ for three different coupling-constants sets shown in the plot. Other parameters are the same as in Fig.\ \ref{fig:HC2q}. As in previous figures, we also show the miniband-edge field lines. The vertical  dotted-dashed marks the location of the first Lifshitz transition (see Fig.\ \ref{fig:LT}).}
	\label{fig:Hc2muhtz045gz03}
\end{figure}
For better understanding of general trends, we plot in  Fig.\ \ref{fig:Hc2muhtz045gz03} the dependences of the low-temperature upper critical field on the shallow-band Fermi energy $\mu_h$. The upper curve is made for the same parameters as in Fig.\ \ref{fig:HC2q}. We can see that this dependence has several salient features. The $H_{C2}$ curve has a clear tendency to follow one of the miniband-edge field lines and it sharply turns away from this line close to crossing points with the second miniband-edge field line. As a consequence, the regions of multiple $H_{C2}$ values corresponding to reentrant behavior appear below some of these crossing points. Near the crossing of  $0\!\downarrow\!-t_z$ and $0\!\uparrow\!+t_z$ lines corresponding to the matching of the opposite van Hove singularities for the zero Landau level, the region of the alternating FFLO state is realized. 

So far, we only considered the purely interband-coupling scenario for which the quantum effects from the shallow band are quite pronounced. The behavior, however, is sensitive to the structure of the coupling-constants matrix reflecting the pairing mechanism. 
In addition to the interband-coupling case, Fig.\ \ref{fig:HC2q} also shows the $H_{C2}$~-~$\mu_h$ dependences at low temperatures for two coupling-constant sets with finite $\Lambda_{ee}$. We can see that with increasing deep-band pairing weight, the upper critical field \emph{decreases} and moves to the region of miniband-edge field lines for higher Landau levels, especially for '$+t_z$' lines corresponding to larger-area cross section of the Fermi surface at $k_z\!=\!\pi$. Nevertheless, the main trends remain: the $H_{C2}$ curves still tend to follow the miniband-edge field lines and the reentrant regions appear near the crossing points (at least, for the case $\Lambda_{ee}=0.2$ and $\Lambda_{eh}=0.3$). Note that the reentrance completely vanishes when the deep-band coupling is too strong (the lowest curve in Fig.\ \ref{fig:HC2q}). The ground state at low temperature is modulated for all studied cases, but the optimal modulation wave vector $Q_{\mathrm{op}}$  progressively decreases with increasing the deep-band weight. 
The middle curve includes the crossing of $0\!\downarrow\!-t_z$ and $1\!\uparrow\!+t_z$ lines,  where the shallow band strongly favors the alternating state. This state, however, is not formed due to the large weight of the deep band. The modulation wave vector is sharply enhanced when $\mu_h$ approaches the crossing-point value $1.125T_C$ but it only reaches $\sim 1.1$ at the maximum.
\begin{figure}[htbp] 
	\centering
	\includegraphics[width=3.2in]{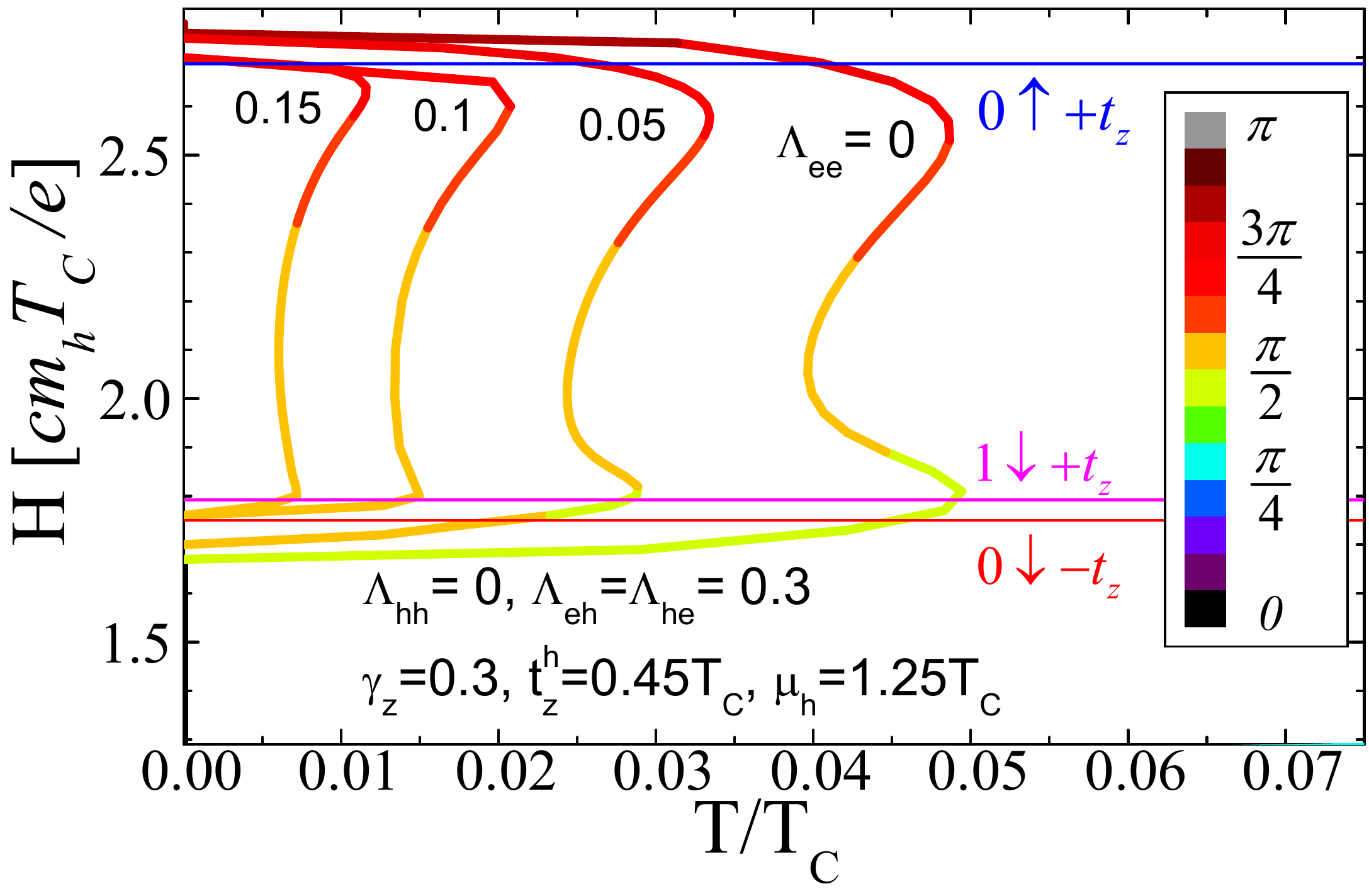}
	\vspace{-0.1in}
	\caption{Shrinking of the reentrant region with increasing deep-band coupling constant $\Lambda_{ee}$. All parameters except $\Lambda_{ee}$ are the same as in Fig.\ \ref{fig:HC2q}(b).}
	\label{fig:Reenr-Lee}
\end{figure}

The temperature range of reentrant states rapidly reduces with increasing deep-band coupling.
We illustrate this trend in Fig.\ \ref{fig:Reenr-Lee} in which we present the evolution of the reentrant region  of Fig.\ \ref{fig:HC2q}(b) with increasing deep-band coupling constant $\Lambda_{ee}$. We can see that the size of this region rapidly shrinks with increasing $\Lambda_{ee}$. 
This counterintuitive behavior is again caused by the reduction of the shallow-band weight in the pairing. In the regime of dominating deep-band coupling, $\Lambda_{ee}>\Lambda_{eh}$, the reentrant region becomes practically invisible for these electronic-spectrum parameters.
 
We discuss in Appendix \ref{app:DeepScen} the scenario in which superconductivity is dominated by the deep-band coupling with $\Lambda_{ee}>\Lambda_{he}$. The resulting $H$-$T$ diagrams are not qualitatively different from the representative case in Figs.\ \ref{fig:HC2R} and \ref{fig:HC2q}, but there are substantial quantitative differences. In the resonant case, the sizes of reentrant regions are much smaller than for the interband-coupling case due to the strong reduction of the shallow-band contribution to the pairing (see also Fig.\ \ref{fig:TC2}). In the nonresonant case, the reentrant behavior only appears either for smaller interlayer tunneling or smaller Fermi energy of the deep band in comparison with the parameters considered in this section. Nevertheless, the FFLO instability in the main superconducting region always appears at low temperatures.

\section{Summary and discussion}\label{Sec:Summary}

In summary, we have investigated the pairing instabilities of a clean two-band layered superconductor in the magnetic field oriented perpendicular to the layers.  
In this system, the interlayer tunneling lifts the degeneracies in all Landau levels and transforms them into the dispersive minibands along the out-of-plane momentum ($k_z$) with two van Hove singularities at $k_z\!=\!0$ and $\pi$. We explored the possible pairing instabilities as the Landau minibands cross the chemical potential in the vicinity of the first Lifshitz transition ($\mu_h\gtrsim 2t_z^h$ in Fig. \ref{fig:LT}).

Solving the linearized gap equation [Eq.\ \eqref{eqn:gap-eqn}], we found that the Landau quantization leads to strong $1/\sqrt{T}$ divergences in the shallow band's pairing kernel if either the same [Eq.\ \eqref{eqn:ResCond}] or opposite [Eqs.\ (\ref{eqn:pi-wc} and \ref{eqn:pi-cond})] van Hove points match at the chemical potential. The former matching condition gives the \textit{uniform} superconducting state and the latter one favors the \textit{alternating} FFLO state with $Q_z\!=\!\pi$ modulation. For general Zeeman spin-splitting energy, the pairing kernel has $\ln T$ divergence in the pairing channel with the optimal wave vector $Q_\text{op}$ [Eq.\ \eqref{eq:Qop}] corresponding to the difference between the spin-up and spin-down Fermi momenta.
These logarithmic divergences are similar to the FFLO instabilities in quasi-one-dimensional superconductors \cite{Suzumura:PoThPhys70.1983,Machida:PRB30.1984,Matsuda:JPSJ76.2007} and they strongly promote the formation of the FFLO states with $Q_z=Q_\text{op}$. 

Furthermore, we studied the magnetic field versus temperature phase diagrams and found that the shallow-band divergences yield a complex reentrant behavior in the high-magnetic-field superconductivity. The properties of these reentrant states are highly sensitive to the shallow-band parameters, $\mu_h$, $t_z^h$, $\gamma_z$, as well as to the coupling-matrix structure. In general, the interlayer tunneling smears the Landau-level densities of states which reduces the quantum effects in the resonant cases. Therefore, the reentrant transition temperatures are smaller than those in the two-dimensional monolayer\cite{Song:PRB95.2017}. On the other hand, the interlayer tunneling enables the formation of the FFLO modulations which mitigate the Zeeman suppression in the reentrant states. As a result, the reentrant region is stretched to a wider magnetic-field range [see Figs. \ref{fig:HC2q}(a) and (b)]. Furthermore, we also found the reentrant \textit{alternating}-FFLO states [see Fig. \ref{fig:HC2q}(c)] due to the interplay between the Landau quantization and interlayer tunneling. 
The similar \textit{alternating}-FFLO state has been also suggested for the layered system in the magnetic field applied parallel to the layers\cite{Budzin:PRL95.2005}. The formation of these \textit{alternating}-FFLO states is the consequence of the compensation between the Zeeman energy and the interlayer hopping energy.

In this study, we only performed the linear stability analysis and, strictly speaking, the computed phase lines describe instability of the normal state. These lines would correspond to true second-order phase transitions only if the coefficient for the quartic order-parameter term in the free energy is positive. Otherwise, superconducting state will emerge via a first-order transition.
Such first-order scenario for the emergence of the FFLO state indeed realizes in quasi-two-dimensional materials with very strong Zeeman effect for the magnetic field perpendicular to conducting layers \cite{Zhuravlev:PRB80.2009}. This analysis, however, has been performed only for the quasiclassical regime and it does not take into account the effect we stress here, promotion of the FFLO instability by the Landau quantization. A full nonlinear consideration with proper accounting for the quantum effects remains to be done. 

The described behavior can only be observed in very clean materials in which the scattering time $\tau$ satisfies the condition $\tau \omega_c \gg 1$. This condition implies that superconducting transition should take place in the region of noticeable quantum magnetic oscillations of magnetization and conductivity (de Haas--van Alphen and Shubnikov--de Haas effects). This is obviously the most stringent requirement for observation of the described anomalous behavior of the upper critical field.
We expect that impurity scattering suppresses such behavior similar to the thermal noise. Indeed, in the interpretation of the quantum-oscillation phenomena scattering is frequently accounted for by introducing a fictitious temperature proportional to the scattering rate, so-called Dingle temperature.  

We only considered the simplest nearest-neighbor interlayer tunneling process. Additional strong pairing instability may arise if new van Hove singularities appear in the Landau band due to the complicated interlayer tunneling processes. Furthermore, for sufficiently strong Zeeman splitting and/or small Fermi surfaces for all bands, the higher-Landau-level gap eigenfunction may be relevant \cite{Buzdin:EPL35.1996,*Buzdin:PLettA218.1996}. Influence of the quantization effects on this gap solution and modification of the reentrant behavior are other interesting topics for future study.

The quantum effects considered in this paper may be relevant to the high-magnetic-field behavior of several iron-based superconductors, in particular, for FeSe and LiFeAs single crystals. The simplest compound FeSe has the transition temperature $\sim 8$K\cite{Medvedev:NatMat8.2009} and the low-temperature upper critical field $\sim17$T\cite{Kasahara:PNAS111.2014,Terashima:PRB90.2014}. The material can be made clean allowing for observation of quantum oscillations at fields $> 20$T \cite{Terashima:PRB90.2014,Watson:PRB91.2015,AudouardEPL2015}. Its band structure is composed of hole and electron bands with rather small Fermi surfaces. In particular, Shubnikov--de Haas oscillations show that the smallest Fermi energy for the electron and hole bands are only 3.9 and 5.4 meV, respectively \cite{Terashima:PRB90.2014} (the first energy corresponds to the ratio $\epsilon_F/\omega_c\sim 4$ at
$H_{c2}$). Similar Fermi energies are also observed by ARPES \cite{Shimojima:PRB90.2014,Nakayama:PRL113.2014,Watson:PRB91.2015,FedorovSciRep2016}. Moreover, the additional field-induced transition inside the superconducting state has been observed in this compound near 13 T by thermal conductivity measurement \cite{Kasahara:PNAS111.2014}. The plausible intriguing interpretation of this transition is the onset of the FFLO modulation along the magnetic field \cite{AdachiJPSJ15}. Such explanation is also consistent with the low-temperature upturn of the upper critical field. If this interpretation is correct, then this FFLO transition is likely influenced by quantum effects. Furthermore, the transition temperature in FeSe can be increased by pressure up to 20 K at 25 kbar and the upper critical field increases above 35 T at 14 kbar. As a consequence, at high pressures, the superconducting transition in the magnetic field takes place in the region of pronounced quantum oscillations \cite{TerashimaPhysRevB93.2016}. As for the most shallow band $\epsilon_F/\omega_c\sim 2$ at $H_{C2}$ in this region, it is clear that the quantum effects strongly influence the superconducting instability.

The 111 compound LiFeAs is one of the few stoichiometric iron-based superconductors with transition temperature $\sim18$ K\cite{Zhang:PRB80.2009} and zero-temperature $c$-axis $H_{C2}$ $\sim24$T\cite{ChoPhysRevB.83.060502,Khim:PRB84.2011}.
The quantum oscillations also have been reported for this material in Refs.\ \cite{PutzkePhysRevLett2012,Zeng:PRB88.2013}. 
It has the hole-like shallow band located at the zone center with 
closed Fermi surface and tiny Fermi energy $\mu_h\sim3$ meV \cite{Zeng:PRB88.2013}. It is important to note that this band actually has the largest superconducting gap \cite{Miao:NatComm6.2015}, meaning strong participation in the formation of the superconducting state.
In addition, this band can be completely depleted by small Co doping \cite{Miao:NatComm6.2015}. The band with so small Fermi energy definitely should cause noticeable quantum effects near the upper critical field if the material can be made sufficiently clean. 
The quantitative predictions for a particular material require a detailed knowledge of the electronic spectrum, spin-splitting factors for all bands, and coupling matrix.

The high-field reentrant superconductivity has been reliably found at least in one material, Eu-doped Chevrel phases, Eu$_x$Sn$_{1\!-\!x}$Mo$_6$S$_8$ \cite{Meul:PhysRevLett1984,*Rossel:JAP1985}, where an isolated semi-elliptical superconducting region has been observed in the $H$-$T$ diagram for $T<1$ K and a very wide magnetic field range, 4T$<\!H\!<$22.5T. The interpretation of this strong reentrance was based on the Jaccarino-Peter effect\cite{Jaccarino:PhysRevLett1962}, the compensation of the Zeeman spin splitting due to the interaction with the local magnetic moments of the $Eu$ dopants. However, such interpretation also required an assumption of the extremely weak orbital effect of the magnetic field (a huge Maki parameter $\alpha_M\approx 4.8$ was used in the theoretical fits). The reason for this assumption is not very clear,  as the material is isotropic. We cannot, therefore, exclude that the quantization effects may also play a role in the formation of the reentrant region.

The newly discovered Dirac/Weyl semimetals \cite{Weng:PRX5.2015,Huang:NatComm6.2015,Xu:Science349.2015,Lv:PRX5.2015,Yang:NatPhys11.2015} (see also review \cite{Armitage:RMP90.2018}) with very small and tunable Fermi energies are another interesting systems to search for the high magnetic field induced superconductivity. Of particular interests to this paper are the superconducting states that are found in some compounds under high pressure\cite{Zhou:PNAS113.2016,Liu:SciRep7.2017,Qi:PRB94.2016,Li:NPJ2.2017}. It has been demonstrated in Ref.\ \onlinecite{Rosenstein:PRB96.2017} that the orbital quantization in superconducting materials with such Dirac-type spectrum can also lead to the reentrant superconductivity in high magnetic fields. Furthermore, the similar reentrant behavior has also been found in the theoretical study on the Dirac-type surface states of a topological insulator \cite{Zhuravlev:EPL120.2017}. In contrast to the conventional quadratic electronic dispersion, the linearly-dispersive band yields nonuniform Landau-level spacing and it has been argued that this intriguing feature makes the quantum limit much easier to attain \cite{Rosenstein:PRB96.2017}.
In addition to this feature, due to the small cyclotron effective mass near the Dirac point, the Landau quantization is more robust against disorders \cite{Zhuravlev:EPL120.2017,Zhuravlev:PRB95.2017}. 
So far, the above studies only considered the Landau-quantization effects in the uniform states. The possible enhancement of the FFLO instabilities in the three-dimensional Dirac/Weyl semimetal due to quantum effects is an interesting problem to be explored in the future. 

Finally, we note that the study in this paper only concentrated on the correlation effects in the BCS pairing channels. For real systems, other correlation effects in spin and charge channels are also likely to play an important role. Particularly, the Landau-level miniband is effectively a one-dimensional system, and the spin and charge fluctuations may have significant influence. Indeed, the strong competition between the pairing and density-wave instabilities may have dramatic effects and give rise to a new class of strongly correlated quantum states\cite{Yakovenko:PRB47.1993}. The rich strongly correlated phenomena due to the interplay between these fluctuations in quantum limit are certainly important problems for future consideration.

\begin{acknowledgements}
The authors would like to thank A. Buzdin, K. Matveev, T. Shibauchi, L. Balicas, Y. Kopelevich, and A. Gurevich for useful discussions. The work was supported by the U.S. Department of Energy, Office of Science, Materials Sciences and Engineering Division. K.W.S. was supported by the Center for Emergent Superconductivity, an Energy Frontier Research Center funded by the U.S. DOE, Office of Science, under Award No. DEAC0298CH1088.
\end{acknowledgements}

\appendix

\section{Calculation of the kernel eigenvalues}\label{App:kernels}
In this appendix, we give details of the calculations of the kernels in zero and finite magnetic field.

\subsection{Zero magnetic field}
\label{App:kernelsH0}

The kernels, Eq.\ \eqref{KernelDef} are composed by the zero-field Green's function, Eq.\ \eqref{G0}. The effective coupling constant $\Lambda_{0,\alpha}$ [Eq.\ \eqref{Lambda0}] is determined by
\begin{align*}
&\Lambda_{0,\alpha}^{-1}\!=\!\frac{T}{N_\alpha}\sum_{\omega_n,j'}\sum_{\substack{\mathbf{k}\mathbf{k}'\\k_zk_z'}}\int_{\mathbf{r}'}\mathrm{e}^{-\mathrm{i}[(\mathbf{k}-\mathbf{k}')\cdot(\mathrm{r}-\mathrm{r}')+(k_z-k'_z)(j-j')]}\notag\\
\!\!&\times\frac{1}{(\mathrm{i}\omega_n\!-\!\xi^\alpha_\mathbf{k}\!+\!2t_z\cos k_z)(\mathrm{i}\omega_n\!+\!\xi^\alpha_{\mathbf{k}'}\!-\!2t_z\cos k'_z)}.
\end{align*}
Since the system is translationally invariant, it is straightforward to integrate out $\mathbf{r}'$ and sum all $j'$. This yields
\begin{equation*}
\Lambda_{0,\alpha}^{-1}=N_\alpha^{-1}T\sum_{\omega_n}\sum_{\mathbf{k}k_z}\frac{1}{\omega^2_n+(\xi^\alpha_\mathbf{k}-2t_z\cos k_z)^2 }.
\end{equation*}
We further integrate out the $(\mathbf{k},k_z)$ by using $\sum_\mathbf{k}=N_e\int^{\Omega}_{-\Omega}\mathrm{d}\xi^e$ for $e$ band ($\sum_\mathbf{k}=N_h\int^{\mu_h}_{-\Omega}\mathrm{d}\xi^h$ for $h$ band) with $N_e=\frac{m_e}{2\pi}$ ($N_h=\frac{m_h}{2\pi}$) and $\sum_{k_z}=\int^{\pi}_{-\pi}\frac{\mathrm{d}k_z}{2\pi}$. We therefore obtain
\begin{align*}
&\Lambda_{0,e}^{-1}
\!=\!\sum_{\omega_n}\int^{\pi}_{-\pi}\frac{\mathrm{d}k_z}{2\pi}\left[\frac{T}{\omega_n}\tan^{-1}\frac{x}{\omega_n}\right]^{x=\Omega}_{x=-\Omega},\\
&\Lambda_{0,h}^{-1}\!=\!\sum_{\omega_n}\int^{\pi}_{-\pi}\frac{\mathrm{d}k_z}{2\pi}\left[\frac{T}{\omega_n}\tan^{-1}\frac{x}{\omega_n}\right]^{x=\mu_h-2t_z\cos k_z}_{x=-\Omega}.
\end{align*}
Note that we assumed $\Omega\gg t_z$. Further simplifying the above expressions, we obtain Eqs.\ \eqref{eqn:L0e} and \eqref{eqn:L0h} in the main text.

\subsection{Finite magnetic field}
\label{App:kernelsFinH}

\subsubsection{Deep band}
\label{App:kernelsFinHDeep}

For the deep $e$ band in which the Landau quantization does not play a role, we have the following quasiclassical result for the kernel eigenvalue:
\begin{align}\label{eqn:le0}
\lambda^e_{\omega_n,Q_z}\!=&\!\frac{2}{N_e}\sum_{\mathbf{k}\mathbf{k}'k_z}\!\int\limits_0^\infty\!\frac{\rho\mathrm{d}\rho\exp\left[-\mathrm{i}(\mathbf{k}-\mathbf{k}')\cdot\bm{\rho}-\frac{\rho^2}{2l^2}\right]}{[\mathrm{i}\omega_n\!-\!\mu_zH\!-\!\xi^e_{\mathbf{k}}\!+\!2t_z^e\cos (k_z\!-\frac{1}{2}Q_z)]}\notag\\
&\times\frac{1}
{[\mathrm{i}\omega_n\!-\!\mu_zH\!+\!\xi^e_{\mathbf{k}'}\!-\!2t_z^e\cos (k_z\!+\!\frac{1}{2}Q_z)]}.
\end{align}
Setting $\mathbf{k}\to\mathbf{k}-\frac{1}{2}\mathbf{q}$ and $\mathbf{k}'\to\mathbf{k}+\frac{1}{2}\mathbf{q}$, and keeping only the linear order in $\mathbf{q}$, we can approximate the dispersion near Fermi level as
$\xi^e_\mathbf{k}\to\xi^e_\mathbf{k}-\tfrac{1}{2}\bm{v}^e\cdot\mathbf{q}$  and $\xi^e_{\mathbf{k}'}\to\xi^e_\mathbf{k}+\tfrac{1}{2}\bm{v}^e\cdot\mathbf{q}$.
We now integrate out $\mathbf{k}$ in Eq.\ \eqref{eqn:le0} with the above approximation by using $\sum_\mathbf{k}=N_e\int^{\infty}_{-\infty}\mathrm{d}\xi^e$. Extending the energy integration to a closed contour in the complex plane, this yields
\begin{align*}
&\lambda^e_{\omega_n,Q_z}\!\approx\!-\pi \mathrm{i}\sum_{k_z\mathbf{q}}
\!\int\limits_0^\infty\!\rho\mathrm{d}\rho\exp\left[-\mathrm{i}\mathbf{q}\cdot\bm{\rho}-\frac{\rho^2}{2l^2}\right]\Big\{\mathrm{i}\omega_n\!-\!\mu_zH\!\notag\\
&
+\!\tfrac{1}{2}\bm{v}^e\!\cdot\mathbf{q}+t_z^e[\cos (k_z\!-\!\frac{1}{2}Q_z)\!-\!\cos (k_z\!+\!\frac{1}{2}Q_z)]\Big\}^{-1}.
\end{align*}
The absolute value of the momentum $\mathbf{k}$ in the Fermi velocity $\bm{v}^e(\mathbf{k})$ is approximately determined by $\xi^e_\mathbf{k}=t_z^e[\cos (k_z\!-\!\frac{1}{2}Q_z)\!+\!\cos (k_z\!+\!\frac{1}{2}Q_z)]$ giving 
\begin{equation}
v^2_e\approx \frac{2}{m_e}\left( \mu+2t_z^e\cos k_z \cos \frac{Q_z}{2}\right).
\label{eq:v2e}
\end{equation}
Exponentiating the denominator as follows,
\begin{align*}
&\lambda^e_{\omega_n,Q_z}\!\approx\!2\pi \!\sum_{k_z\mathbf{q}}\!\int\limits_0^\infty\!\rho\mathrm{d}\rho
\exp\left[-\mathrm{i}\mathbf{q}\cdot\bm{\rho}-\frac{\rho^2}{2l^2}\right]
\!\int\limits_0^\infty\!\mathrm{d}s\exp\Big\{2\zeta_\omega s\!
\notag\\
&\times\Big[\omega_n\!+\!\mathrm{i}[\mu_zH\!-\!\tfrac{1}{2}\bm{v}^e\cdot\mathbf{q}\!-\!2t_z^e\sin (k_z)\sin (\tfrac{1}{2}Q_z)]\Big]\Big\},
\end{align*}
with $\zeta_\omega=\text{sign}(\omega_n)$, we integrate out $\mathbf{q}$, which gives
\begin{align*}
\lambda^e_{\omega_n,Q_z}\!&\approx\!2\pi \!\sum_{k_z}\!\int\limits_0^\infty\rho\mathrm{d}\rho\!\int\limits_0^\infty\!\mathrm{d}s
\,\delta(\bm{\rho}+\zeta_\omega s\bm{v}^e)\exp\left[-\frac{\rho^2}{2l^2}\right]
\notag\\
&\times\exp\Big\{2\zeta_\omega s\!\Big[\omega_n\!+\!\mathrm{i}[\mu_zH\!-\!2t_z^e\sin (k_z)\sin (\tfrac{1}{2}Q_z)]\Big]\Big\}.
\end{align*}
Finally, taking the $\rho$ integral, we obtain the eigenvalue of the kernel 
\begin{equation}\label{eqn:le-ve}
\lambda^e_{\omega_n,Q_z}\!=\!2\!\int\limits^\infty_0\!\mathrm{d}s\left\langle\exp\Big[\!-\!2\zeta_\omega s(\omega_n\!+\!\mathrm{i}\tilde{\gamma}_z^e)\!-\!\frac{v^2_es^2}{2l^2}\Big]\right\rangle_z\!,
\end{equation}
where $\tilde{\gamma}_z^e=\gamma_z-2t_z^e\sin k_z\sin\frac{Q_z}{2}$.
Substituting the Fermi velocity from Eq.\ \eqref{eq:v2e}, we obtain Eq.\ \eqref{eqn:le}.

\subsubsection{Shallow band}
\label{App:kernelsFinHSh}

In the case of shallow band, the quantum kernel eigenvalue $\mathcal{J}_{1}$ is defined
by Eqs.\ \eqref{lh} and \eqref{Jh}. The summation over the Matsubara frequencies in Eq.
\eqref{lh} can be performed using the relation
$T\!\sum_{\omega_{n}}\!\frac{1}{(i\omega_{n}\!+z)(i\omega_{n}\!-z')}\!
=\!-\frac{1}{2}\left(\tanh\frac{z}{2T}\!+\!\tanh\frac{z'}{2T}\right)\!/(z\!+\!z')$, 
which gives
\begin{align*}
\pi T\!\sum_{\omega_{n}\!=\!-\infty}^{\infty}\!\lambda_{\omega_{n}}^{h}&=\frac{1}{4}\sum_{m=0}^{M_{\Omega}}\sum_{\ell=0}^{m}\frac{m!}{2^{m}\left(m\!-\!\ell\right)!\ell!}\\
&\times\left\langle \frac{\mathcal{T}(\ell\!+\!\tilde{\gamma}_{z}-\!\tilde{\mu}_{h})+\mathcal{T}(m\!-\!\ell\!-\!\tilde{\gamma}_{z}-\!\tilde{\mu}_{h})}{m+1-2\tilde{\mu}_{h}}\right\rangle _{z},
\end{align*}
where $\ensuremath{\mathcal{T}(x)\!\equiv\!\tanh[\omega_{c}(x\!+1/2)/2T]}$ and the
functions $\tilde{\mu}_{h}(k_z,Q_z)$ and $\tilde{\gamma}_{z}(k_z,Q_z)$ are defined by
Eqs.\ \eqref{tmh} and \eqref{tg} correspondingly. Here, we introduced a new summation index
$m=\ell+\ell'$ and cut off the diverging sum over $m$ at $M_{\Omega}=2\Omega/\omega_{c}$.
The converging function $\mathcal{J}_{1}$ is obtained by subtracting the zero-field limit
of this sum at $Q_{z}=0$ which yields
\begin{widetext}
\begin{equation}
\mathcal{J}_{1}\!=\frac{1}{4}\!\sum_{m=0}^{M_{\Omega}}\sum_{\ell=0}^{m}\frac{m!}{2^{m}\left(m\!-\!\ell\right)!\ell!}\left\langle \frac{\mathcal{T}(\ell+\tilde{\gamma}_{z}-\!\tilde{\mu}_{h})+\mathcal{T}(m\!-\!\ell-\tilde{\gamma}_{z}-\!\tilde{\mu}_{h})}{m+1-2\tilde{\mu}_{h}}\right\rangle _{\!z}\!
-\frac{1}{2}\int\limits _{0}^{M_{\Omega}}dx\left\langle \frac{\mathcal{T}(\frac{x-1}{2}-\tilde{\mu}_{h0})}{x-2\tilde{\mu}_{h0}}\right\rangle _{\!z}
\label{J1cut}
\end{equation}
\end{widetext}
 with $\tilde{\mu}_{h0}(k_{z})\equiv\tilde{\mu}_{h}(k_{z},0)=\bar{\mu}_{h}-2\bar{t}_{z}^{h}\cos k_{z}$.
This quantity remains finite in the limit $\Omega\to\infty$. Note that the subtracted term can also be represented as
\begin{align}
&\frac{1}{2}\int\limits _{0}^{M_{\Omega}}dx\left\langle \frac{\mathcal{T}(\frac{x-1}{2}-\tilde{\mu}_{h0})}{x-2\tilde{\mu}_{h0}}\right\rangle _{\!z} =\pi T\!\sum_{0<\omega_{n}<\Omega}\!\frac{1}{\omega_{n}}\!+\!\Upsilon_{T}\nonumber \\
&\approx\frac{1}{2}\ln\frac{\mathit{A}\Omega}{T}+\!\Upsilon_{T}.
\label{SubTerm}
\end{align}
To derive a presentation better suited for numerical evaluation, we split the integration
over $x$ as $\int _{0}^{\infty}dxF(x)=\int
_{0}^{1/2}dxF(x)\!+\!\sum_{m=0}^{\infty}\int_{-1/2}^{1/2}dxF(m\!+\!1\!+\!x)$ and subtract the
term
$\left\langle \frac{\mathcal{T}(\frac{m}{2}-\tilde{\mu}_{h})}{m+1-2\tilde{\mu}_{h}}\right\rangle _{z}$ from both $m$ sums making them converging independently. This results in Eq.\ \eqref{J1llSumPres} of the main text.

\section{Low-temperature asymptotics of $\mathcal{J}_1(H,T,Q_z)$}
\label{app:LowTJ1}

In this appendix, we analyze the low-temperature behavior of the function $\mathcal{J}_1(H,T,Q_z)$. In most cases it behaves $\propto \ln(T)$ which gives the finite zero-temperature limit of the total kernel eigenvalue $\mathcal{J}_1-\mathcal{A}_1$. This zero-temperature limiting value has square-root singularities when magnetic field crosses the miniband-edge fields [Eq.\ \eqref{wc-vH}]. In addition,
in several resonance cases $\mathcal{J}_1$ diverges as $1/\sqrt{T}$. Such behavior may be realized only in the uniform state with $Q_z\!=\!0$ or in the alternating FFLO state with $Q_z\!=\!\pi$ and for special values of the spin-splitting factor $\gamma_z$.

\subsection{Derivation of the pairing kernel near the miniband-edge fields }
\label{app:J1sqrtCross}

In this appendix, we derive the square-root contribution to the pairing kernel eigenvalue $\mathcal{J}_{1}$, Eq.\ \eqref{eq:SqRJ1}, which appears when the chemical potential enters the Landau-level miniband at $H\!=\!H_{\ell_{0},\sigma,\delta_t}$ [see Eq.\ \eqref{wc-vH}]. The singular behavior is coming from the terms with $\ell=\ell_{0}$ for $\sigma=1$ ($m-\ell=\ell_{0}$ for $\sigma=-1$) in the Landau-level sum, Eq.\ \eqref{J1cut}. The corresponding singular term can be written as
\[
r_{m,\ell_{0},\sigma}(\omega_{c})\!=\!\left\langle\! \frac{\tanh\Big[\frac{\omega_{c}(\ell_{0}-\sigma\gamma_{z}\!+\!\frac{1}{2})-\!\mu_{h}+2t_{z}^{h}\cos(k_{z}-\tfrac{1}{2}Q_{z})}{2T}\Big]}{m\!+\!1\!-\!2\left(\mu_{h}-2t_{z}^{h}\cos k_{z}\cos\frac{Q_{z}}{2}\right)/\omega_{c}}\!\right\rangle _{z}.
\]
At $T\to0$ we can replace $\tanh(A/T)$ with the sign function $\mathrm{sign}(A)$.
For $\omega_{c}$ near $\omega_{\ell_{0},\sigma,\delta_{t}}$, we
use the presentation 
\begin{align*}
&\omega_{c}(\ell_{0}-\sigma\gamma_{z}\!+\!\frac{1}{2})-\!\mu_{h}+2t_{z}^{h}\cos\left(k_{z}-\tfrac{Q_{z}}{2}\right)\\
&=\left(\mu_{h}\!+\!2\delta_{t}t_{z}^{h}\right)\left(\frac{\omega_{c}}{\omega_{\ell_{0},\sigma,\delta_{t}}}\!-\!1\right)\!+\!2t_{z}^{h}\left(\delta_{t}\!+\!\cos\left(k_{z}\!-\!\tfrac{Q_{z}}{2}\right)\right),
\end{align*}
which indicates that the main contribution to the difference $r_{m,\ell_{0},\sigma}(\omega_{c})-r_{m,\ell_{0},\sigma}(\omega_{\ell_{0},\sigma,\delta_{t}})$
comes from the miniband-edge region (i.e., near $k_{z}=\pi+\tfrac{Q_{z}}{2}$
for $\delta_{t}=1$ and $k_{z}=\tfrac{Q_{z}}{2}$ for $\delta_{t}=-1$),
where the quadratic expansion of the cosine can be used. This allows us
to evaluate
\begin{align*}
&r_{m,\ell_{0},\sigma}(\omega_{c})-r_{m,\ell_{0},\sigma}(\omega_{\ell_{0},\sigma,\delta_{t}})\\
&  \approx\frac{-2\delta_{t}/\pi}{m\!+\!1\!-\!2\left(\mu_{h}\!+\!2\delta_{t}t_{z}^{h}\cos^{2}\tfrac{Q_{z}}{2}\right)/\omega_{c}}\\
&\times  \sqrt{\frac{\mu_{h}\!+\!2\delta_{t}t_{z}^{h}}{t_{z}^{h}}}\sqrt{\left|1\!-\!\frac{\omega_{c}}{\omega_{\ell_{0},\sigma,\delta_{t}}}\right|}\theta\left[\delta_{t}\!\left(\!1\!-\!\frac{\omega_{c}}{\omega_{\ell_{0},\sigma,\delta_{t}}}\right)\right],
\end{align*}
where $\theta\left(x\right)$ is step function. We see that $r_{m,\ell_{0},\sigma}(\omega_{c})$
has a singular square-root behavior when $\omega_{c}$ approaches
$\omega_{\ell_{0},\sigma,\delta_{t}}$ from the side, at which the
chemical potential is inside the miniband. Collecting the singular
terms, we obtain the corresponding result for the kernel eigenvalue
\begin{align*}
&\mathcal{J}_{1}(\omega_{c})\!-\mathcal{J}_{1}(\omega_{\ell_{0},\sigma,\delta_{t}})\\
&\approx\frac{1}{4}\!\sum_{m=\ell_{0}}^{\infty}\frac{m!}{2^{m}\left(m\!-\!\ell_{0}\right)!\ell_{0}!}\left[r_{m,\ell_{0},\sigma}(\omega_{c})-r_{m,\ell_{0},\sigma}(\omega_{\ell_{0},\sigma,\delta_{t}})\right].
\end{align*}
Substituting the difference $r_{m,\ell_{0},\sigma}(\omega_{c})-r_{m,\ell_{0},\sigma}(\omega_{\ell_{0},\sigma,\delta_{t}})$ from the previous equation, we arrive at Eq.\ \eqref{eq:SqRJ1} of the main text.

\subsection{Uniform state ($Q_z=0$) and  $2\gamma_z=j_z$}
\label{App:J1Q0smallT}

We analyze the low-temperature divergence of the function $\mathcal{J}_1(H,T,Q_z)$ in the
uniform case when the spin-splitting energy matches the Landau-level separation, i.\ e.,
$\gamma_z\!=\!j_z/2$. In this case, a singular behavior takes place when the chemical potential
matches the Landau level energy at $k_z\!=\!0$ or $\pi$ corresponding to the condition
$(\ell_0+\frac{1}{2}j_z+\frac{1}{2})\omega_c=\mu_h\mp2t_z^h$. To extract the leading
low-$T$ divergence, we will use Eqs.\ \eqref{J1cut} and \eqref{SubTerm}. First, near $T=0$, the
second term of Eq.\ \eqref{J1cut} diverges as $\sim\ln T$. The dominating  low-$T$ divergent contribution
is coming from the first term with $m=2\ell_0+j_z$ and $\ell=\ell_0$,
\begin{align}
\mathcal{J}_1  \simeq &\frac{\omega_c}{8t_z^h}
\frac{(2\ell_0+j_z)!}{2^{2\ell_0+j_z}(\ell_0+j_z)!\ell_0!}
\int^\pi_{0}\frac{\mathrm{d}k_z}{\pi}
\frac{\tanh\frac{t_z^h(\cos k_z\mp1)}{T}
	}{\cos k_z\mp1}.
\end{align}
Making the substitution 
$x=\frac{t_z^h}{T}(\cos k_z\mp1)$, we obtain
\begin{align}\label{J100}
&\mathcal{J}_1\simeq\frac{\omega_c}{8\pi\sqrt{Tt_z^h}}\frac{(2\ell_0+j_z)!}{2^{2\ell_0+j_z}(\ell_0+j_z)!\ell_0!}\nonumber\\
\times&\int^{2t_z^h/T}_{0}\!\mathrm{d}x
\frac{\tanh x
	}{x^{3/2}\sqrt{2-Tx/t_z^h}}.
\end{align}
In the limit $T\ll t_z^h$, using
$\int_{0}^{\infty}\mathrm{d}xx^{-3/2}\tanh x=\sqrt{2}(2\sqrt{2}-1)\zeta(3/2)/\sqrt{\pi}$, 
we obtain Eq.\ \eqref{eqn:J1Resonant} in the main text.

We derive a more accurate asymptotics for the strongest resonance, $\ell_0=0$ and $j_z=0$. 
Substituting $\omega_{c}=2(\mu_{h}\mp 2t_{z})$ for $k_z=0$/$\pi$ and $Q_{z}=0$ into the Eq.\ \eqref{J1cut}, we obtain 
\begin{align}
\mathcal{J}_{1}\! & =\frac{1}{2}\!\sum_{m=0}^{2\Omega/\omega_{c}}\sum_{\ell=0}^{m}\frac{m!}{2^{m}\left(m\!-\!\ell\right)!\ell!}\left\langle \frac{\tanh\frac{\ell\!-2\bar{t}_{z}^h\left(\pm1-\cos k_{z}\right)}{2T/\omega_{c}}}{m-4\bar{t}_{z}^h\left(\pm1-\cos k_{z}\right)}\right\rangle _{z}\nonumber\\
-&\frac{1}{2}\ln\frac{\mathit{A}\Omega}{T}-\Upsilon_{T}.\label{J1LLSum-App-1}
\end{align}
At low temperatures we can replace $\tanh \to 1$ in all terms except $m=0$. 
This gives,
\begin{align*}
\mathcal{J}_{1}\!  
\simeq & \frac{1}{2}\!\left\langle\! \frac{\tanh\frac{t_{z}^h\left(1\mp\cos k_{z}\right)}{T}}{4\bar{t}_{z}^h(1\mp\cos k_{z})}\!+\!\!\sum_{m=1}^{\infty}\!\!\left[\!\frac{1}{m\!+\!4\bar{t}_{z}^h(\cos k_{z}\!\mp\!1)}\!-\!\frac{1}{m}\right]\!\right\rangle _{\!\!z}\!\\
&+ \frac{1}{2}\left(\!\sum_{m=1}^{2\Omega/\omega_{c}}\frac{1}{m}-\ln\frac{\mathit{A}\Omega}{ T}\right)\!-\!\Upsilon_{T}.
\end{align*}

Using  $\lim_{N\to\infty}\left(\sum_{m=1}^{N}\frac{1}{m}-\ln N\right)=\gamma_{\mathrm E}$ and performing averaging with respect to $k_z$, we finally obtain
\begin{align}\label{eq:J100Ap}
\mathcal{J}_{1} & \approx\frac{\mathit{C}\omega_{c}}{\sqrt{t_{z}^hT}}\!+\!\mathcal{R}_{\mp}(\bar{t}_{z}^h)\!-\frac{1}{2}\ln\frac{\omega_{c}}{\pi T}\!-\!\Upsilon_{T},\\
\mathcal{R}_{\mp}(\bar{t}_{z}^h) & =\frac{1}{2}\!\sum_{m=1}^{\infty}\left(\frac{1}{\sqrt{m\left(m\mp8\bar{t}_{z}^h\right)}}-\frac{1}{m}\right),\nonumber
\end{align}
where $\mathit{C}\!=\!(2\sqrt{2}\!-\!1)\zeta(\tfrac{3}{2})/(8\pi^{3/2})\!\approx\! 0.1072$ and $\zeta(x)$ is the Riemann zeta function. Subtracting $\mathcal{A}_{1}$ gives  Eq.\ \eqref{eqn:J1T0R} of the main text.

\subsection{Alternating state ($Q_{z}=\pi$)}
\label{App-J1pi}

For the alternating case, $Q_{z}=\pi$, the resonance conditions are given
by Eqs.\ \eqref{eqn:pi-wc} and \eqref{eqn:pi-cond} corresponding
to $\bar{t}_{z}^{h}=\frac{1}{4}\left(\ell_{\pi}-\ell_{0}\pm2\gamma_{z}\right)$
and $\bar{\mu}_{h}=\frac{1}{2}\left(\ell_{0}+\ell_{\pi}+1\right)$.
Substituting these relations into the first term of Eq.\ \eqref{J1cut}
and using $\tilde{\mu}_{h}(k_{z},\pi)=\bar{\mu}_{h}$, $\tilde{\gamma}_{z}(k_{z},\pi)=\gamma_{z}-2\bar{t}_{z}^{h}\sin k_{z}$,
we obtain 
\begin{widetext}
\begin{align*}
& \mathcal{J}_{1}(H,T,\pi)\simeq\frac{1}{4}\!\sum_{m=0}^{\infty}\sum_{\ell=0}^{m}\frac{m!}{2^{m}\left(m\!-\!\ell\right)!\ell!\left(m\!-\!\ell_{0}\!+\!\ell_{\pi}\right)}\\
& \times\left\langle \tanh\frac{2\ell-\ell_{0}-\ell_{\pi}\pm2\gamma_{z}+\left(\ell_{\pi}-\ell_{0}\pm2\gamma_{z}\right)\sin k_{z}}{4T/\omega_{c}}\!+\!\tanh\frac{2\left(m-\ell\right)-\ell_{0}-\ell_{\pi}\mp2\gamma_{z}-\left(\ell_{\pi}-\ell_{0}\pm2\gamma_{z}\right)\sin k_{z}}{4T/\omega_{c}}\right\rangle _{z}.
\end{align*}
\end{widetext}
The singular term $m=\ell_{0}+\ell_{\pi}$ requires resolution of
a ``zero over zero'' uncertainty which leads to 
\begin{align*}
&\mathcal{J}_{1}(H,T,\pi)\simeq\frac{\omega_{c}}{8T}\!\sum_{\ell=0}^{\ell_{0}+\ell_{\pi}}\frac{\left(\ell_{0}+\ell_{\pi}\right)!}{2^{\ell_{0}\!+\!\ell_{\pi}}\left(\ell_{0}\!+\!\ell_{\pi}\!-\!\ell\right)!\ell!}\\
&\times\left\langle \mathrm{sech}^{2}\frac{\ell_{0}\!+\!\ell_{\pi}\!-\!2\ell\!\mp\!2\gamma_{z}\!-\!\left(\ell_{\pi}\!-\!\ell_{0}\!\pm\!2\gamma_{z}\right)\sin k_{z}}{4T/\omega_{c}}\right\rangle _{z}.
\end{align*}
The terms for which $\left|\ell_{0}+\ell_{\pi}-2\ell\mp2\gamma_{z}\right|=\left|\ell_{\pi}-\ell_{0}\pm2\gamma_{z}\right|$
are divergent. This condition is always satisfied for $\ell=\ell_{\pi}$
and this term is
\[
\mathcal{J}_{1}(H,T,\pi)\!\simeq\!\frac{\omega_{c}}{16\pi T}\!\frac{\left(\ell_{0}+\ell_{\pi}\right)!}{2^{\ell_{0}+\ell_{\pi}}\ell_{0}!\ell_{\pi}!}\int\limits_{-\pi}^{\pi}\!\mathrm{d}k_{z}\mathrm{sech}^{2}\frac{t_{z}^{h}\left(1\!+\!\sin k_{z}\right)}{T}.
\]
At low temperatures, $T\ll t_{z}^{h}$, the dominating contribution
to the integral comes from the region near the inflection point $k_{z}=-\pi/2$
and the integration can be approximately evaluated using the substitution
$k_{z}\!=\!-\pi/2\!+\!\sqrt{2Tx/t_{z}^{h}}$, giving 
\begin{align*}
\int_{-\pi}^{\pi}\mathrm{d}k_{z}\mathrm{sech}^{2}\frac{t_{z}^{h}\left(1+\sin k_{z}\right)}{T} & \approx\sqrt{\frac{2T}{t_{z}^{h}}}\int_{0}^{\infty}\frac{\mathrm{d}x}{\sqrt{x}\cosh^{2}x}\\&=\sqrt{\frac{2T}{t_{z}^{h}}}\frac{(2\sqrt{2}-1)\zeta(\frac{3}{2})}{\sqrt{2\pi}}.
\end{align*}
Therefore, we obtain
\begin{equation}
\mathcal{J}_{1}(H,T,\pi)\simeq\frac{\mathit{C}}{2}\!\frac{\left(\ell_{0}+\ell_{\pi}\right)!}{2^{\ell_{0}+\ell_{\pi}}\ell_{0}!\ell_{\pi}!}\frac{\omega_{c}}{\sqrt{Tt_{z}^{h}}},
\end{equation}
where the constant $\mathit{C}$ is defined after Eq.\ \eqref{eq:J100Ap}. We also mention that in the exceptional cases when $2\gamma_{z}$ is integer,  the term with $\ell=\ell_{0}-2\gamma_{z}$ also diverges and this gives additional diverging contribution
\begin{align*}
\mathcal{J}_{1}(H,T,\pi) & \simeq\frac{\mathit{C}}{2}\!\frac{\left(\ell_{0}+\ell_{\pi}\right)!}{2^{\ell_{0}+\ell_{\pi}}\left(\ell_{\pi}\!+\!2\gamma_{z}\right)!\left(\ell_{0}\!-\!2\gamma_{z}\right)!}\frac{\omega_{c}}{\sqrt{Tt_{z}^{h}}}.
\end{align*}

We again derive a more accurate asymptotic for the strongest resonance of this kind with $t_{z}^{h}/\mu_{h}\!=\!\gamma_{z}$ and $\omega_{c}\!=\!2\mu_{h}$. Near the first Lifshitz transition, $\mu_{h}>2t_{z}^{h}$ which implies that $\gamma_{z}\!<\!0.5$. For these relations, the function $\mathcal{J}_{1}$ becomes
\begin{align}
&\mathcal{J}_{1}\!  =\frac{1}{4}\!\sum_{m=0}^{M_{\Omega}}\sum_{\ell=0}^{m}\frac{m!}{2^{m}\left(m\!-\!\ell\right)!\ell!m}\left\langle \tanh\frac{\ell\!+\!\gamma_z\left(1\!+\!\sin k_{z}\right)}{2T/\omega_{c}}\right.\notag\\
&\left.+\!\tanh\frac{m\!-\!\ell\!-\!\gamma_{z}\left(1\!+\!\sin k_{z}\right)}{2T/\omega_{c}}\right\rangle _{z}\!\! -\frac{1}{2}\ln\frac{\mathit{A}\Omega}{T}-\!\Upsilon_{T}.
\end{align}
Note that the $m\!=\!0$ term again has ``zero over zero'' uncertainty. 
All terms in the sum remain finite for $T\to0$
except the $m=\ell=0$ term. Therefore, we separate this diverging
term from the sum and take the limit $T\to0$ for the rest of the
terms. This yields 
\begin{align}
\mathcal{J}_{1}\!  \simeq&\frac{\omega_{c}}{8T}\left\langle \mathrm{sech}^{2}\frac{\gamma_{z}\left(1+\sin k_{z}\right)}{2T/\omega_{c}}\right\rangle _{z}\nonumber\\
&-\frac{1}{2}\!\sum_{m=1}^{\infty}\frac{1}{2^{m}m}+\frac{1}{2}\!\sum_{m=1}^{M_{\Omega}}\frac{1}{m}-\frac{1}{2}\ln\frac{\mathit{A}\Omega}{ T}-\!\Upsilon_{T}\nonumber \\
 \approx&\frac{\mathit{C}}{2}\frac{\omega_{c}}{\sqrt{Tt_{z}^{h}}}-\frac{1}{2}\ln\frac{2\omega_{c}}{\pi T}-\!\Upsilon_{T}.
\end{align}
This corresponds to Eq.\ \eqref{eqn:J1T0pi} in the main text.

\section{The phase diagrams for the deep-band dominating scenario}\label{app:DeepScen}
\begin{figure}[htbp] 
	\centering
	\includegraphics[width=3.2in]{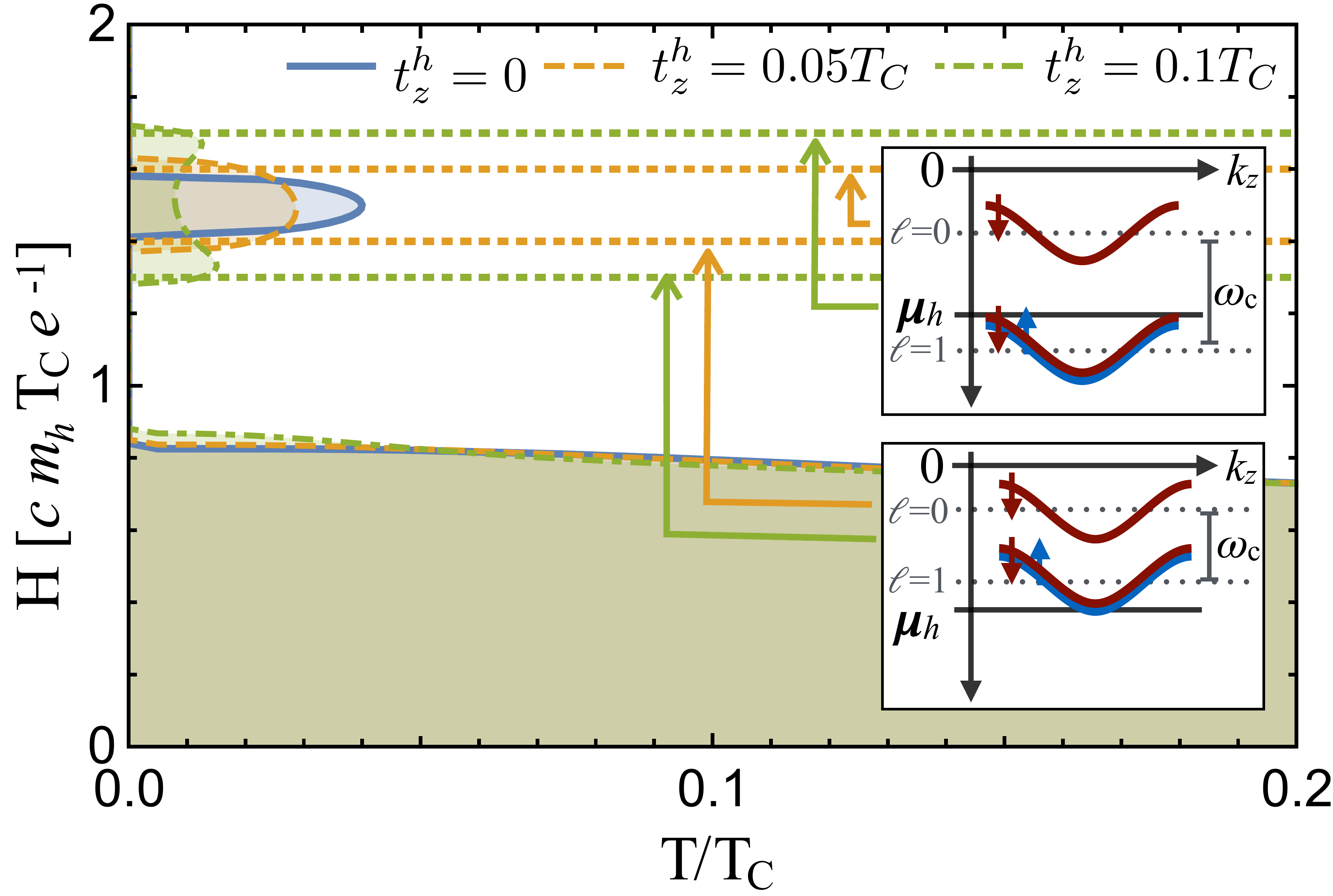} 
	\caption{The representative $H$-$T$ diagram for the  resonant case, $\gamma_z\!=\!0.5$, within the deep-band coupling dominating scenario. 
		We have used the coupling constants $\Lambda_{ee}\!=\!0.3$, $\Lambda_{he}\!=\!\Lambda_{eh}\!=\!0.2$, and $\Lambda_{hh}\!=\!0$. Other parameters are $\mu_h\!=\!1.5T_C$, $\mu\!=\!10.5T_C$, and $m_e\!=\!m_h$. 
		The interlayer hopping energies $t_z^h$ are shown in the plot.}
	\label{fig:HC2Deep}
\end{figure}
\begin{figure*}[htbp] 
	\centering
	\includegraphics[width=3.2in]{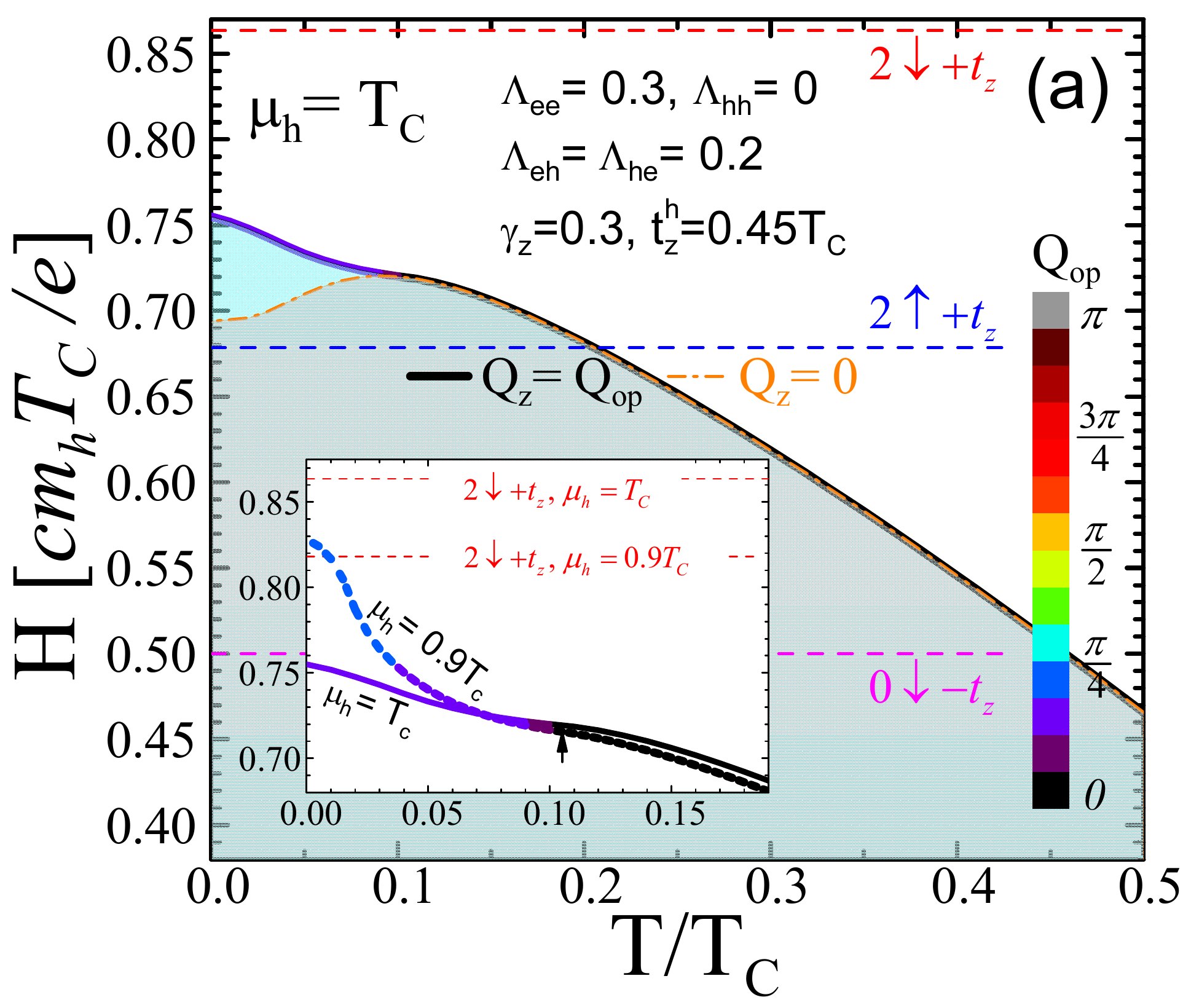}
	\includegraphics[width=3.2in]{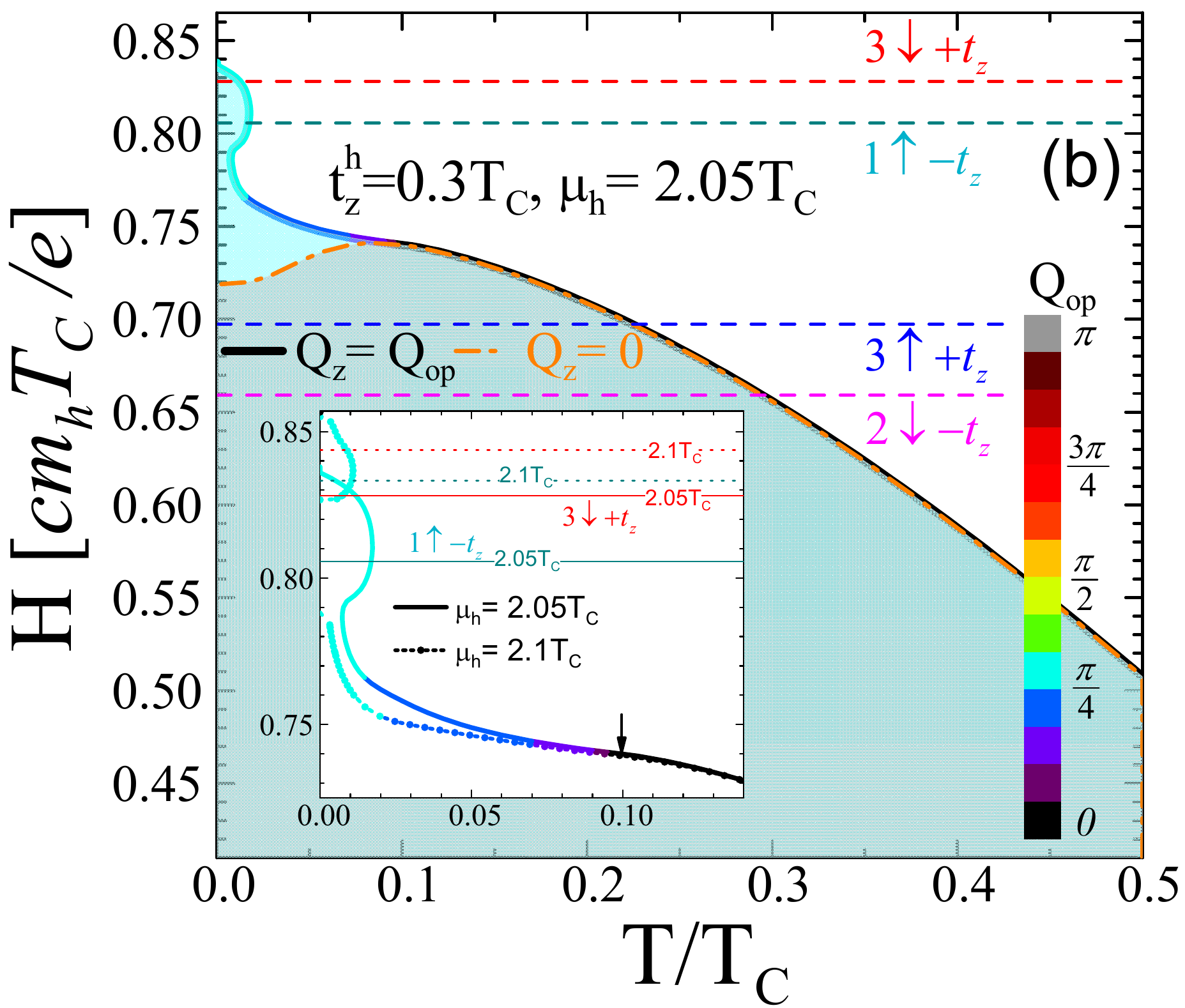} 
	\vspace{-0.1in}
	\caption{The magnetic field--temperature phase diagrams for the non-resonant case, $\gamma_z=0.3$, and dominating deep-band coupling, $\Lambda_{ee}\!=\!0.3>\Lambda_{eh}\!=\!\Lambda_{he}\!=\!0.2$. (a)The diagram for the same spectrum parameters as in Fig.\ \ref{fig:HC2q}(a). The inset zooms into the low-temperature region for the same Fermi energy as in the main plot and for a smaller value, $\mu_h=0.9 T_{C}$. In the latter case, the $H_{C2}$ line has a pronounced upturn at low temperatures caused by the proximity to the van Hove singularity. (b) The example of the $H$-$T$ phase diagrams with reentrant regions for the same parameters as in the previous figure except smaller $t_z^h=0.3T_C$ and different $\mu_h$. The inset shows the low-temperature region for two values of $\mu_h$ for which the reentrant region is connected with and separated from the main domain.}
	\label{fig:H-TDeepgz03}
	\vspace{-0.1in}
\end{figure*}

In the main text, we mostly presented results for the interband-coupling scenario ($\Lambda_{ee}\!=\!\Lambda_{hh}\!=\!0$), for which the shallow band is essential for the formation of the superconducting state and the quantization effects caused by this band are very pronounced. In this appendix, we consider the phase diagrams for the alternative scenario when the deep band itself has strong pairing strength and the interband coupling induces superconductivity into the shallow band, i.~e., $\Lambda_{ee}>\Lambda_{he},\Lambda_{eh},\Lambda_{hh}$. In such scenario, the Lifshitz transition weakly affects superconductivity as is indeed realized in several iron-based superconductors. On the other hand, the influence of the shallow band is only noticeable if it has substantial contribution to pairing, i.~e., the interband coupling constants  $\Lambda_{he}$ and $\Lambda_{eh}$ should be comparable with $\Lambda_{ee}$. Note that we focus in this paper on the scenario when there is no intraband pairing in the shallow band, $\Lambda_{hh}\!=\!0$. Such pairing would further enhance shallow-band quantization effects. For illustration, we use the coupling constants $\Lambda_{ee}\!=\!0.3$ and $\Lambda_{he}\!=\!\Lambda_{eh}\!=\!0.2$ in this appendix. This corresponds to the ratio $\Lambda_{he}/\Lambda_{0,e} \sim 0.54$ in Fig.\ \ref{fig:TC2}.

We consider the resonant case first. Figure \ref{fig:HC2Deep} shows the representative $H$-$T$ diagrams for $\gamma_z\!=\!0.5$. The diagrams are presented for three values of the interlayer hopping energies $t_z^h$ shown in the plot. We can see that the overall behavior is similar to the interband-coupling scenario shown in Fig.\ \ref{fig:HC2R}. The quantitative difference is that the size of the reentrant region is rather small even in the 2D case and it further diminishes with increasing the interlayer tunneling. Already for small interlayer hopping energy, $t_z^h\!=\!0.1T_C$, the maximum $T_{C2}$ drops below $0.02 T_{C}$.

In a more generic nonresonant case (noninteger $2\gamma_z$), two features of the $H$-$T$ diagrams have been emphasized in the main text: (i)the interlayer FFLO instability accompanied by the enhancement of the upper critical field at low temperatures and (ii)the emergence of reentrant states for the magnetic fields slightly below the crossings of two miniband-edge field lines.
We find that the FFLO instability in the main superconducting region is robust. For example, Fig.\ \ref{fig:H-TDeepgz03}(a) shows the magnetic field -- temperature phase diagram for the same electronic parameters as in Fig.\ \ref{fig:HC2q}(a) but for large pairing strength in the deep band 
(the same coupling constants $\Lambda_{ee}\!=\!0.3$, $\Lambda_{he}\!=\!\Lambda_{eh}\!=\!0.2$, and $\Lambda_{hh}\!=\!0$).
For these parameters, the shallow band induces the FFLO instability at $T\approx 0.1T_{C}$ accompanied by the upturn of the upper critical field at lower temperatures. This upturn is much smaller than the one for the interband-coupling scenario presented in Fig.\ \ref{fig:HC2q}(a). It is enhanced, however, when the miniband-edge field approaches the zero-temperature $H_{C2}$. This can be seen from the inset which also shows the low-temperature behavior of the $H_{C2}$ line for the smaller Fermi energy $\mu_h\!=\!0.9T_C$, where
the $H_{C2}$ line approaches the miniband-edge field ``$2\!\downarrow\!+\!t_z$'' at $T\to 0$.
Such enhancement is only observed for few special values of the Fermi energy, while the behavior illustrated in the main figure is typical. Another difference from the interband-coupling scenario is that the modulation wave vector remains rather small, $Q_{\mathrm{op}}<1$, much smaller than the optimal wave vector favored by the shallow band. 

We find that in the dominating deep-band regime the reentrant behavior is not realized for the electronic parameters used in Fig.\ \ref{fig:HC2q}. Nevertheless, this behavior does appear for smaller interlayer hopping energy and/or smaller Fermi energy of the deep band. For example, Fig.\ \ref{fig:H-TDeepgz03}(b) presents of the phase diagram for $t_z^h=0.3T_C$ and $\mu_h=2.05T_C$. In this case, the reentrant region connected with the main domain exists. As one can see from the inset, it separates at slightly larger $\mu_h$. This region is caused by the closely located miniband-edge fields for $\ell\!=\!3$/spin down/$k_z\!=\!\pi$ and $\ell\!=\!1$/spin up/$k_z\!=\!0$. Such reentrance only exists within narrow range of the Fermi energies.

\begin{figure}[!] 
	\centering
	\includegraphics[width=3.2in]{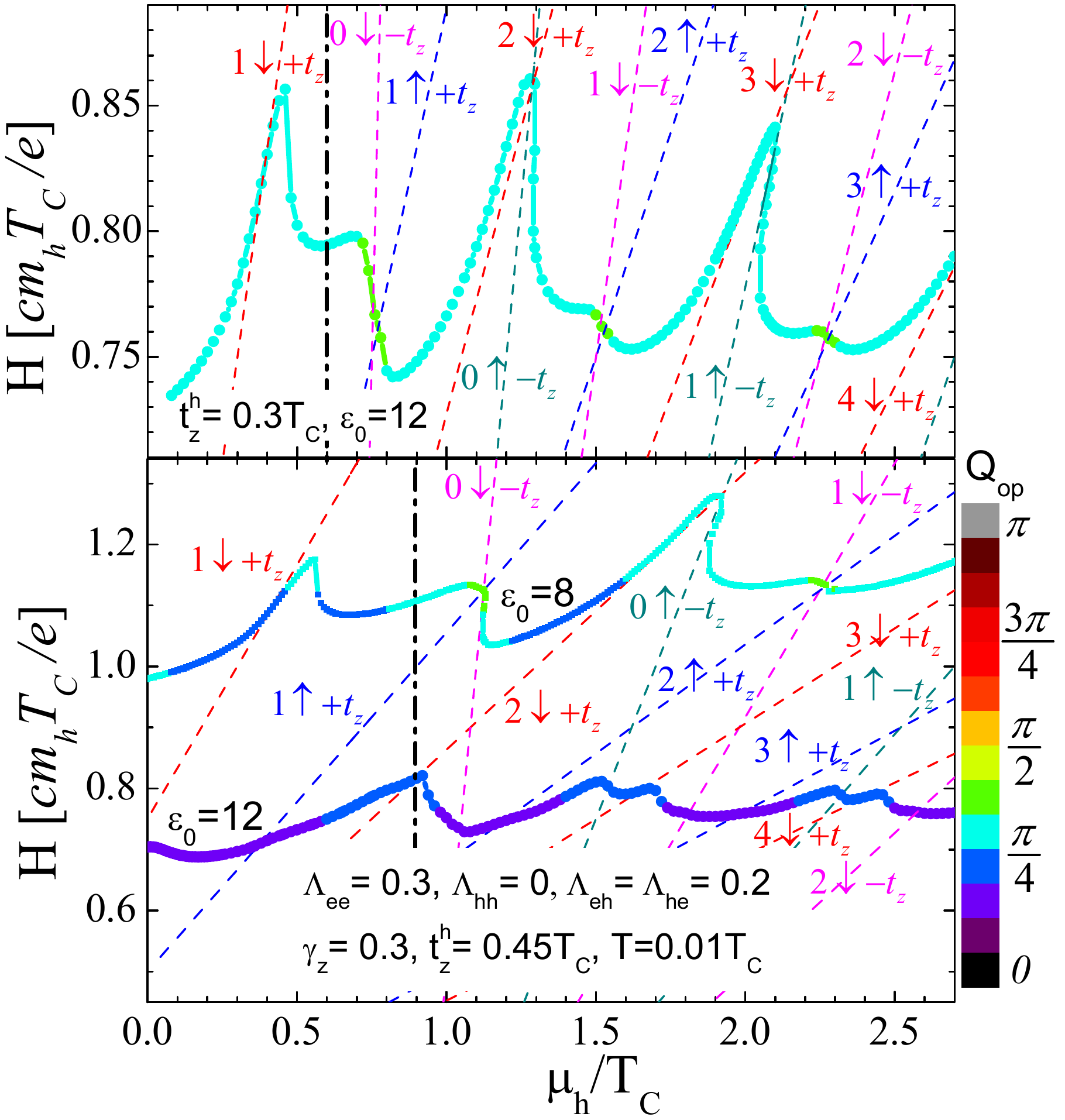} 
	\vspace{-0.1in}
	\caption{The dependences of the upper critical field on the Fermi energy $\mu_h$ for the case of dominating deep-band pairing.
		\textit{The lower plots} are computed for the interlayer hopping energy $t_z^h\!=\!0.45T_C$ and two values of separation between the band edges $\varepsilon_0\!=\!8T_C$ and $12T_C$. \textit{The upper plot} is computed for $t_z^h\!=\!0.3T_C$ and  $\varepsilon_0\!=\!12T_C$. The dashed lines show the miniband-edge fields defined by Eq.\ \eqref{wc-vH}. The vertical dotted-dashed lines mark locations of the first Lifshitz transition at zero magnetic field.}
	\label{fig:Hc2muhLee03Lhe02gz03}
\end{figure}
For better representation of the overall behavior, we show in Fig.\ \ref{fig:Hc2muhLee03Lhe02gz03} the dependences of the low-temperature upper critical field on the Fermi energy $\mu_h$. 
The lower curve in the lower plot is made for parameters $t_z^h\!=\!0.45T_C$ and $\varepsilon_0\!=\!12T_C$ (first set). The same plot is also shown in Fig.\ \ref{fig:Hc2muhtz045gz03}.  There are no reentrant regions for this parameter set. To illustrate the emergence of such regions and other general trends, we also present $H_{C2}(\mu_h)$ dependences for (i) smaller energy separation between the band edges  $\varepsilon_0\!=\!8T_C$ (the upper curve in the lower plot, second set) and (ii) smaller interlayer hopping energy  $t_z^h\!=\!0.3T_C$ (upper plot, third set).
Note that the parameter $\varepsilon_0$ controls the Fermi energy of the deep band $\mu=\varepsilon_0-\mu_h$.
In all cases, the $H_{C2}(\mu_h)$ curves have pronounced oscillations and have a clear tendency to follow the miniband-edge field lines. As we already mentioned, for the first set $H_{C2}(\mu_h)$ is a single-valued function reconfirming the absence of the reentrant behavior. We see, however, that the $H_{C2}(\mu_h)$ line has the pronounced kinklike features near the crossing of the miniband-edge lines. For the Fermi energies at the maxima, the $H_{C2}(T)$ line has a pronounced low-temperature upturn, as illustrated in the inset of Fig.\ \ref{fig:H-TDeepgz03} for $\mu_h=0.9T_{C}$. For the second and third sets multiple solutions for $H_{C2}(\mu_h)$ appear below some crossing points corresponding to existence of the reentrant behavior in these regions. For example, for the $t_z^h=0.3T_C$ (upper plot), such regions are located near $\mu_h\!=\!1.28$ and $2.1T_{C}$.  We also see that for these sets a kinklike feature also exists below the first Lifshitz transition at $\mu_h\!=\!0.9T_C$, in the region of closed Fermi surface, where the $H_{C2}$ curve separates from the ``$1\!\downarrow\!+\!t_z$'' line. It is clear that the reentrant regions will become more pronounced with further decreasing of either $t_z^h$ or $\varepsilon_0$.

\bibliography{2Lifshitz}

\end{document}